\documentclass[a4paper,12pt]{article}
\usepackage{hyperref}
\usepackage[a4paper,left=2.5cm,right=2.5cm,top=3cm,bottom=2.5cm]{geometry}
\usepackage{authblk}
\usepackage{amsmath,bm}
\usepackage{graphicx}
\usepackage{mathrsfs}
\usepackage{amssymb}
\usepackage{physics}
\usepackage{tensor}
\usepackage{xcolor}
\usepackage{tikz}
\usepackage{pgfplots}
\usepackage{amsthm}
\usepackage{booktabs,dcolumn,caption}
\usepackage{multirow}
\usepackage[english]{babel}
\usepackage{soul}

\usepackage[title]{appendix}

\usepackage{xstring}

\theoremstyle{remark}

\title{A novel section-section potential for short-range interactions between plane beams}

\author[1,2]{A. Borković}
\author[1]{M.H. Gfrerer}
\author[3,4,5]{R.A. Sauer}
\author[1]{B. Marussig}
\author[6]{T.Q. Bui}

\affil[1]{Institute of Applied Mechanics, Graz University of Technology, Technikerstraße 4/II, 8010 Graz, Austria, aborkovic@tugraz.at, aleksandar.borkovic@aggf.unibl.org}
\affil[2]{University of Banja Luka, Faculty of Architecture, Civil Engineering and Geodesy, Department of Mechanics and Theory of Structures, 78000 Banja Luka, Bosnia and Herzegovina}
\affil[3]{Institute for Structural Mechanics, Ruhr University Bochum, Universitätsstraße 150, 44801 Bochum }

\affil[4]{Faculty of Civil and Environmental Engineering, Gdansk University of Technology, ul. Narutowicza 11/12, 80-233 Gdansk, Poland} 

\affil[5]{Dept. of Mechanical Engineering, Indian Institute of Technology Guwahati, Assam 781039, India }

\affil[6]{Duy Tan Research Institute for Computational Engineering (DTRICE), Duy Tan University, 6 Tran Nhat Duat, 1 distc, HCMC 700000, Vietnam}
\date{}                     
\begin{document}
	\newcommand{\red}[1]{\textcolor{red}{#1}}

	\newcommand{\ssub}[2]{{#1}_{#2}} 
	\newcommand{\vsub}[2]{\textbf{#1}_{#2}} 
	\newcommand{\ssup}[2]{{#1}^{#2}} 
	\newcommand{\vsup}[2]{\textbf{#1}^{#2}} 
	\newcommand{\ssupsub}[3]{{#1}^{#2}_{#3}} 
	\newcommand{\vsupsub}[3]{\textbf{#1}^{#2}_{#3}} 
	
	\newcommand{\veq}[1]{\bar{\boldsymbol{#1}}} 
	\newcommand{\veqn}[1]{\bar{\textbf{#1}}} 
	\newcommand{\seq}[1]{\bar{#1}} 
	\newcommand{\ve}[1]{\boldsymbol{#1}} 
	\newcommand{\ven}[1]{\textbf{#1}} 
	\newcommand{\vepre}[1]{\boldsymbol{#1}^\sharp} %
	\newcommand{\sdef}[1]{#1^*} 
	\newcommand{\vdef}[1]{{\boldsymbol{#1}}^*} 
	\newcommand{\vdefeq}[1]{{\bar{\boldsymbol{#1}}}^*} 
	\newcommand{\trans}[1]{\boldsymbol{#1}^\mathsf{T}} 
	\newcommand{\transn}[1]{\textbf{#1}^\mathsf{T}} 
	\newcommand{\transmd}[1]{\dot{\boldsymbol{#1}}^\mathsf{T}} 
	\newcommand{\mdvdef}[1]{\dot{\boldsymbol{#1}}^*} 
	\newcommand{\mdsdef}[1]{\dot{#1}^*} 
	\newcommand{\mdv}[1]{\dot{\bm{#1}}} 
	\newcommand{\mdvni}[1]{\dot{\textbf{#1}}} 
	\newcommand{\mds}[1]{\dot{#1}} 
	
	\newcommand{\loc}[1]{\hat{#1}} 
	\newcommand{\iloc}[3]{\hat{#1}^{#2}_{#3}} 
	\newcommand{\ilocmd}[3]{\dot{\hat{#1}}^{#2}_{#3}} 
	\newcommand{\md}[1]{\dot{#1}} 

	\newcommand{\ii}[3]{{#1}^{#2}_{#3}} 
	\newcommand{\iv}[3]{\boldsymbol{#1}^{#2}_{#3}} 
	\newcommand{\idef}[3]{{#1}^{* #2}_{#3}} 
	\newcommand{\ivdef}[3]{\boldsymbol{#1}^{* #2}_{#3}} 
	\newcommand{\ipre}[3]{{#1}^{\sharp #2}_{#3}} 
	\newcommand{\ivpre}[3]{\textbf{#1}^{\sharp #2}_{#3}} 

	\newcommand{\ieq}[3]{\bar{#1}^{#2}_{#3}} 
	\newcommand{\ic}[3]{\tilde{#1}^{#2}_{#3}} 
	\newcommand{\icdef}[3]{\tilde{#1}^{* #2}_{#3}} 
	\newcommand{\icpre}[3]{\tilde{#1}^{\sharp #2}_{#3}} 
	\newcommand{\iveq}[3]{\bar{\textbf{#1}}^{#2}_{#3}} 
	\newcommand{\ieqdef}[3]{\bar{#1}^{* #2}_{#3}} 
	\newcommand{\iveqdef}[3]{\bar{\textbf{#1}}^{* #2}_{#3}} 
	\newcommand{\ieqmddef}[3]{\dot{\bar{#1}}^{* #2}_{#3}} 
	\newcommand{\icmddef}[3]{\dot{\tilde{#1}}^{* #2}_{#3}} 
	\newcommand{\iveqmddef}[3]{\dot{\bar{\textbf{#1}}}^{* #2}_{#3}} 
		\newcommand{\iveqmdddef}[3]{\ddot{\bar{\textbf{#1}}}^{* #2}_{#3}} 
	
	\newcommand{\ieqpre}[3]{\bar{#1}^{\sharp #2}_{#3}} 
	\newcommand{\iveqpre}[3]{\bar{\textbf{#1}}^{\sharp #2}_{#3}} 
	\newcommand{\ieqmdpre}[3]{\dot{\bar{#1}}^{\sharp #2}_{#3}} 
	\newcommand{\icmdpre}[3]{\dot{\tilde{#1}}^{\sharp #2}_{#3}} 
	\newcommand{\iveqmdpre}[3]{\dot{\bar{\textbf{#1}}}^{\sharp #2}_{#3}} 
	
	\newcommand{\ieqmd}[3]{\dot{\bar{#1}}^{#2}_{#3}} 
	\newcommand{\icmd}[3]{\dot{\tilde{#1}}^{#2}_{#3}} 
	\newcommand{\iveqmd}[3]{\dot{\bar{\textbf{#1}}}^{#2}_{#3}} 
	\newcommand{\iveqmdd}[3]{\ddot{\bar{\textbf{#1}}}^{#2}_{#3}} 
	
	\newcommand{\imddef}[3]{\dot{#1}^{* #2}_{#3}} 
	\newcommand{\ivmddef}[3]{\dot{\textbf{#1}}^{* #2}_{#3}} 
		\newcommand{\ivmdddef}[3]{\ddot{\textbf{#1}}^{* #2}_{#3}} 
	
		\newcommand{\imdpre}[3]{\dot{#1}^{\sharp #2}_{#3}} 
	\newcommand{\ivmdpre}[3]{\dot{\textbf{#1}}^{\sharp #2}_{#3}} 
	
	\newcommand{\imd}[3]{\dot{#1}^{#2}_{#3}} 
	\newcommand{\imdd}[3]{\ddot{#1}^{#2}_{#3}} 
	
	\newcommand{\ivmd}[3]{\dot{\textbf{#1}}^{#2}_{#3}} 
	
	\newcommand{\iii}[5]{^{#2}_{#3}{#1}^{#4}_{#5}} 
	\newcommand{\iiv}[5]{^{#2}_{#3}{\boldsymbol{#1}}^{#4}_{#5}} 
	\newcommand{\iivn}[5]{^{#2}_{#3}{\tilde{\boldsymbol{#1}}}^{#4}_{#5}} 
	\newcommand{\iiieq}[5]{^{#2}_{#3}{\bar{#1}}^{#4}_{#5}} 
	\newcommand{\iiieqt}[5]{^{#2}_{#3}{\tilde{#1}}^{#4}_{#5}} 
	
	\newcommand{\eqqref}[1]{Eq.~\eqref{#1}} 
	\newcommand{\fref}[1]{Fig.~\ref{#1}} 

	\maketitle
	
\section*{Abstract}

We derive a novel formulation for the interaction potential between deformable fibers due to short-range fields arising from intermolecular forces. The formulation improves the existing section-section interaction potential law for in-plane beams by considering an offset between interacting cross sections. The new law is asymptotically consistent, which is particularly beneficial for computationally demanding scenarios involving short-range interactions like van der Waals and steric forces. The formulation is implemented within a framework of rotation-free Bernoulli-Euler beams utilizing the  isogeometric paradigm. The improved accuracy of the novel law is confirmed through thorough numerical studies. We apply the developed formulation to investigate the complex behavior observed during peeling and pull-off of elastic fibers interacting via the Lennard-Jones potential.

\textbf{Keywords}: Section-section interaction potential; Lennard-Jones potential; van der Waals forces; steric exclusion; plane Bernoulli-Euler beam; isogeometric analysis

\section{Introduction}

Many observable phenomena are governed by the interactions of physical fields that give rise to the forces between material bodies. Motivated by the desire to model and study the underlying mechanisms defining the form and function of biological and biomimetic materials and structures, the present research deals with the interactions between molecular assemblies that resemble the shape of fibers. Some examples of biological fiber-like macromolecules are proteins such as filamentous actin \cite{2012murrell} and collagen \cite{2021slepukhin}, nucleic acids such as DNA and RNA \cite{2018franquelim}, cellulose \cite{2018nishiyama}, and hyphae \cite{2017islama}. Furthermore, the development of new materials and technologies is often motivated by nature. The technology for the production of composites based on glass fibers \cite{2008alavinasab}, silicon nanotubes \cite{2012yoo}, carbon nanotubes \cite{2021canadija}, and mycelium \cite{2017islama} is constantly developing. Among the many general types of interactions between fibers, important examples are adhesion, peeling, and pull-off. The understanding of these is central to many important applications in coating, bonding, and adhesion technology. One of the main challenges in these applications is the large stress that can occur in a very narrow zone at the peeling front.

Modeling of interactions between fibers is challenging due to the interplay of many forces and the involved time and length scales at the nano- and micro-level. Since experimental and theoretical methods are limited in scope, computational methods have become the main -- and often only -- reliable tool for the analysis of interactions at small scales. Two commonly utilized computational models for the interaction of material bodies are phenomenological (continuum) and physical (molecular) models. The continuum approach considers the interaction between bodies as \emph{mechanical contact} by requiring their impenetrability \cite{2006wriggers}. 

\begin{figure}[h]
	\centering
	\includegraphics[width=\textwidth]{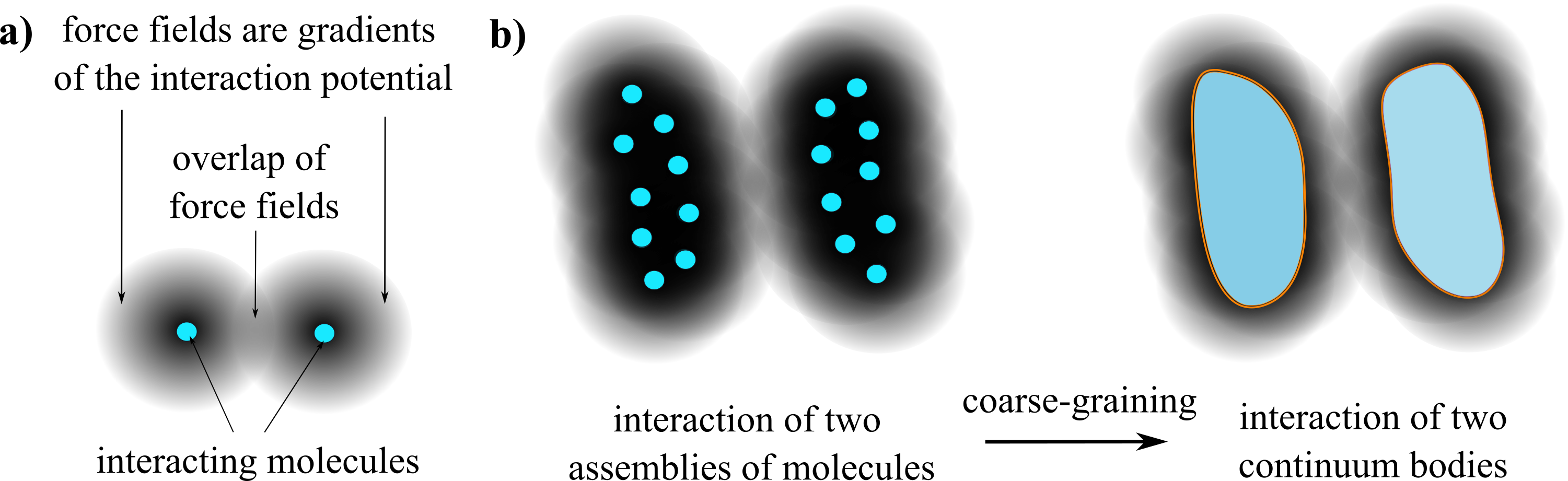}
	\caption{Model of intermolecular interaction. a) Interaction of two molecules via interaction potential. b) Coarse-graining of interaction between molecular assemblies.}
	\label{CG}
\end{figure}

Although suitable for the contact of macroscopic bodies, the continuum approach is not directly applicable to the interactions at nano- and micro-scales where potential-based molecular interaction models are usually preferable \cite{2011israelachvili}. For example, the electrostatic and van der Waals (vdW) interaction potentials give rise to forces that are fundamental for the processes at these small scales. A schematic sketch of a potential-based interaction between two molecules is shown in Fig.~\ref{CG}a. For the assemblies of molecules, a common assumption is that the interaction potentials of all interacting pairs sum up, see Fig.~\ref{CG}b (left).
The computation of interactions by the molecular approach is often based on either Monte Carlo (statistical) or molecular dynamics (application of the laws of motion on each molecule) simulations. A shortcoming of the Monte Carlo approach is that the actual movement of molecules cannot be traced, while molecular dynamics is computationally expensive \cite{2011israelachvili}.

A compromise between the continuum and molecular modeling of interactions at small scales is the \emph{coarse-grained} approach. This model utilizes the physics of molecular interactions with the elegance and numerical efficiency of a continuum formulation, which is achieved by the coarse-graining and homogenization of the molecular model, see Fig.~\ref{CG}b. The main difficulty when modeling potential-based interactions is that its solution yields an integral with respect to (w.r.t.) the volumes of all interacting bodies. 

The first coarse-grained computational model for potential-based interactions between deformable 3D bodies has been developed in \cite{1997argento} by using a generalization of the Derjaguin approximation. The approach has later been successfully extended to the nonlinear setting in \cite{2007sauer, 2008sauer, 2016fan}. The idea is to separate the interactions that occur within the body (intrasolid) and between the bodies (intersolid). This led to the definition of the inter-surface stress tensor that can be used to convert the more involved potential-based contact problem to a conventional boundary value problem. A potential that characterizes surface interactions between two bodies is formulated as a function of a gap vector. The idea is further utilized for the development of an unbiased approach for the interaction of 3D bodies that unifies various contact models \cite{2013sauerb, 2019duonga}. In these references, a 6D integral is reduced to a 4D integral by transforming body-body interactions to surface-surface ones. Additional dimensional reduction is attained in \cite{2007sauer} by assuming that an interacting body can be approximated as a homogeneously deforming flat half-space in the vicinity of the neighboring body.

Further reduction of interaction integral is possible when dealing with slender bodies. The interaction between beams and rigid half-spaces represents a special case since the influence of the rigid half-space can be integrated analytically, and numerical integration is required only w.r.t.~the beam. Nonlinear planar beam-substrate adhesion was considered in \cite{2009sauera, 2010sauer} by assuming that the adhesive forces are acting at the centroid. The extension to thickness-varying adhesion was formulated \cite{2014sauer} and applied to study thin film peeling \cite{2011sauerd} and adhesive shape optimization \cite{2014mergel,2014mergela}.

The first general model for the 2D interactions between two fiber-like macromolecules using the finite element method is given in \cite{2009cyron}. The formulation is extended to study the structural polymorphism, phase transitions, and rheology of 3D biopolymer networks \cite{2013cyron, 2013cyronb, 2014muller}. The first potential-based formulation for the interactions between deformable spatial beams has been developed \cite{2020grill, 2021grill, 2023meier}. In these works the authors introduce the concept of a section-section interaction potential based on \emph{a priory} evaluation of the interaction potential between two cross sections. The analytical integration of a point-pair interaction law over areas of two cross sections allows the reduction of the 6D integral to a 2D one. Due to the inaccuracy of their implementation of the section-section interaction potential for short-range interactions, a section-beam approach is developed by the same group of authors \cite{2023grill, 2022grill}, practically reducing the 6D integral to 1D.

The computational modeling of potential-based interactions between curved deformable fibers is challenging and time consuming. The present contribution aims to derive a coarse-grained computational model that is more accurate than the existing ones. It enables the extension of this efficient concept to applications such as the adhesion, peeling, snapping, and buckling of fibers due to intermolecular forces. 

The contribution of the paper is twofold: First, we are pointing out that, for planar beams, the inconsistency of the existing section-section interaction potential law \cite{2020grill} can be attributed to an offset between cross-sectional planes. To tackle this issue, we derive a new improved section-section interaction potential law and prove that it provides an accurate scaling for short-range interactions. Second, the implementation issues of this new law are thoroughly investigated, and a new formulation providing a good balance between accuracy and efficiency is proposed. Numerical experiments confirm our theoretical findings and give detailed insight into the interaction of deformable fibers due to the Lennard-Jones potential.

The remainder of the paper is structured as follows. Origins and types of interaction potentials, along with appropriate assumptions are discussed in the next section. Section \ref{besec} briefly presents the utilized beam model. It is followed by the main section \ref{secsec} where the new section-section interaction potential law is derived and verified. The weak form of equilibrium is considered in Section \ref{secvar}, which is followed by thorough numerical considerations and appropriate conclusions.

\section{Intermolecular interactions between bodies}

Before discussing details of the computational formulation, a brief overview of the origin and types of intermolecular interactions is given. It is accompanied by appropriate definitions and assumptions.

\subsection{Intermolecular vs.~intramolecular}

The present research examines \emph{intermolecular interactions} between assemblies of molecules, which are treated as solid bodies in our study. It is important to differentiate these interactions from \emph{intramolecular} ones, such as various chemical bonds. Intermolecular interactions primarily arise from the electromagnetic force field. One of the most apparent intermolecular interactions is the electrostatic force that is observed in everyday life between charged bodies. These interactions are defined at the elementary level as interactions between particles, represented as atoms or molecules.

\subsection{Interaction potential between assemblies of particles}

\label{ipassem}
To formally define an interaction potential, let us observe two particles that are at infinite separation, and therefore do not interact. An interaction potential $\hat{\Phi}(r)$ is the amount of energy required to move these particles to the finite distance $r$. It is often modeled as an inverse power law w.r.t.~the distance $r$,
\begin{equation}
	\label{eq: ip01}
	\begin{aligned}
		\hat{\Phi}_m \left(r\right) = k_m \, r^{-m},
	\end{aligned}
\end{equation}
where $k_m$ is a physical constant \cite{2011israelachvili}. The gradient of $\hat{\Phi}_m$ w.r.t.~the position gives a pair of forces that act on the two particles, and have the same magnitudes but opposite directions. 

Due to the reciprocal dependence on the distance, there is a fundamental difference between the interaction potentials w.r.t.~the exponent $m$. Let us assume that the interaction potential between two assemblies of particles, which we will consider as 3D bodies, is obtained by the summation of \eqqref{eq: ip01} over all point-pairs of interacting bodies. Without losing generality, let us consider the special case of interaction between two spherical bodies for which the analytical solutions are available. It can be shown that, for $m>3$, the potential between the two bodies is a function of the gap between their surfaces, since the closest point-pairs dominate the interaction. On the other hand, for $m \le 3$, the influence of all the point-pairs is of a similar order and the interaction is a function of the distance between the spheres' centers \cite{2020grill}. For this reason, we refer to the potentials with $m\le 3$ as \emph{long-range potentials} and those with $m>3$ as \emph{short-range potentials}. This classification allows us to tailor our formulations to a problem at hand. For example, the electrostatic and the gravitational fields belong to the long-range potentials. The focus of the present research is on short-range fields, such as vdW and steric potentials. Although not the subject of this paper, the final aim is to define interaction laws that are valid for all ranges.

\subsection{Lennard-Jones potential}

While the existence of electrostatic fields requires charged bodies, vdW and steric interactions exist for practically all bodies, making them one of the most common forces in nature.
Actually, the origin of vdW forces is essentially electrostatic, arising from the fluctuating dipole field that occurs in basically every molecule. It can be shown that the vdW potential is well-approximated by the inverse-sixth power law.
Although vdW interactions are generally nonadditive, it has been proven that pairwise summation allows good approximation if the Hamaker constant is appropriately selected \cite{2011israelachvili,2020grill}.  
The vdW force between two equal molecules in a medium is always attractive, while it can be repulsive for other cases.

For the modeling of adhesion due to vdW forces, it is necessary to include repulsive effects in the model. When atoms or molecules come into proximity, a strong repulsive force develops as a result of overlapping electron clouds. We observe this effect as \emph{contact} from a macroscopic point of view. The repulsive force follows from the steric potential that is commonly modeled as an inverse power law with a high exponent. By adopting an exponent of $m=12$ for the steric potential and adding it to the vdW potential we obtain the well-known Lennard-Jones law
\begin{equation}
	\hat{\Phi}_{LJ}(r)=\hat{\Phi}_{6}(r) + \hat{\Phi}_{12}(r) = k_{6} \, r^{-6} + k_{12} \, r^{-12}.
\end{equation}
At large separations, the Lennard-Jones law yields very small attraction. As the separation decreases, the attraction increases, and eventually transforms into repulsion. The important characteristics of the Lennard-Jones potential are (i) the equilibrium distance and (ii) the maximal adhesive force, commonly referred to as the \emph{pull-off} force. These values fundamentally influence the peeling phenomenon considered in Section \ref{numeric}.

The spatial distribution of interacting particles inside the body can vary. For our analysis, it is important that both vdW and steric interactions are distributed over the whole volume of the interacting bodies. For example, this is not the case with electrostatic fields where the interaction particles are located on the body surface.

\subsection{Assumptions and restrictions}

It is necessary to introduce additional restrictions to develop a feasible computational formulation. Analogous to \cite{2020grill}, we will assume that:
\begin{itemize}
	\item There is no redistribution of particles or charges inside the bodies, that is, we are dealing with dielectric or nonconducting materials. Also, there is no flow of particles inside or outside of the considered bodies. This assumption guarantees that an interacting property (such as mass, electric charge, etc.) of elementary volumes does not change during deformation.
	\item To allow a pre-integration of a potential, it is necessary to assume constant density distributions of particles and physical constants over cross-sectional areas.
	\item The total interaction between assemblies of particles is the mere sum of individual pair-interactions. This allows us to apply a \emph{coarse-graining} procedure and focus on interactions between two bodies only. Later, we can generalize the formulation by simply adding other interacting bodies.
\end{itemize}
Furthermore, we will solely consider elastic fibers with constant circular cross sections.

\subsection{Coarse-grained interaction potential between two bodies}

Let us consider the interaction between two assemblies of molecules $a$ and $b$ that occupy volumes $V_a$ and $V_b$, and have particle densities $\beta_a$ and $\beta_b$ at initial configuration. The distance between two interacting particles, with positions $\ve{a}_i$ and $\ve{b}_j$, is $r_{ij}=\norm{\ve{a}_i - \ve{b}_j} $, and the total interaction is assumed as the sum of all point-pair contributions,
\begin{equation}
	\label{eq: ip01sum}
	\begin{aligned}
	\Phi_{m} &=  \sum_{i \in a}^{} \sum_{j \in b}^{} \hat{\Phi}_m \left(r_{ij}\right).
	\end{aligned}
\end{equation}
By applying the coarse-graining procedure \cite{2007sauer}, this interaction between assemblies of molecules can be approximated as a volume integral over both bodies,
\begin{equation}
	\label{eq: ip01x}
	\begin{aligned}
		\Phi_{m} &\approx  \int_{\idef{V}{}{a}} \int_{\idef{V}{}{b}} \beta^*_a \left(\ve{a}\right) \beta^*_b \left(\ve{b}\right) \hat{\Phi}_m \left(r\right) \dd{V^*_a} \dd{V^*_b}= \int_{\idef{V}{}{a}} \int_{\idef{V}{}{b}} \beta^*_a \, \beta^*_b \, k_m \, r^{-m}  \dd{V^*_a} \dd{V^*_b}
	\end{aligned}
\end{equation}
where $V^*_i$ are current volumes while $\beta^*_i$ are current particle volume densities. The quantity $\beta^*_a \beta^*_b k_m r^{-m} \dd{V^*_a} \dd{V^*_b}$ can be considered as a potential between two differential volumes at the current configuration. Since the interacting property of an elementary volume does not change during deformation, we have $\beta_{i} \dd{V} = \beta^*_{i} \dd{V^*_i}$.
This fact allows us to calculate an interaction potential at the current configuration by integrating over the reference volume $V_i$, using the reference particle densities $\beta_{i}$, i.e.
\begin{equation}
	\label{eq: ip01xhh}
	\begin{aligned}
		\int_{\idef{V}{}{a}} \int_{\idef{V}{}{b}} \beta^*_a \,  \beta^*_b \, k_m \, r^{-m}  \dd{V^*_a} \dd{V^*_b} = \int_{V_a} \int_{V_b} \beta_a  \, \beta_b \, k_m \, r^{-m} \dd{V_a} \dd{V_b}.
	\end{aligned}
\end{equation}
Calculating this integral for practical time and space resolutions is extremely time-consuming and often impossible. Therefore, some additional simplifications and restrictions are necessary. Since the subject of our research is fibers, it is reasonable to apply the mechanical model of a beam to describe interactions between these slender bodies.

\section{Bernoulli-Euler beam model}
\label{besec}

This section introduces the beam model developed in \cite{2022borkovic,2022borkovica,2023borkoviće}, which is well-suited to represent arbitrarily curved slender bodies. In particular, this work focuses on in-plane beams. 
In the subsequent notation, lowercase and uppercase boldface letters are used for vectors and tensors or matrices, respectively. An overbar designates quantities at an equidistant line, i.e., a line with a fixed distance to the beam's axis, and an asterisk indicates values of the current configuration.

\subsection{Metric of a beam}
A beam is defined as a body with one dominant dimension. It consists of a beam axis, i.e., an arbitrary smooth curve, and an infinite number of cross sections, i.e., plane figures that are attached at their centroids to a beam axis. Each cross section is assumed to be rigid and perpendicular to the beam axis in all configurations, which is known as the Bernoulli-Euler (BE) hypothesis. These assumptions allow a degeneration of a 3D continuum beam model into an arbitrarily shaped line -- the beam axis. It is convenient to parametrize the axis with both the arc-length coordinate $s$ and a parametric coordinate $\xi$. As a representative example, we will consider the interaction between two beams, $x$ and $y$. In this section, we are exemplary considering the beam $x$, and refer to the beam $y$ at the end. 

\begin{figure}[h]
	\includegraphics[width=\linewidth]{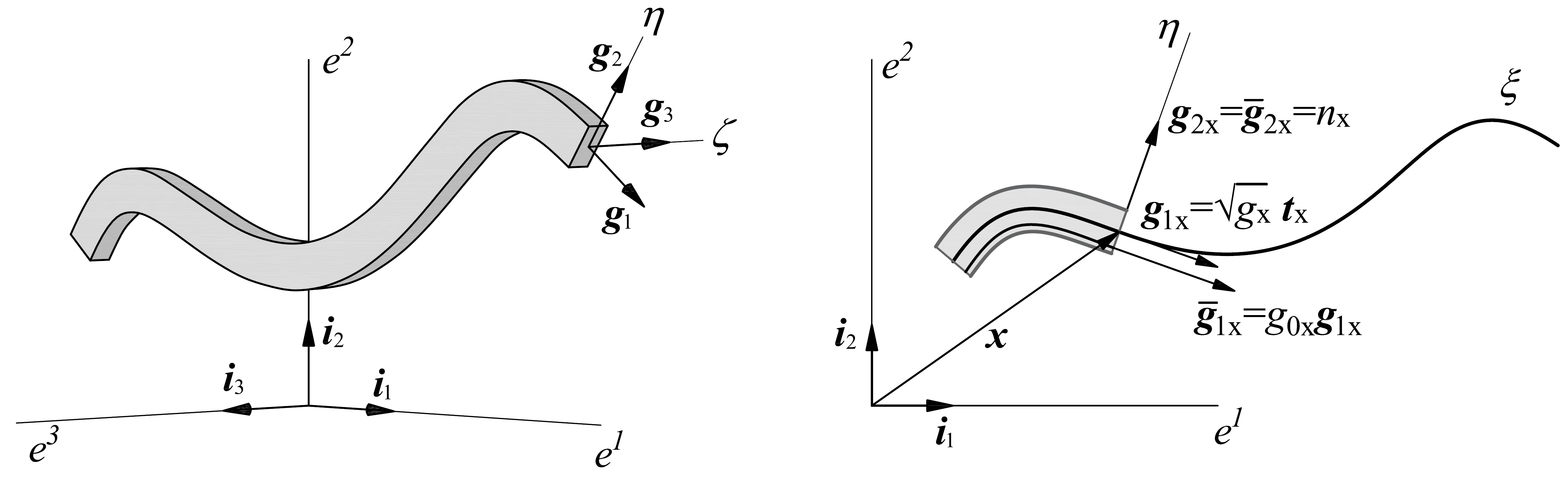}
	\caption{Degeneration of a planar 3D beam into an in-plane 1D beam model. Base vectors at the centroid and at an equidistant line are depicted.}
	\label{fig:Figure 1}
\end{figure}

The position vector of the beam axis in Cartesian coordinates is $\ve{x}=x^\alpha \ve{i}_\alpha$ where $\ve{i}^\alpha = \ve{i}_\alpha$ are the base vectors of the Cartesian coordinate system, see \fref{fig:Figure 1}. 
The tangent base vectors of a beam axis, w.r.t.~parametric and arc-length coordinate are
\begin{equation}
	\begin{aligned}
	\iv{g}{}{1x}&=\frac{\partial \ve{x}}{\partial \xi} = \iv{x}{}{,1}=\ii{x}{\alpha}{,1}\iv{i}{}{\alpha}\,, \\
	\ve{t}_x &=\frac{\partial \ve{x}}{\partial s} = \iv{x}{}{,s}=\iv{g}{}{1x}/\norm{\iv{g}{}{1x}}, \quad \norm{\iv{g}{}{1x}} = \sqrt{\iv{g}{}{1x}\cdot\iv{g}{}{1x}} = \sqrt{g_{11x}} = \sqrt{g_x}.	
\end{aligned}	
\end{equation}
Here, $g_x$ is a component of the metric that relates differentials of the arc-length and parametric coordinate as $\dd{s}=\sqrt{g_x}\dd{\xi}$. 

The other base vector, $\iv{g}{}{2x}=\ve{n}_x$, is perpendicular to $\iv{t}{}{x}$ and has a unit length. It is defined here through an anti-clockwise rotation of the vector $\iv{t}{}{x}$, i.e.
\begin{equation}
	\label{eq: def: g2}
	\iv{g}{}{2x}= \iv{n}{}{x} = \Lambda \iv{t}{}{x} \quad \text{with}  \quad 
	\Lambda =
	\begin{bmatrix}
		0 & -1\\
		1 & 0
	\end{bmatrix}.
\end{equation}
The metric of the beam axis is completely defined by the introduction of the Christoffel symbols,
\begin{equation}
	\label{eq: def: derivatives of base vectors}
	\begin{bmatrix}
		\iv{g}{}{1,1x}\\
		\iv{g}{}{2,1x}
	\end{bmatrix}
	=
	\begin{bmatrix}
		\ii{\Gamma}{1}{11x} & \ii{\Gamma}{2}{11x}\\
		\ii{\Gamma}{1}{21x} & \ii{\Gamma}{2}{21x}
	\end{bmatrix}
	\begin{bmatrix}
		\iv{g}{}{1x}\\
		\iv{g}{}{2x}
	\end{bmatrix}
	=
	\begin{bmatrix}
		\ii{\Gamma}{1}{11x} & \ic{K}{}{x}\\
		-K_x & 0
	\end{bmatrix}
	\begin{bmatrix}
		\iv{g}{}{1x}\\
		\iv{g}{}{2x}
	\end{bmatrix},
\end{equation}
where $K_x$ and $\ic{K}{}{x} = g_x K_x$ are the so-called \textit{signed curvatures} w.r.t.~arc-length and parametric coordinates, respectively.

Since we are dealing with plane beams, all quantities are constant along the $\zeta$ coordinate, as illustrated in \fref{fig:Figure 1}. 
Thus, the degeneration from a 3D to a 2D beam model is straightforward. If we define an \emph{equidistant line} as a set of points with $\eta=const$, its position and base vectors are
\begin{align}
	\label{eq:def:r_eq}
	\begin{aligned}
		\veq{x} &= \ve{x} + \eta \, \ve{g}_{2x}, \\
		\veq{g}_{1x} &= \veq{x}_{,1} = \ve{g}_{1x} - \eta \, K_x \,  \iv{g}{}{1x} = g_{0x} \, \iv{g}{}{1x}, \quad \textnormal{with} \quad g_{0x}=1-\eta \, K_x, \\
		\veq{g}_{2x} &=  \ve{g}_{2x} = \ve{n}_{x}.
	\end{aligned}
\end{align}
In this way, the complete metric of a beam in the reference configuration is defined by the metric of the beam axis. Due to the BE hypothesis, the shear strain vanishes and the position of the cross section is completely determined by the position of the beam axis. Since the current configuration of both beams is obtained by adding the displacement field to the initial position, $\vdef{x}=\ve{x}+\ve{u}$ and $\vdef{y}=\ve{y}+\ve{v}$, the relations given in \eqqref{eq:def:r_eq} are valid for every configuration. This fact gives rise to so-called \emph{rotation-free} beam theories \cite{2018borkovicb}.

\subsection{Strain energy}

The only relevant component of the Green-Lagrange strain tensor for a BE beam is the axial strain
\begin{equation}
	\label{eq:def:strain}
	\ieq{\epsilon}{}{11x} = \frac{1}{2} \left( \ieqdef{g}{}{11x} - \ieq{g}{}{11x} \right).
\end{equation}
By inserting \eqqref{eq:def:r_eq} into \eqqref{eq:def:strain}, we obtain \cite{2023borkoviće}
\begin{equation}
	\label{eq:e11eq}
	\begin{aligned}
		\ieq{\epsilon}{}{11x}&= g_{0x} \left[\left(1-\eta K_x \right) \ii{\epsilon}{}{11x}+ \eta \kappa_x \right] + \eta^2 \chi_x \left(\frac{1}{2} \kappa_x - K_x \ii{\epsilon}{}{11x} \right),
	\end{aligned}
\end{equation}
where
\begin{equation}
	\label{eq:kapa}
	\begin{aligned}
		\ii{\epsilon}{}{11 x} = \frac{1}{2} \left( \idef{g}{}{11 x} - \ii{g}{}{11x } \right), \quad \ii{\kappa}{}{x} =\icdef{K}{}{x} - \ic{K}{}{x}, \quad \ii{\chi}{}{x}=\idef{K}{}{x} - \ii{K}{}{x}. 
	\end{aligned}
\end{equation}
$\ii{\epsilon}{}{11 x}$ is the axial strain of the beam axis, while $\ii{\kappa}{}{x}$ and $\ii{\chi}{}{x}$ are the changes of bending curvatures of the beam axis w.r.t.~the parametric and arc-length convective coordinates, respectively.
The obtained expression \eqref{eq:e11eq} for the axial strain at an arbitrary point shows strong coupling between stretching and bending. The resulting formulation is often referred to as the \emph{strongly curved beam model} \cite{2023borkoviće}.

By utilizing the hyperelastic St.~Venant--Kirchhoff material model, the second Piola-Kirchhoff stress component is
\begin{equation}
	\label{eq:const}
	\begin{aligned}
		\ieq{S}{11}{x} = E \ieq{g}{11}{x} \ieq{g}{11}{x} \ieq{\epsilon}{}{11x}\,,
	\end{aligned}
\end{equation}
where $E$ is the Young's modulus of elasticity. Now, we can define the strain energy of beam $x$ as
\begin{equation}
	\label{eq:internalener}
	\begin{aligned}
		\Phi_{\mathrm{int}x} = \frac{1}{2} \int_{V_x}^{} \ieq{S}{11}{x}  \ieq{\epsilon}{}{11 x} \dd{\ieq{V}{}{x}}.
	\end{aligned}
\end{equation}
In this way, the main ingredients required for the implementation of the nonlinear strongly curved BE beam model are defined. For further details on the variation, spatial discretization and linearization of the equilibrium equation see \cite{2023borkoviće}.

\section{Section-section interaction potential laws}
\label{secsec}

In this section we consider the concept of section-section interaction potentials as introduced in \cite{2020grill,2021grill}. In these papers, the important difference between long- and short-range fields is emphasized, cf. Subsection \ref{ipassem}. Long-range potentials are much easier to deal with due to the small value of the exponent, $m\le 3$, in comparison to short-range ones, $m>3$, which are the focus of this work.  
To improve the accuracy of the existing approach, we propose a new section-section interaction potential law and verify its accuracy via analytical and numerical calculations. 

In order to relax notation, we will remove asterisks in the remainder of the paper since all quantities will be defined at an arbitrary configuration, while the integration is done w.r.t.~the initial configuration.

\subsection{Reduction from 6D to 2D}

Starting from \eqqref{eq: ip01xhh}, we aim to reduce the complexity of solving two nested 3D integrals in order to find an interaction potential between beams $x$ and $y$. As aforementioned, we assume rigid cross sections and constant density distributions over cross-sectional areas. Since $\dd{V}=\dd{A}\dd{s}$, we can simplify the double volume integral into
\begin{equation}
	\label{eq: ip01xy}
	\begin{aligned}
		\Phi_{m} &=  \int_{L_x} \int_{L_y} \int_{A_x} \int_{A_y}\beta_{x} \, \beta_{y} \, \hat{\Phi}_m \dd{A_x} \dd{A_y} \dd{s_x} \dd{s_y}= \int_{L_x} \int_{L_y} \phi_{m,\mathrm{ss}} \dd{s_x} \dd{s_y}, \\
		\phi_{m,\mathrm{ss}} &=\beta_{x} \, \beta_{y} \int_{A_x} \int_{A_y} \hat{\Phi}_m \dd{A_x} \dd{A_y} = 
		\beta_{x} \, \beta_{y} k_m I_{\Phi_m} \left(r\right), \\
		I_{\Phi_m} \left(r\right) &= \int_{A_x} \int_{A_y}  r^{-m} \dd{A_x} \dd{A_y},				
	\end{aligned}
\end{equation}
where $\phi_{m,\mathrm{ss}}$ represents an interaction potential between two cross sections, while $\phi_{m,\mathrm{ss}} \dd{s_x} \dd{s_y}$ is an interaction potential between two differential line segments. Essentially, if $\phi_{m,\mathrm{ss}}$ is integrated analytically, the 6D integral is reduced to 2D and a significant gain in computational efficiency is obtained. This approach is pioneered in \cite{2020grill} and the integral $\phi_{m,\mathrm{ss}}$ is named \emph{section-section interaction potential}. However, the preintegration of $r^{-m}$ over the areas of two cross sections is far from trivial, and it is the main subject of this paper.

 \subsection{Improved section-section interaction potential}
 \label{subexistingSSIP}
 
For two circular disks that belong to the same plane, the integral $I_{\Phi_m}$ in \eqqref{eq: ip01xy} was derived by Langbein \cite{1972langbein} and then used by Grill et al.~\cite{2020grill} as a section-section interaction potential law. We will refer to this law as the LSSIP in the following. Although the case of in-plane circular disks is one of the simplest, the integration is again neither trivial nor exact. Additional insight into this integration approach is shown in Appendix \ref{appendix:a}, while the main expressions of the LSSIP are given in Appendix \ref{appendix:b} for the sake of completeness. 

The LSSIP returns a reasonably accurate approximation for circular cross sections that lie in the same plane. Unfortunately, small change in an offset between the cross-sectional planes changes the interacting forces significantly due to the strong gradients of short-range potentials.
We aim to improve the existing LSSIP by making it explicitly dependent on the cross-sectional offset.

\begin{figure}[h]
	\includegraphics[width=\textwidth]{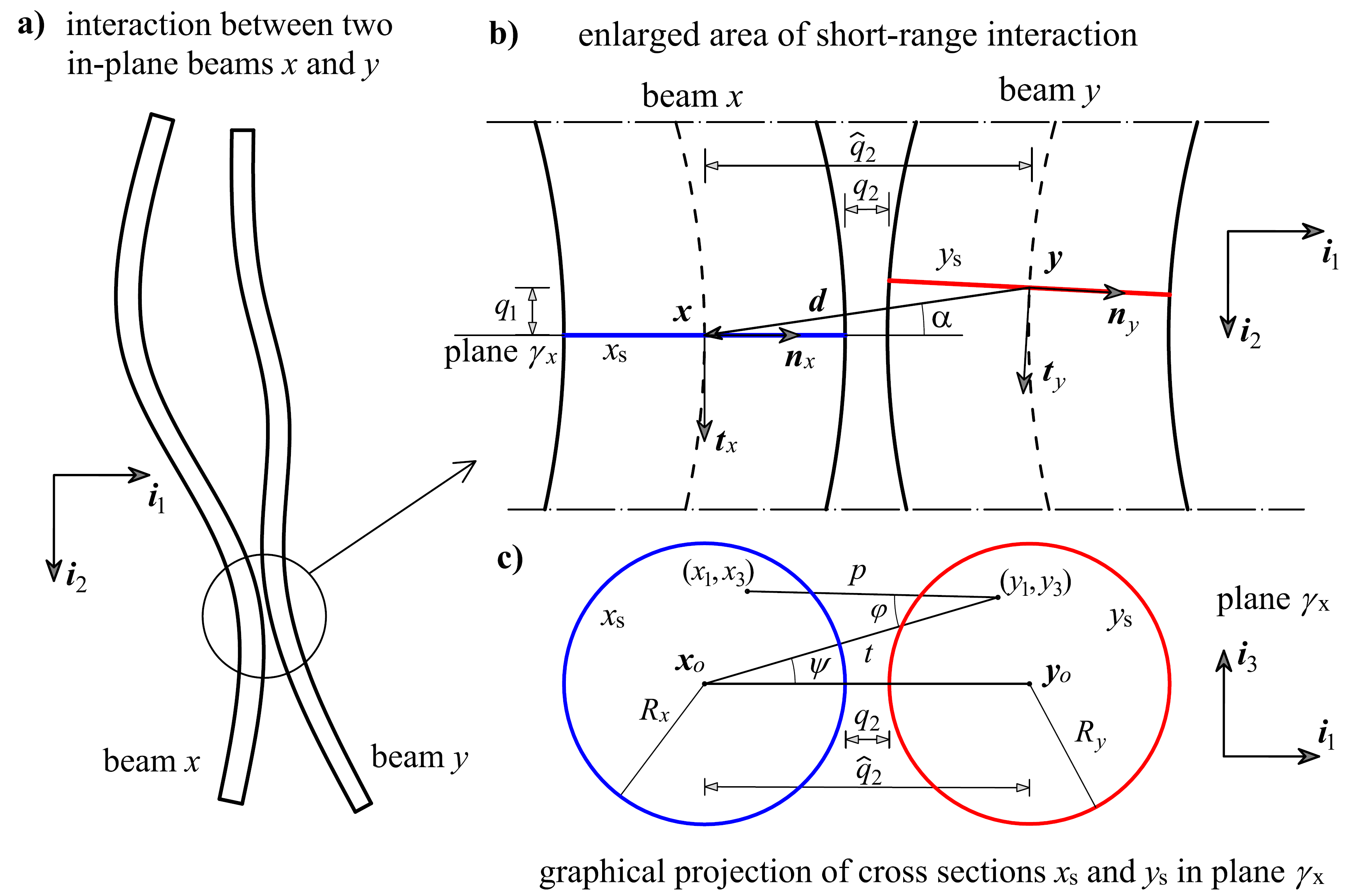}\centering
	\caption{Interaction of two in-plane beams. a) Schematic global point-of-view. b) Enlarged area of two interacting parts of the beams, with colored interacting cross sections. c) Graphical projections of interacting cross sections, $x_s$ and $y_s$, into the plane $\gamma_x$ with the normal $\ve{t}_x$. }
	\label{fig:beams}
\end{figure}

Let us consider two interacting beams, $x$ and $y$, and observe their two cross sections $x_s$ and $y_s$, see Fig.~\ref{fig:beams}.
The positions of the sections are $\ve{x}$ and $\ve{y}$, and their distance is $\ve{d}=\ve{x}-\ve{y}$. The cross sections are close to each other, $\norm{\ve{d}}-R_x-R_y \ll R_x,R_y$, but they do not belong to the same plane. In order to find an analytical expression for the interaction potential, we will first assume that the interacting cross sections are parallel
\begin{equation}
	\label{eq: assum}
	\begin{aligned}
		\ve{n}_x \parallel \ve{n}_y.
	\end{aligned}
\end{equation}
Although inaccurate in general, this is a sound approximation for short-range interactions between in-plane beams, especially when the interaction is dominated by smooth internal parts of beams. Evidently, the assumption is erroneous for interaction cases where at least one beam's end is included, due to the possible arbitrary orientation between the interacting cross sections.

Let us designate the cross-sectional plane of the section $x_s$ by $\gamma_x$. Assumption \eqref{eq: assum} implies that the graphical projection of cross section $y_s$ onto plane $\gamma_x$ is a circle with radius $R_y$. This fact is a key ingredient that will allow us to utilize and extend the integration approach from \cite{1972langbein}. Next, we need to adopt a reference local coordinate system (LCS). We have several choices here. By choosing the LCS of one beam as the reference LCS, we introduce a bias that is well-known from standard master-slave approaches in computational contact mechanics. This bias can be eliminated by implementing specific procedures, such as the two-half-pass approach \cite{2013sauerb}. An alternative is to choose an averaged LCS as the reference LCS, see Subsection \ref{averaged}.

To simplify the derivation of the new law, we will first pursue a straightforward approach here, by choosing the LCS of beam $x$ as the reference LCS, $\left(\ve{t}_{\mathrm{ref}},\ve{n}_{\mathrm{ref}}\right)=\left(\ve{t}_x,\ve{n}_x\right)$. The gap $q_1$ and the offset $q_2$ between the interacting cross sections are defined as
\begin{equation}
	\label{eq: gapdef}
	\begin{aligned}
		q_1= \ve{d} \cdot \ve{t}_{\mathrm{ref}}, \;\; q_2= \abs{\ve{d} \cdot \ve{n}_{\mathrm{ref}}} -R_x -R_y = \hat{q}_2-R_x -R_y.
	\end{aligned}
\end{equation}
The offset $q_1$ is the projection of the distance vector to $\ve{t}_{\mathrm{ref}}=\ve{t}_x$ and takes both negative and positive values. The gap $q_2$ is the projection of the distance vector to $\ve{n}_{\mathrm{ref}}=\ve{n_{x}}$, $\hat{q}_2$, reduced by the radii of interacting cross sections, and takes only positive values. The gap $q_2$ actually represents the distance between graphical projections of both cross sections in the plane $\gamma_x$. With these definitions, the distance between arbitrary point-pairs of interacting cross sections becomes $r=\left(p^2 + q_1^2\right)^{1/2}$, and the integral to calculate is
\begin{equation}
	\label{eq: ip77bb}
	\begin{aligned}
		I_{\Phi_m} &= \int_{A_x} \int_{A_y} r^{-m}  \dd{A_y} \dd{A_x} =\int_{A_x} \int_{A_y}  \left(p^2+q_1^2\right)^{-m/2}  \dd{A_y} \dd{A_x}.
	\end{aligned}
\end{equation}
To evaluate this integral, we need to choose a coordinate system. If we start from Cartesian coordinates with the origin at the center of cross section $x_s$, the integral is
\begin{equation}
	\label{eq: ip77car}
	\begin{aligned}
		I_{\Phi_m} &= \int_{-R_x}^{Rx} \int_{-\sqrt{R_x^2-x_1^2}}^{\sqrt{R_x^2-x_1^2}} \int_{-R_y}^{Ry} \int_{-\sqrt{R_y^2-\left(\hat{q_2}-y_1\right)^2}}^{\sqrt{R_y^2-\left(\hat{q_2}-y_1\right)^2}} r^{-m}  \dd{y_3} \dd{y_1} \dd{x_3} \dd{x_1}.
		\end{aligned}
\end{equation}
Since this form does not have an evident analytical solution, we seek a more suitable coordinate system. 
Let us take the coordinate transformation from the Cartesian coordinates to the coordinates $\left(t,p,\psi,\varphi\right)$ proposed in \cite{1972langbein}, see Fig.~\ref{fig:beams}c, so that
\begin{equation}
	\label{eq: ip77car88ppp}
	\begin{aligned}
	I_{\Phi_m}	&=  \int_{q_2+R_x}^{q_2+R_x+2 R_y} \int_{t-R_x}^{t+Rx} \int_{-\bar{\psi}}^{\bar{\psi}} \int_{-\bar{\varphi}}^{\bar{\varphi}} r^{-m} t \, p \dd{\varphi} \dd{\psi}  \dd{p}\dd{t} \quad \text{with} \\
		\bar{\psi} &=\arccos \frac{t^2+\hat{q}_2^2-R_y^2}{2t\hat{q}_2} \quad \text{and} \quad \bar{\varphi}=\arccos \frac{t^2+p^2-R_x^2}{2tp}, 
	\end{aligned}
\end{equation}
where the Jacobian determinant of this coordinate transformation is $t  p$, while the limits $\bar{\varphi}$ and $\bar{\psi}$ follow from the cosine theorem. For more details on this parametrization and limits of integration, see Appendix \ref{appendix:a}. Since the integrand is not an explicit function of coordinates $\varphi$ and $\psi$, the two innermost integrals can be easily solved, and the 4D integral thus reduces to 2D, i.e.
\begin{equation}
	\label{eq: ip77car88}
	\begin{aligned}
		I_{\Phi_m}	&=  4 \int_{q_2+R_x}^{q_2+R_x+2R_y} \int_{t-R_x}^{t+Rx} r^{-m} t \, p \, \bar{\varphi} \, \bar{\psi}  \dd{p} \dd{t}. 
	\end{aligned}
\end{equation}
Although the integral is reduced to two variables, $t$ and $p$, it does not have an analytical solution since the functions $\bar{\varphi}$ and $\bar{\psi}$ are quite complicated. In order to find an analytical expression, the following approximations are made in \cite{1972langbein} for small separations $q_2 \ll R_x,R_y$,
\begin{equation}
	\label{eq: ip7799}
	\begin{aligned}
		&\frac{t^2+p^2-R_x^2}{2tp} \approx 
		\frac{t-R_x}{p}, \\
		 &t\bar{\psi} \approx  \left[\frac{2 R_x R_y}{R_x+R_y}\left(t-R_x-q_2\right)\right]^{1/2}, \\
	\end{aligned}
\end{equation}
see Appendix \ref{appendix:a} for details. With these approximations and by introducing the reduced variables $\hat{p}=p/q_2$ and $\hat{t}=\left(t-R_x\right)/q_2$, the integral transforms to
\begin{equation}
	\label{eq: ip77xttnew}
	\begin{aligned}
		I_{\Phi_m} &=  4 \sqrt{\frac{2 R_x R_y}{R_x+R_y}}\int_{1}^{\frac{q_2+2R_y}{q_2}} \int_{\hat{t}}^{\frac{t+R_x}{q_2}} \left[\left(\hat{p} \,  q_2\right)^2+q_1^2\right]^{-m/2} \hat{p} \, q_2^{7/2} \arccos \frac{\hat{t}}{\hat{p}} \sqrt{\hat{t}  - 1} \dd{\hat{p}} \dd{\hat{t}}.
	\end{aligned}
\end{equation}
The final assumptions are that the upper limits can be approximated with infinity, $\left(t+R_x\right)/q_2 \approx \infty$ and $\left(q_2+2R_y\right)/q_2 \approx \infty$. Now, the integral becomes
\begin{equation}
	\label{eq: ip77xtt}
	\begin{aligned}
		I_{\Phi_m} &=  4 \sqrt{\frac{2 R_x R_y}{R_x+R_y}} q_2^{7/2} \int_{1}^{\infty} \int_{\hat{t}}^{\infty} \left[\left(\hat{p} q_2\right)^2+q_1^2\right]^{-m/2} \hat{p} \, \arccos \frac{\hat{t}}{\hat{p}} \, \sqrt{\hat{t}  - 1}  \dd{\hat{p}} \dd{\hat{t}}.
	\end{aligned}
\end{equation}
This form of the integral has the analytical solution
\begin{equation}
	\label{eq: ip13jkk}
	\begin{aligned}
		I_{\Phi_{m}}& =  2^{\frac{5}{2}-m} \pi^{\frac{3}{2}} \sqrt{\frac{R_x R_y}{R_x+R_y}} \frac{\Gamma \left(m-\frac{7}{2}\right)}{\Gamma \left(m/2\right)^2} \; \ii{q}{-m+\frac{7}{2}}{2} \; _2F_1 \left[ \frac{2m-7}{4},\frac{2m-5}{4};\frac{m}{2};-\left(\frac{q_2}{q_1}\right)^2\right],
	\end{aligned}
\end{equation}
where $\Gamma\left(z\right)= \int_{0}^{\infty} p^{z-1} e^{-w} \dd{w}$ is the gamma function, $_2F_1 \left(a,b;c;z\right)$ is the Gaussian hypergeometric function
\begin{equation}
	\label{eq: ip13hyp}
	\begin{aligned}
		_2F_1 \left(a,b;c;z\right)= \sum_{k=0}^{\infty} \frac{\left(a\right)_k \left(b\right)_k}{\left(c\right)_k} \frac{z^k}{k!},
	\end{aligned}
\end{equation}
while $\left(a\right)_k$ is the Pochhammer symbol
\begin{equation}
	\label{eq: ip13poch}
	\begin{aligned}
		\left(a\right)_k=\frac{\Gamma \left(a+k\right)}{\Gamma \left(k\right)}.
	\end{aligned}
\end{equation}

To summarize, the closed-form expression \eqref{eq: ip13jkk} for integral $I_{\Phi_m}$ allows us to represent an interaction potential between two cross sections as
\begin{equation}
	\label{eq: ip13oi}
	\begin{aligned}
		&\phi_{m,\mathrm{ss}} = \beta_{x} \,  \beta_{y} \int_{A_x} \int_{A_y}  k_m \,  r^{-m} \dd{A_y} \dd{A_x} = c_{m,\mathrm{ss}} q_2^{-m+\frac{7}{2}} F\left(q_1,q_2,m\right) , \;\; m>\frac{7}{2}, \;\; \textrm{with} \\ 
		&c_{m,\mathrm{ss}} = k_m \,  \beta_{x} \,  \beta_{y} \,  2^{\frac{5}{2}-m} \pi^{\frac{3}{2}} \sqrt{\frac{R_x R_y}{R_x+R_y}} \frac{\Gamma \left(m-\frac{7}{2}\right)}{\Gamma \left(m/2\right)^2},  \\
		&F\left(q_1,q_2,m\right) =\; _2F_1 \left[ \frac{2m-7}{4},\frac{2m-5}{4};\frac{m}{2};-\left(\frac{q_2}{q_1}\right)^2\right],	\;\; \textrm{and} \\
		&q_1 = \ve{d} \cdot \ve{t}_{\mathrm{ref}}, \;\; q_2=   \hat{q}_2-R_x -R_y, \; \; \hat{q}_2=\abs{\ve{d} \cdot \ve{n}_{\mathrm{ref}}}, \;\;
		\ven{d}=\iv{x}{}{} - \iv{y}{}{}.
	\end{aligned}
\end{equation}
We will refer to this new law \eqref{eq: ip13oi} as the \emph{improved section-section interaction potential} (ISSIP). It represents a generalization of the LSSIP \cite{2020grill} that follows if we let $q_1\rightarrow0$, i.e.
\begin{equation}
	\label{eq: ip13oi66}
	\begin{aligned}
	\phi_{m,\mathrm{ss}}\stackrel{q_1=0}{=}  c_{m,\mathrm{ss}} q_2^{-m+\frac{7}{2}} F\left(0,q_2,m\right) = c_{m,\mathrm{ss}} q_2^{-m+\frac{7}{2}}.
	\end{aligned}
\end{equation}
The constant $c_{m,\mathrm{ss}}$ was introduced and correctly derived in \cite{2020grill}. Our expression is the same, just in a different form w.r.t.~the gamma functions. Finally, the interaction potential between beams $x$ and $y$ follows as
\begin{equation}
	\label{eq: ip03}
	\begin{aligned}
		\Phi_{m} &= \int_{L_x} \int_{L_y} \phi_{m,\mathrm{ss}} \dd{s_y} \dd{s_x} =\int_{0}^{1} \int_{0}^{1} \phi_{m,\mathrm{ss}}  \sqrt{\ii{g}{}{y}} \sqrt{\ii{g}{}{x}}\dd{\xi_y} \dd{\xi_x}.
	\end{aligned}
\end{equation}

\subsection{Verification of the new law by considering the interaction between two parallel rigid cylinders}

Here, we consider the fundamental case of short-range interaction between two parallel rigid cylinders. Due to the simple setting, this example has a well-known analytical solution. Yet, the existing LSSIP fails to predict the interaction potential in this case. The inaccuracy is attributed to the orientation between interacting cross sections, see \cite{2020grill,2021grill}, which is significant for spatial beams. For in-plane beams, the error is mainly caused by the offset between cross-sectional planes, as we argue in the following.

Let us consider two parallel rigid cylinders of infinite length, $x$ and $y$, that interact via some short-range interaction potential. If we integrate the novel ISSIP over the cylinder $y$, from negative to positive infinity, we obtain  
the interaction potential between one section of cylinder $x$ and the whole cylinder $y$, i.e.
\begin{equation}
	\label{eq: ip1399}
	\begin{aligned}
		\phi_{m,\mathrm{ss},\parallel \mathrm{cyl}}& = \int_{-\infty}^{\infty} \phi_{m,\mathrm{ss}} \dd{s_y}.
	\end{aligned}
\end{equation}
For parallel cylinders, $\dd{s_y}=\dd{q_1}$, and the resulting potential is
\begin{equation}
	\label{eq: ip13}
	\begin{aligned}
		\phi_{m,\mathrm{ss},\parallel \mathrm{cyl}}& = \int_{-\infty}^{\infty} \phi_{m,\mathrm{ss}} \dd{s_y} = k_m \, \beta_{x} \, \beta_{y} \,  2^\frac{3}{2}\pi^\frac{3}{2} \sqrt{\frac{R_x R_y}{R_x+R_y}} \frac{\Gamma \left(m-\frac{9}{2}\right)}{\Gamma \left(m-1\right)} \; \ii{q}{-m+\frac{9}{2}}{2} , \; m > \frac{9}{2}.
	\end{aligned}
\end{equation}
The obtained expression for the case of parallel cylinders is the same as in \cite{2023grill}. Importantly, the scaling factor of $\left(-m+9/2\right)$ w.r.t.~the gap $q_2$ between the cylinders agrees with the analytical predictions. Note that the authors in \cite{2023grill} had to derive a section-beam interaction potential law in order to find this proper scaling factor, while our solution follows from the integration of the ISSIP.

Finally, the potential between two cylinders can be found by integrating \eqqref{eq: ip13} over the cylinder $x$, but this would result in an infinite value. Therefore, an interaction potential per unit length of cylinder $x$ is usually used. For the fundamental vdW case of $m=6$, we obtain a scaling factor of $-3/2$, which is the well-known analytical prediction \cite{1972langbein}. This result confirms that the derived ISSIP provides correct scaling for short-ranged intermolecular interactions between planar beams. 

\subsection{Verification of the new law by comparing section-section potentials}

\begin{figure}[h]
	\includegraphics[width=\textwidth]{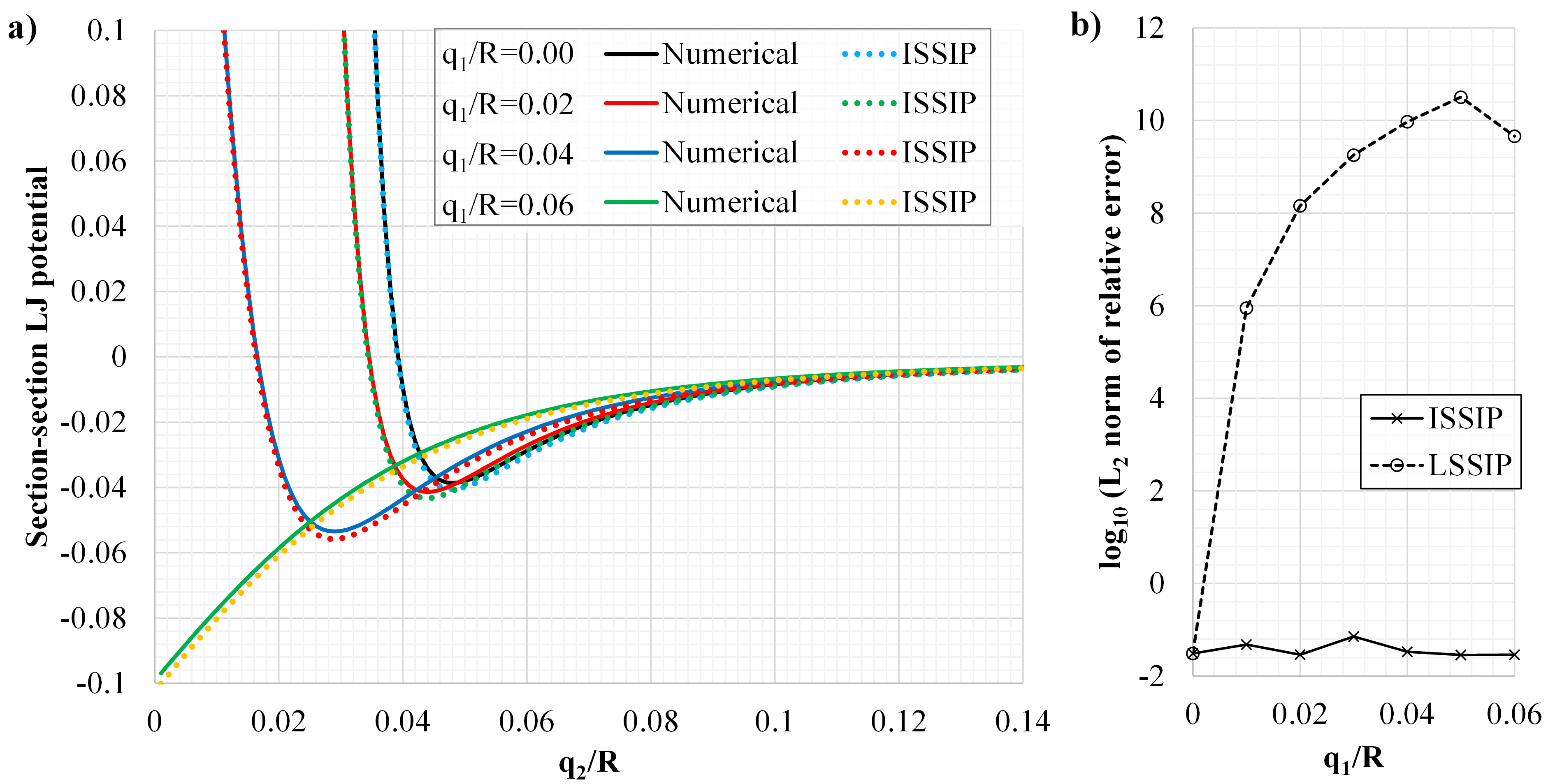}\centering
	\caption{Comparison of the section-section Lennard-Jones interaction potential. a) The results obtained with numerical integration and ISSIP for four values of offset $q_1$ vs.~gap $q_2$. b) $L_2$--norm of the relative errors of LSSIP and ISSIP w.r.t.~numerical integration vs. offset $q_1$.}
	\label{fig:LJpoten}
\end{figure}

Let us consider two parallel straight beams and observe one section of the first beam and its Lennard-Jones interaction potential with different sections of the second beam. Three approaches are considered: The first is a straightforward numerical integration of the interaction potential between sections, without any further approximations. The other two are the LSSIP and our ISSIP. Densities are set as $\beta_{x}=\beta_{y}=1$, while the radii are $R_x=R_y=R=0.02$. The material constants are $k_{6}=-10^{-7}$ and $k_{12}=5\times 10^{-25}$. The numerical integration is done using \eqqref{eq: ip77car88} since the numerical integration w.r.t.~Cartesian or polar coordinates fails for small gaps \cite{2020grill}. The results obtained with numerical integration and the ISSIP for four values of the offset $q_1$ are displayed in Fig.~\ref{fig:LJpoten}a as a function of the gap $q_2$.
All graphs of the proposed ISSIP are in good agreement with the numerical integration ones. Small differences evidently exist, but are acceptable as a necessary trade-off between accuracy and efficiency. As anticipated, the LSSIP returns erroneous results for all offsets $q_1 \ne 0$ and its graphs are excluded from Fig.~\ref{fig:LJpoten}a for clarity. The $L_2$ norm of the relative error returned by LSSIP and ISSIP w.r.t.~the numerical integration are plotted in Fig.~\ref{fig:LJpoten}b as a function of the offset $q_1$. For the offset $q_1 = 0$, the LSSIP and ISSIP return the same results, and the relative error is of the order $10^{-2}$. For all the other values of offset, the error produced by the ISSIP is of the similar order, while the error of the LSSIP blows up. 
All in all, analytical integration of the section-section potential requires assumptions that reduce accuracy. 
Nevertheless, this analysis shows that the new ISSIP law returns very good accuracy, considering the complexity of the problem.

\section{Variation of the improved section-section interaction potential}
\label{secvar}

In this section we consider the main ingredient required for obtaining the equilibrium equations of interacting beams -- the variation of the interaction potential. Since the ISSIP is the function of the distance vector components w.r.t.~a reference LCS, the gradients of the potential w.r.t.~to both the position and the tangent vectors must be found. Two formulations w.r.t.~the choice of reference LCS are considered, and the introduction of the interaction moment is discussed.

\subsection{Weak form of equilibrium}

The total potential energy here consists of the strain energy $\Phi_{\mathrm{int}}$, the potential of external forces $\Phi_{\mathrm{ext}}$, and the interaction potential $\Phi_{\mathrm{IP}}$. The equilibrium of a system is guaranteed by the minimum of total potential energy,
\begin{equation}
	\label{eq: ip14xtt}
	\begin{aligned}
		\delta \Phi_{\mathrm{tot}} = \delta \Phi_{\mathrm{int}} +\delta \Phi_{\mathrm{ext}}+\delta \Phi_{\mathrm{IP}}=0,
	\end{aligned}
\end{equation}
where the variation of interaction potential follows from \eqqref{eq: ip01xy},
\begin{equation}
	\label{eq: ip14xbbtt}
	\begin{aligned}
\delta \Phi_{\mathrm{IP}} =  \int_{L_x} \int_{L_y} \delta \phi_{m,\mathrm{ss}} \dd{s_x} \dd{s_y}.
	\end{aligned}
\end{equation}
The variation of strain energy and external potential can be readily found in the literature \cite{2023borkoviće} and is skipped here for brevity, since the focus is on the variation of interaction term $\Phi_{\mathrm{IP}}$.

\subsection{Gradient of the interaction potential}

The new ISSIP \eqref{eq: ip13oi} explicitly depends on the gap between cross sections, $q_2$, and the offset between cross-sectional planes, $q_1$, i.e.
\begin{equation}
	\label{eq: ip14x}
	\begin{aligned}
		\phi_{m,\mathrm{ss}} &= c_{m,\mathrm{ss}} q^{-m+\frac{7}{2}}_2 F\left(q_1,q_2,m\right) = c_{m,\mathrm{ss}} \phi \left( q_1 , q_2 \right), \\
		\phi \left( q_1 , q_2 \right) &= q_2^{-m+\frac{7}{2}} \; _2F_1  \left[ \frac{2m-7}{4} ,\frac{2m-5}{4} ;\frac{m}{2};-\left(\frac{q_1}{q_2}\right)^2\right] .
	\end{aligned}
\end{equation}
To simplify the following derivation, we set the constant $c_{m,\mathrm{ss}}$ to unity such that we can write the ISSIP function as $\phi_{m,\mathrm{ss}}= \phi \left( q_1 , q_2 \right)= \phi$. The variation of the interaction potential requires gradients of $\phi$ w.r.t.~the configuration of both beams, i.e.
\begin{equation}
	\label{eq: for}
	\begin{aligned}
		\delta \phi = \nabla_{x} \, \phi \delta x + \nabla_{y} \, \phi \delta y.
	\end{aligned}
\end{equation}
The gradient of the bivariate ISSIP law w.r.t.~both beams is
\begin{equation}
	\label{eq: ip20hj}
	\begin{aligned}
		\nabla_{j} \phi = \frac{\partial \phi }{\partial q_1} \nabla_{j} q_1 + \frac{\partial \phi  }{\partial q_2} \nabla_{j} q_2 = \phi_{,1} \nabla_{j} q_1 + \phi_{,2} \nabla_{j} q_2, \quad j=x,y,
	\end{aligned}
\end{equation}
where
\begin{equation}
	\label{eq: ip15}
	\begin{aligned}
\phi_{,1} &= -\frac{\left(2m-7\right)\left(2m-5\right)}{4m} q_2^{3/2-m} q_1 \; _2F_1  \left[ \frac{2m-3}{4} ,\frac{2m-1}{4} ;\frac{2+m}{2};-\left(\frac{q_1}{q_2}\right)^2\right], \\
\phi_{,2} &= \left(\frac{7}{2}-m\right) q_2^{5/2-m} \; _2F_1  \left[ \frac{2m-7}{4} ,\frac{2m-5}{4} ;\frac{m}{2};-\left(\frac{q_1}{q_2}\right)^2\right] \\
&+ \frac{\left(2m-7\right)\left(2m-5\right)}{4m} q_2^{1/2-m} q_1^2 \; _2F_1  \left[ \frac{2m-3}{4} ,\frac{2m-1}{4} ;\frac{2+m}{2};-\left(\frac{q_1}{q_2}\right)^2\right] \\
&= \frac{7-2m}{2 q_2}\phi - \frac{q_1}{q_2}\phi_{,1}.
\end{aligned}
\end{equation}
These expressions are utilized in their full form throughout the paper.

Since the offset $q_1$ and gap $q_2$ are functions the distance vector components w.r.t.~a reference LCS $\left(\ve{t}_{\mathrm{ref}},\ve{n}_{\mathrm{ref}}\right)$, their gradients are
\begin{equation}
	\label{eq: g2intro}
	\begin{aligned}
		\nabla_{j} q_\alpha &= \nabla_{\ve{j}} q_\alpha + \nabla_{\iv{j}{}{,1}} q_\alpha.
	\end{aligned}
\end{equation}
Here, $\nabla_{\iv{j}{}{}}$ is the gradient w.r.t.~the position of the beam $j$ axis, while $\nabla_{\iv{j}{}{,1}}$ is the gradient w.r.t.~to the tangential basis vectors $\ve{x}_{,1}=\ve{g}_{1x}$ and $\ve{y}_{,1}=\ve{g}_{1y}$. In general, the gradients of gap and offset are intricate quantities, see, e.g., \eqqref{eqC: ipxxf}. Implementation of their full form and consistent linearization strongly affect the efficiency. In order to find the balance between accuracy and efficiency, we consider two approaches w.r.t.~to the choice of reference LCS and discuss possible simplifications.

First, we define a formulation that is based on the LCS of beam $x$ in Subsection \ref{straight}. We refer to it as \emph{the straightforward formulation} since the same reference LCS is used in Subsection \ref{subexistingSSIP} for the derivation of the ISSIP, $\left(\ve{t}_{\mathrm{ref}},\ve{n}_{\mathrm{ref}}\right)=\left(\ve{t}_x,\ve{n}_x\right)$. Although it lacks robustness, this approach is convenient as a first step towards an optimal formulation because it returns simple equilibrium equations and allows an elegant insight into the interaction moments, see Subsection \ref{intmoment}. Second, in Subsection \ref{averaged} we propose a formulation based on the averaged LCS that turns out to be both robust and accurate, but at the cost of increased complexity. For this formulation, we consider simplifications related to the interaction moment, tangential part of the interaction force, and linearization.

\subsection{Straightforward implementation}
\label{straight}

A straightforward approach is obtained by adopting the LCS of beam $x$ as the reference LCS, $\left(\ve{t}_{\mathrm{ref}},\ve{n}_{\mathrm{ref}}\right)=\left(\ve{t}_x,\ve{n}_x\right)$. In this case, the variations of the interaction potential w.r.t.~beams $x$ and $y$ are
\begin{equation}
	\label{eq: full3}
	\begin{aligned}
		\delta_{x} \phi &=\ve{f} \cdot \delta \iv{u}{}{} +\hat{\ve{f}} \cdot \delta \iv{u}{}{,1}, \\
		\delta_{y} \phi &= -\ve{f} \cdot \delta \iv{v}{}{}, 
	\end{aligned}
\end{equation}
where
\begin{equation}
	\label{eq: full4}
	\begin{aligned}
		\ve{f} &=  \left(\phi_{,1}  \ve{t}_x + \phi_{,2} s_\alpha \ve{n}_x \right), \\
		\hat{\ve{f}}&= \frac{s_\alpha}{\sqrt{g_x}} \left(   \phi_{,1} \hat{q}_2 - \phi_{,2} q_1 \right)  \ve{n}_x,
	\end{aligned}
\end{equation}
and $s_\alpha = \textrm{sgn} \left(\ve{n}_x \cdot \ve{d} \right)$. A detailed derivation of these expressions is given in Appendix \ref{appendix:c}.

The variation of the potential w.r.t.~beam $y$ differs from the one w.r.t.~beam $x$ by the term involving $\hat{\ve{f}}$, because we have assumed that the potential is not a function of the LCS of beam $y$. This term represents the variation of the potential due to the interaction force couple that acts on both beams, i.e.
\begin{equation}
	\label{eq: full11}
	\begin{aligned}
		\delta_x \phi_M &= \ve{M}_x \cdot \delta \ve{\varphi}_x = -\left(\ve{d} \cross \ve{f}\right) \cdot \left(\frac{1}{\sqrt{g_x}}\ve{n}_x \cdot \delta \ve{u}_{,1}\right) \ve{b}\\
		&= -\left(q_1 \phi_{,2} s_\alpha  - s_\alpha \hat{q}_2 \phi_{,1}  \right) \ve{b} \cdot \left(\frac{1}{\sqrt{g_x}}\ve{n}_x \cdot \delta \ve{u}_{,1}\right) \ve{b} = \hat{\ve{f}} \cdot \delta \ve{u}_{,1}.
	\end{aligned}
\end{equation}
The minus sign appears because the term follows from the work of the interaction moment on beam $x$ due to the force that acts on beam $y$.

The straightforward approach is simple, consistent with the assumptions of ISSIP, and its linearization returns a symmetric tangent stiffness. However, there are two issues due to the bias w.r.t.~the chosen reference LCS: First, consider the case of interaction between two symmetric curved beam segments. Due to geometrical symmetry, the interaction forces should be symmetric as well. However, since not all of the interacting cross sections are parallel, this biased reference LCS gives asymmetric interaction forces and equilibrium is lost. Second, we have an interaction moment that acts only on one cross section and further violates the symmetry of interaction and the equilibrium. These issues are not significant, and the numerical implementation based on the straightforward formulation returns reasonably accurate results. However, such a formulation lacks robustness, and we will improve it in Subsection \ref{averaged} by adopting an averaged LCS as the reference LCS.

\subsection{Interaction moment}
\label{intmoment}

\begin{figure}[h]
	\includegraphics[width=10cm]{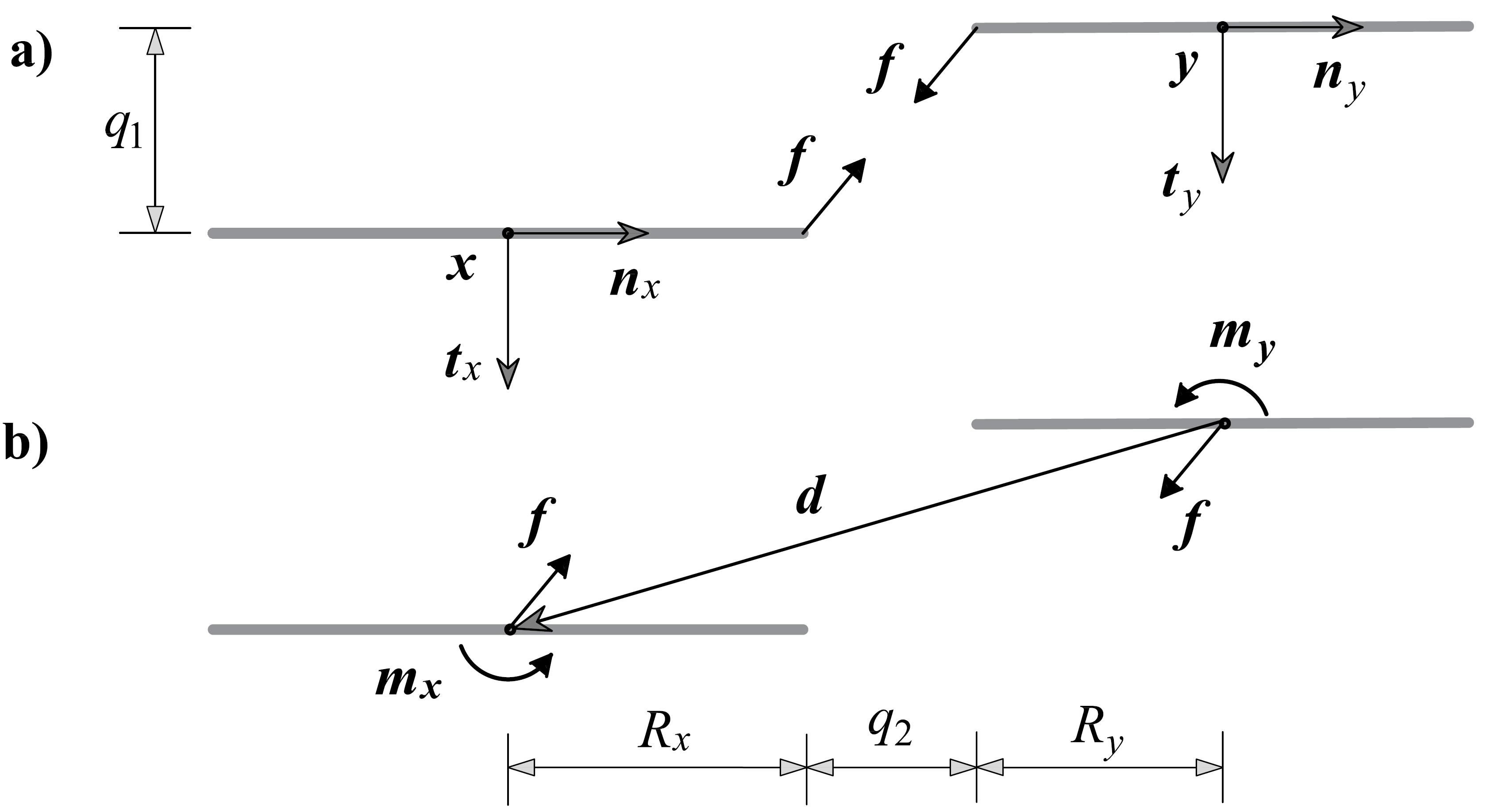}\centering
	\caption{Replacement of section-section interaction forces by equivalent resultant forces at the centroids. a) Approximate resultant forces for short-range interactions ($q_2,q_1 \ll R_x,R_y$). b) Equivalent approximate resultant forces at the centroids. }
	\label{fig:moments}
\end{figure}

An analytical derivation of the interaction moment through a procedure such as the one given in Subsection \ref{subexistingSSIP} is quite difficult, if not impossible. One approach to alleviate the bias in \eqqref{eq: full3} is to distribute the total force couple $\ve{d}\cross \ve{f}$ over both interacting cross sections, by adopting
\begin{equation}
	\label{eq: full3xx}
	\begin{aligned}
		\delta_{x} \phi &=\ve{f} \cdot \delta \iv{u}{}{} + w_x\hat{\ve{f}} \cdot \delta \iv{u}{}{,1}, \\
		\delta_{y} \phi &= -\ve{f} \cdot \delta \iv{v}{}{} + w_y\hat{\ve{f}} \cdot \delta \iv{v}{}{,1},
	\end{aligned}
\end{equation}
where $w_x$ and $w_y$ are moment distribution weights that can be determined by the approximation shown in Fig.~\ref{fig:moments}. 
First, we assume that the interaction force resultants act at the closest points of interacting cross sections, Fig.~\ref{fig:moments}a. The resultant couples and their relationship are
\begin{equation}
	\label{eq: full3xxmom}
		\ve{m}_x = R_x \ve{n}_x \cross \ve{f} \;\; \land \;\; \ve{m}_y = -R_y \ve{n}_y \cross \ve{f} \;\;\implies \;\;
	 \norm{\ve{m}_x} = \frac{R_x}{R_y}\norm{\ve{m}_y},
\end{equation}
and it follows that their values are proportional to the cross-sectional radii. From the equilibrium in Fig.~\ref{fig:moments}b, these two couples must equal the total force couple
\begin{equation}
	\label{eq: full3xxmom1}
	\begin{aligned}
		\ve{m}_x +\ve{m}_y &= \ve{d}\cross\ve{f}_x=-\ve{d}\cross\ve{f}_y,  \\
	\end{aligned}
\end{equation}
and Eqs.~\eqref{eq: full3xxmom} and \eqref{eq: full3xxmom1} thus give us distribution weights as
\begin{equation}
	\label{eq: disw}
	\begin{aligned}
		w_x+w_y&=1 \quad \textrm{and} \quad w_x=R_x/\left(R_x+R_y\right).
	\end{aligned}
\end{equation}
The proposed procedure provides a good approximation of the interaction moments between two cross sections for short-rang interactions.

\subsection{Formulation based on averaged local coordinate system}
\label{averaged}

Even by disregarding the interaction moment, the bias of the straightforward approach remains. The problem arises if the ISSIP is defined w.r.t.~the LCS of one beam. Such a choice of the reference LCS leads to the loss of equilibrium since the assumption of parallel interacting cross sections is violated in general, see Subsection \ref{straight}.

To alleviate this issue, we have considered several remedies and the formulation based on an averaged LCS turns out to be the most robust. The averaged LCS $\left(\hat{\ve{t}}_{xy},\hat{\ve{n}}_{xy}\right)$ is found by adding and normalizing the basis vectors of both beams, i.e.
\begin{equation}
	\label{eq: g1x2}
	\begin{aligned}
		\hat{\ve{t}}_{xy} &= \frac{\ve{t}_{xy}}{\sqrt{t_{xy}}}, \quad \ve{t}_{xy} = \ve{t}_x+\ve{t}_y, \quad t_{xy} = \ve{t}_{xy} \cdot \ve{t}_{xy}, \\
		\hat{\ve{n}}_{xy} &= \ve{\Lambda} \hat{\ve{t}}_{xy} =\frac{\ve{n}_{xy}}{\sqrt{n_{xy}}}, \quad \ve{n}_{xy} = \ve{n}_x+\ve{n}_y, \quad n_{xy} = \ve{n}_{xy} \cdot \ve{n}_{xy}. \\
	\end{aligned}
\end{equation}
In the following, the new averaged LCS is adopted as the reference LCS $\left(\ve{t}_{\mathrm{ref}},\ve{n}_{\mathrm{ref}}\right)=\left(\hat{\ve{t}}_{xy},\hat{\ve{n}}_{xy}\right)$. Note that, for the case of $\ve{t}_x \cdot \ve{t}_y <0$, we should rotate the LCS of one beam by the angle $\pi$.

Let us find the offset $q_1$ and gap $q_2$ by projecting the distance vector to the axes of the new reference LCS. This gives
\begin{equation}
	\label{eq: g1x}
	\begin{aligned}
		q_1 &= \ve{d}\cdot \hat{\ve{t}}_{xy},   \\
		q_2 &= \hat{q_2} - R_1 - R_2 , \quad \hat{q_2} = \abs{\ve{d}\cdot \hat{\ve{n}}_{xy}} = s_{\alpha} \left(\ve{d}\cdot \hat{\ve{n}}_{xy}\right), \quad s_\alpha = \textrm{sgn} \left(\ve{d} \cdot \hat{\ve{n}}_{xy} \right).
	\end{aligned}
\end{equation}
The gradients of the gap and offset w.r.t.~positions of both beams are
\begin{equation}
\label{eq: full3x66}
\begin{aligned}
	\nabla_{\ve{x}} q_1 &=  \nabla_{\ve{x}}  \left(\ve d \cdot \hat{\ve{t}}_{xy}\right) =   \hat{\ve{t}}_{xy} = - \nabla_{\ve{y}} q_1, \\
	\nabla_{\ve{x}} q_2 &= \nabla_{\ve{x}}   \abs{\ve {d} \cdot \hat{\ve{n}}_{xy}} = s_{\alpha} \hat{\ve{n}}_{xy} = -\nabla_{\ve{y}} q_2.
	\end{aligned}
\end{equation}
With these definitions at hand, the variations of interaction potential between two cross sections, neglecting the moment, are
\begin{equation}
	\label{eq: full3x}
	\begin{aligned}
		\delta_{x} \phi &= \ve{f}_x \cdot \delta \iv{u}{}{}=\ve{f} \cdot \delta \iv{u}{}{} , \\
		\delta_{y} \phi &= \ve{f}_y \cdot \delta \iv{v}{}{}= -\ve{f} \cdot \delta \iv{v}{}{} ,
	\end{aligned}
\end{equation}
where
\begin{equation}
	\label{eq: full3xff}
	\begin{aligned}
	\ve{f} &=  \phi_{,1}  \hat{\ve{t}}_{xy} + \phi_{,2} s_\alpha \hat{\ve{n}}_{xy} = f_1  \hat{\ve{t}}_{xy} + f_2  \hat{\ve{n}}_{xy} .
	\end{aligned}
\end{equation}
Now, we can define the total interaction forces on both beams as
\begin{equation}
	\label{eq: ip04FF}
	\begin{aligned}
		\iv{P}{}{x} = -\iv{P}{}{y} =  \int_{L_x} \int_{L_y} \ve f \dd{s_x} \dd{s_y}= \int_{\xi_x} \int_{\xi_y} \ve f \sqrt{\ii{g}{}{x}} \sqrt{\ii{g}{}{y}} \dd{\xi_x} \dd{\xi_y}.
	\end{aligned}
\end{equation}
Regarding the influence of interaction moments, we can define force couples and reformulate the variations of the potential as in Subsections \ref{straight} and \ref{intmoment} to obtain
\begin{equation}
	\label{eq: full4xx}
	\begin{aligned}
		\delta_{x} \phi &= \ve{f} \cdot \delta \iv{u}{}{} + w_x \hat{\ve{f}}_x \cdot \delta \iv{u}{}{,1}, \\
		\delta_{y} \phi &= -\ve{f} \cdot \delta \iv{v}{}{} + w_y \hat{\ve{f}}_y \cdot \delta \iv{v}{}{,1},
	\end{aligned}
\end{equation}
where
\begin{equation}
	\label{eq: full4xxyy}
	\begin{aligned}
		\hat{\ve{f}_x}&= \frac{s_\alpha}{\sqrt{g_x}} \left(   \phi_{,1} \hat{q_2} - \phi_{,2} q_1 \right)  \ve{n}_{x}, \\
		\hat{\ve{f}_y}&= \frac{s_\alpha}{\sqrt{g_y}} \left(   \phi_{,1} \hat{q_2} - \phi_{,2} q_1 \right)  \ve{n}_{y}.
	\end{aligned}
\end{equation}

The most important benefit of the formulation based on the averaged LCS is the preservation of symmetry and equilibrium, which improves the robustness of the formulation and meliorates the consistency of results.  The main drawback is the reduced efficiency, primarily due to the complicated linearization. To tackle this issue, we have considered simplifications regarding the interacting moment and the components of the interaction force. Regarding the interaction moment, we will compare the formulations given by \eqqref{eq: full3x} and \eqqref{eq: full4xx} in Subsection \ref{verif}. The possibility of neglecting the tangential component of the interaction force in \eqqref{eq: full3xff} is discussed in Subsection \ref{simulation}. 

We have carefully considered the available options, and it turns out that the formulation based on the averaged LCS without the interacting moments and with the tangential component of the interaction force provides the best balance between accuracy and efficiency. If not explicitly stated otherwise, all the results in the following section are obtained with this approach.

\section{Numerical example}
\label{numeric}

In order to verify and benchmark the proposed ISSIP, we thoroughly investigate the example of peeling and pull-off of two slender elastic fibers. The example was introduced in \cite{2021grill} and modeled with the LSSIP.  
The same authors have later developed an asymptotically consistent section-beam interaction potential (SBIP) \cite{2022grill, 2023grill} that promises much better accuracy for short-range interactions than LSSIP. 
The example is defined without referring to specific units, so any consistent set of units is appropriate.

\subsection{Problem setup}

\begin{figure}[!htb]
	\includegraphics[width=7cm]{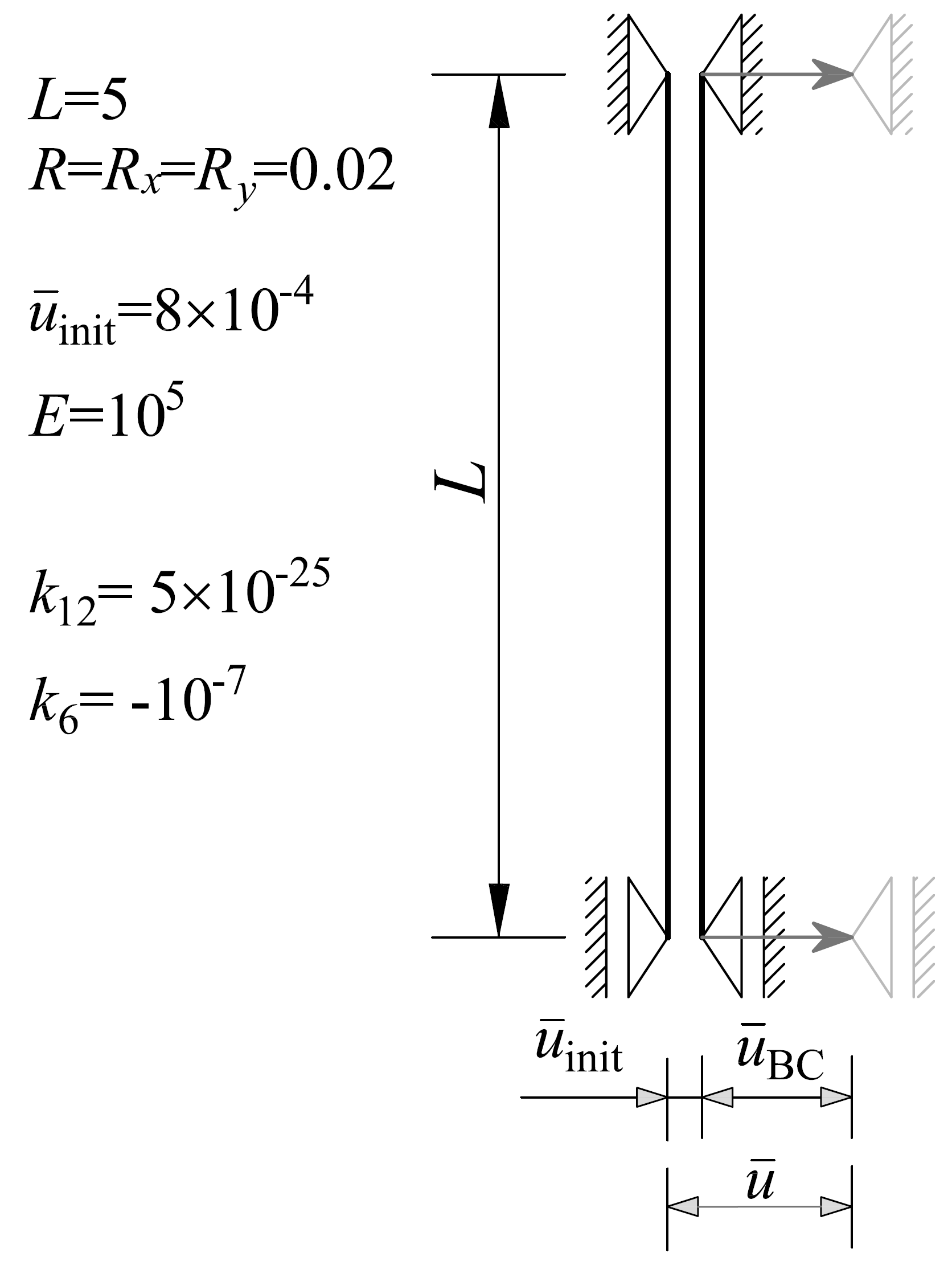}\centering
	\caption{Peeling and pull-off of two elastic fibers. Problem setup. }
	\label{fig:setup}
\end{figure}

The problem setup is given in Fig.~\ref{fig:setup}.
Two elastic fibers, modeled as simply supported straight beams, interact via the Lennard-Jones (LJ) potential. For convergence in nonlinear static analysis, it is advantageous to start from a configuration close to equilibrium. The equilibrium distance, i.e.~gap, between two rigid cylinders is derived in \cite{2021grill} as $q_{2,\mathrm{eq}}=0.00017L$, which can be obtained from \eqqref{eq: ip13}. In this example, the supports are initially placed a bit below this equilibrium distance, $\bar{u}_{\mathrm{init}}=0.00016L=0.0008$, so that there is a repulsion near the supports in the initial configuration. Then, the fibers are separated by prescribed non-homogeneous boundary conditions $\bar{u}_{\mathrm{BC}}$ to the right fiber. The separation between the supports of the beams is $\bar{u}=\bar{u}_{\mathrm{init}}+\bar{u}_{\mathrm{BC}}=Lt=5t$ at each load step, where the load proportionality factor (LPF) or quasi-time takes the values $t \in \left[0.00016,1\right]$.

As in \cite{2021grill, 2022grill, 2023grill}, the horizontal reaction force is followed during the simulation. Actually, in these works, the sum of reaction forces on one beam is followed. Our numerical experiments show that these reactions in fact differ, but the relative difference is of the order $10^{-5}$.

\subsection{Implementation details}

An isogeometric approach is utilized for the spatial discretization \cite{2018borkovicb}. To be precise, quartic B-spline elements ($p=4$) with $C^3$ interelement continuity are employed. A model with 161 control points is adopted in all simulations, see Subsection \ref{convergence} for the spatial convergence analysis.

Due to the strong gradient of the LJ potential, the integration of the interaction part of the stiffness and force requires a dense distribution of integration points. We will designate the total number of integration points with $n_{\mathrm{GP}}$. 
The integration of the interaction forces and stiffnesses is performed by dividing the finite elements into a set of integration segments. After thorough investigations, we have concluded that the mid-point rule allows the most robust simulations. Therefore, this rule is employed in all our analyses and a value of 3200 integration points per unit length is adopted. For a discussion on integration issues see Subsection \ref{sectionIntegration}. On the other hand, the standard strain energy and external potential contributions are integrated with Gaussian quadrature using $p+1$ integration points per element.

An important feature of formulations that deal with short-range interaction fields is a cutoff distance. Here, we have employed the function \emph{Nearest[]} implemented in \emph{Mathematica} \cite{2024wolframresearch} to find point-pairs that are inside the cutoff radius. If not specified otherwise, the cutoff distance of $c=2.5 R=0.05$ is used, see Subsection \ref{cutoff}.

Regarding the nonlinear solver, we have implemented the mixed integration point method to improve the convergence of the Newton-Raphson solution, \cite{2017magisano}. The method uses linear increments of the stress for the geometric part of the stiffness matrix and often results in improved convergence. The linearization of the $\delta \phi$ term is given in Appendix \ref{appendix:d}, while the testing procedure for the accuracy of the tangent stiffness is presented in Appendix \ref{appendix:e}. We utilize an adaptive load--stepping scheme that adjusts load increments based on the number of Newton-Raphson iterations in the previous load step.

For static analysis, it is difficult to obtain the final separated configuration just after pull-off since the two configurations are quite far from each other \cite{2021grill,2022grill,2023grill}. Here, the final separated configuration is successfully calculated by removing the interaction part from the stiffness at the final increment, which allowed the solver to find the new equilibrium. This snap-off is actually a dynamical phenomenon and not crucial for the present quasi-static analysis. However, it is instructive to include this point of the equilibrium path for the comparison of results. Therefore, all presented results include the unstressed configuration just after pull-off.

A regularization of the repulsive part of the interaction potential law is often implemented in the literature to improve convergence \cite{2011sauerb}. For example,  without regularization, the authors in \cite{2021grill} had to significantly restrict the size of iterative displacement in order to avoid the crossing of beams and failure at a load step. The implementation of such a strict iterative displacement limit has led to a simulation requiring 1600 increments with an average of 45 iterations. By implementing a regularization, the authors have reported a significant improvement \cite{2021grill}. It is worth noting that we did not run into similar problems in our simulations, presumably since the beams are not in a state of very large repulsion. If no restriction on iterative displacement is applied, the average number of increments is around 80, with approximately 8 iterations per increment. Therefore, no regularization of the ISSIP law is applied here. 

The incremental solution of the Newton-Raphson procedure is accepted when the error norm w.r.t.~the internal and external forces is satisfied. The value of $10^{-5}$ is employed as the threshold since further refinement did not make any perceptible difference. 

As an initial step for the presented study, we have implemented the LSSIP in our in-house code, see Appendix \ref{appendix:b} for basic details. The results obtained with this code are briefly shown in Subsection \ref{verif} for verification purposes.

\subsection{Simulation results}
\label{simulation}

\begin{figure}[!htb]
	\includegraphics[width=\textwidth]{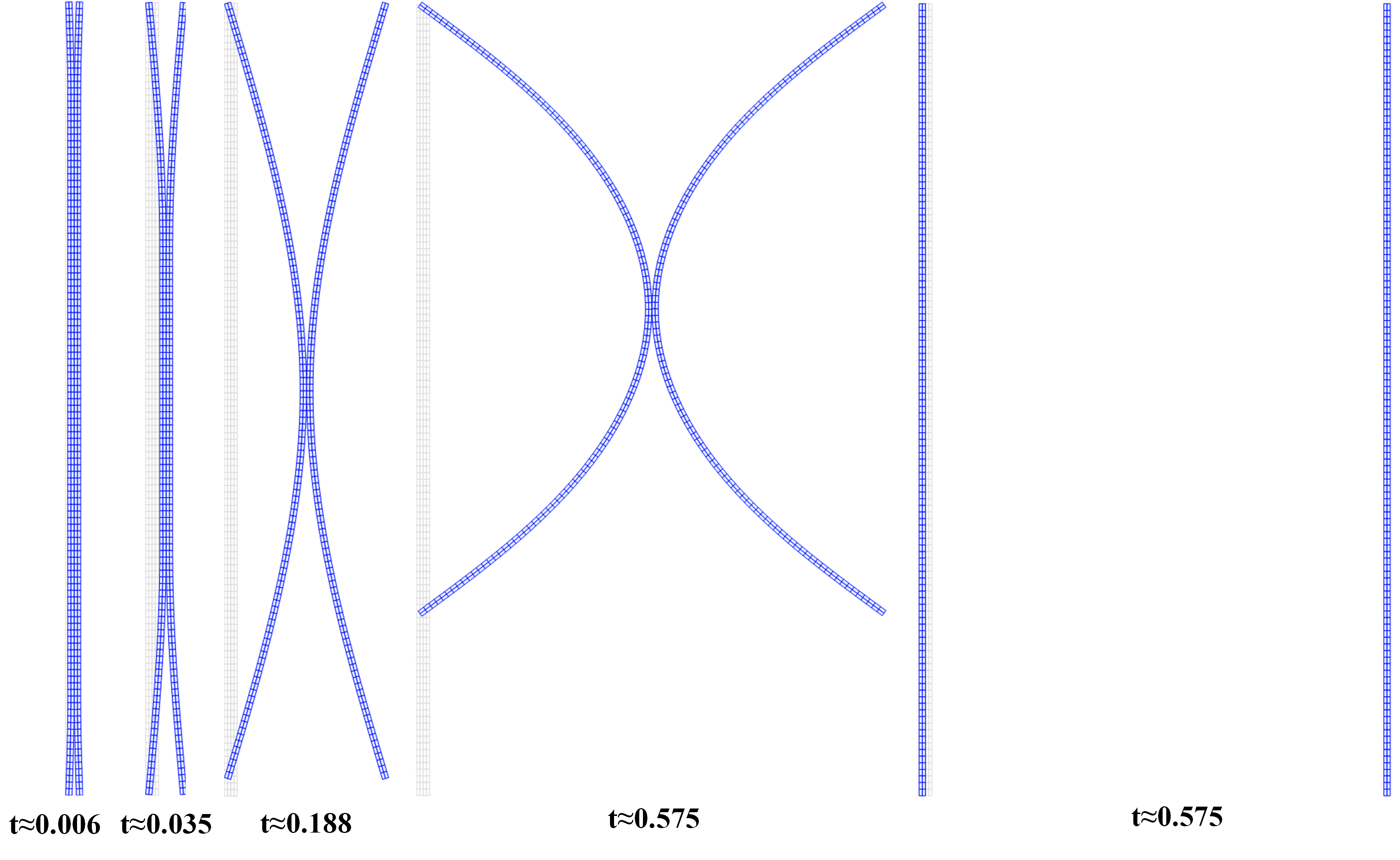}\centering
	\caption{Peeling and pull-off of two fibers. Snapshots of five characteristic configurations. The blue grid is used for visualization purposes only. The animation is available in the online dataset, cf. \nameref{supsec}.}
	\label{fig:snap}
\end{figure}

Let us start this overview of numerical results with the positions of both fibers. Simulation snapshots at five specific increments are shown in Fig.~\ref{fig:snap}.
The first stage of the simulation is characterized by the peeling of the fibers which stops at about $t\approx0.188$. 
In the next stage, the fibers remain connected at a small central area and start to bend significantly due to the pull. Finally, at $t\approx0.575$, the internal forces overwhelm the vdW attraction, and the fibers snap-off.

\begin{figure}[!htb]
	\includegraphics[width=\textwidth]{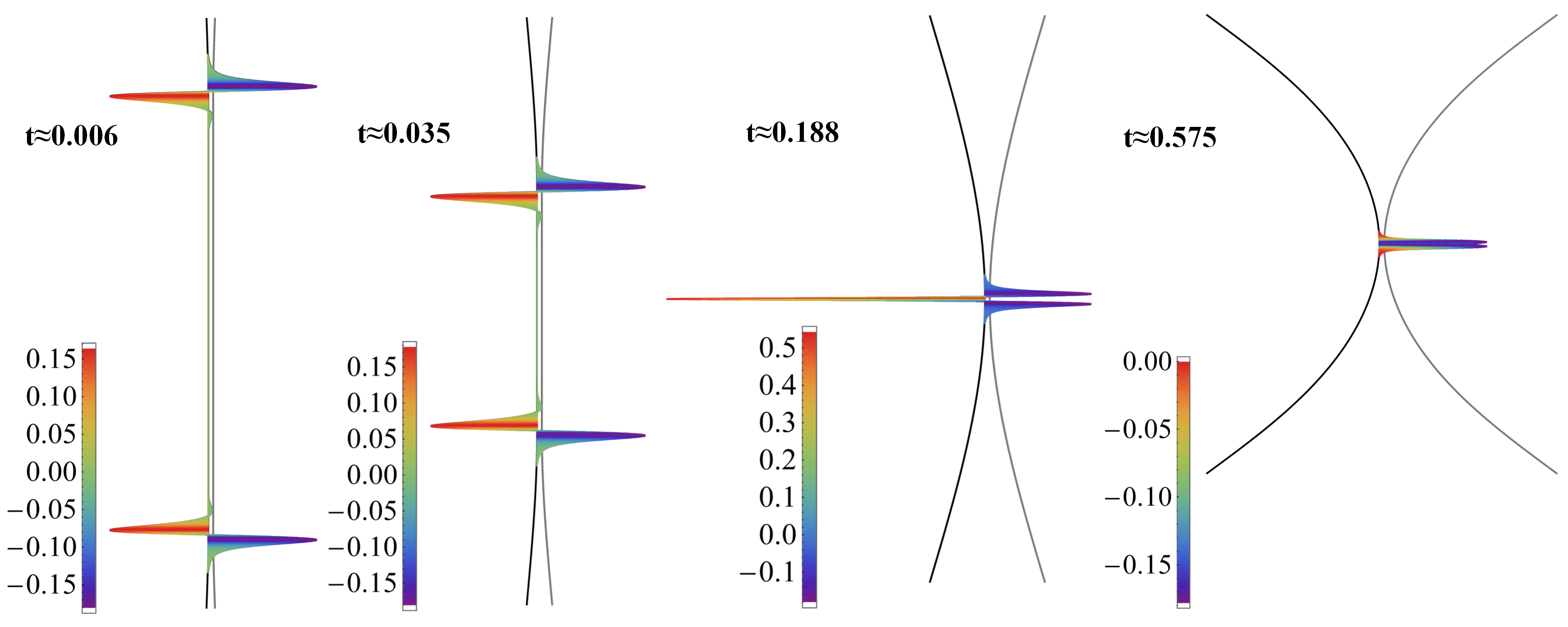}\centering
	\caption{Normal component $f_2$ of the interaction force plotted on the left beam. Four characteristic configurations are shown. The animation is available in the online dataset, cf. \nameref{supsec}. }
	\label{fig:forces}
\end{figure}

It has already been emphasized that the modeling of peeling between two adhesive fibers based on the LJ potential is a very challenging task due to the high gradients of the attractive vdW and repulsive steric forces. This is evident from the graphical representation of the interaction forces in Fig.~\ref{fig:forces}, where the distribution of the normal component of the interaction force, denoted $f_2$, on the left beam is shown for four configurations. 
Although seemingly counterintuitive, both attraction and repulsion exist during peeling due to the bending stiffness of the fibers. At the end of the peeling process ($t\approx 0.188$), a large peak in the repulsive force is observed. Afterward, the remaining LJ force converts into attraction until pull-off. It is interesting to note that the maximum value of the attractive force before pull-off is not in the middle of the beam. Instead, there are two maxima left and right of the middle, which will be shown in more detail in Section \ref{cutoff}.

In order to check equilibrium at the first load step, we have integrated the horizontal component of the interaction force and compared it with the sum of the horizontal reaction forces. The relative difference of $0.0018\%$ is found and attributed to the numerical integration error. Our simulations show that doubling the number of integration points halves this difference.

\begin{figure}[!htb]
	\includegraphics[width=\textwidth]{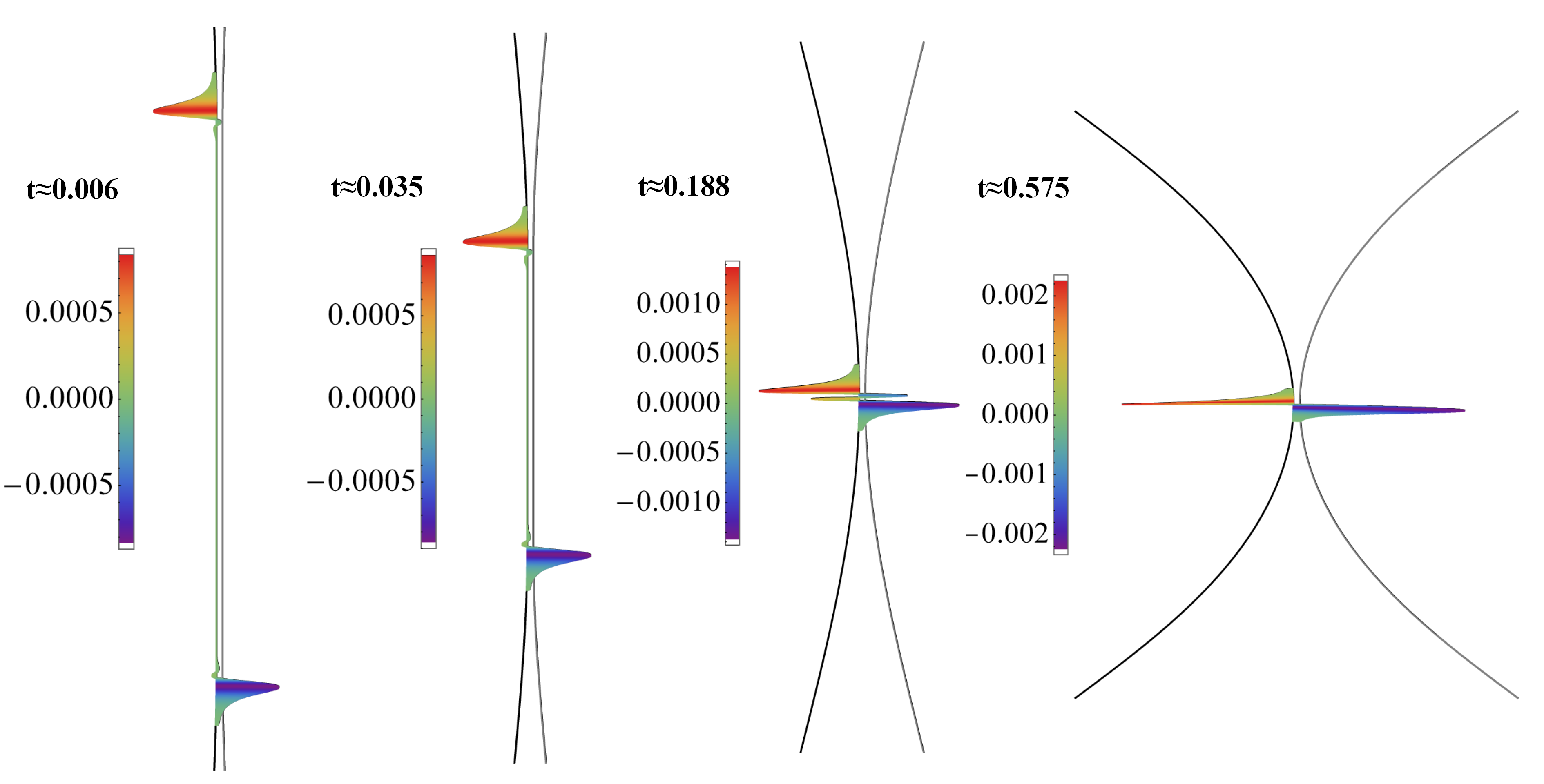}\centering
	\caption{Tangential component of the interaction force, $f_1$, on the left beam. Four characteristic configurations are plotted. The animation is available in the online dataset, cf. \nameref{supsec}.}
	\label{fig:forcesT}
\end{figure}

The tangential component of the interaction force, denoted $f_1$, is approximately two orders of magnitude smaller than the normal component and it is shown in Fig.~\ref{fig:forcesT}.
There is a significant peak of the tangential component during the initial debonding that is omitted here for brevity, but it is visible in the animation, cf. \nameref{supsec}. It is not evident from these figures, but tangential components are oriented towards the center of the beam, which causes a compressive axial stress resultant in both beams.

The present formulation includes both tangential, $f_1$, and normal, $f_2$, components of the interaction force. Due to the orders of magnitudes involved, it is arguable if a reduced formulation that disregards the tangential component $f_1$ of the interaction force is a good choice. It is certainly more efficient since the force vector and tangent stiffness are simplified. Our initial investigations show that both approaches return quite similar results, but the formulation that neglects the component $f_1$ is not as robust as the complete one. 

\begin{figure}[!htb]
	\includegraphics[width=\textwidth]{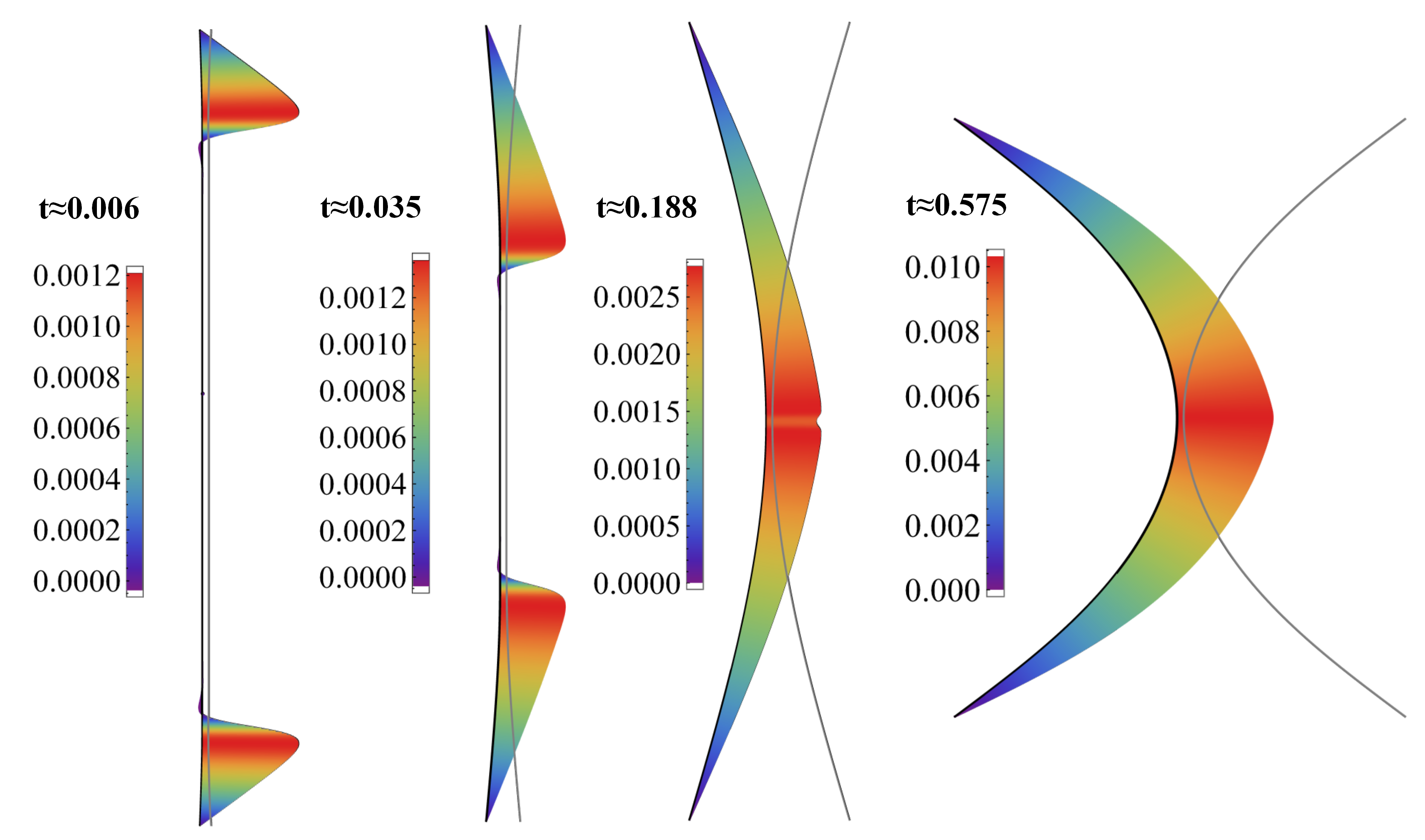}\centering
	\caption{Distributions of the resultant stress couple for four characteristic configurations. The animation is available in the online dataset, cf. \nameref{supsec}.}
	\label{fig:bmoment}
\end{figure}

\begin{figure}[!htb]
	\includegraphics[width=\textwidth]{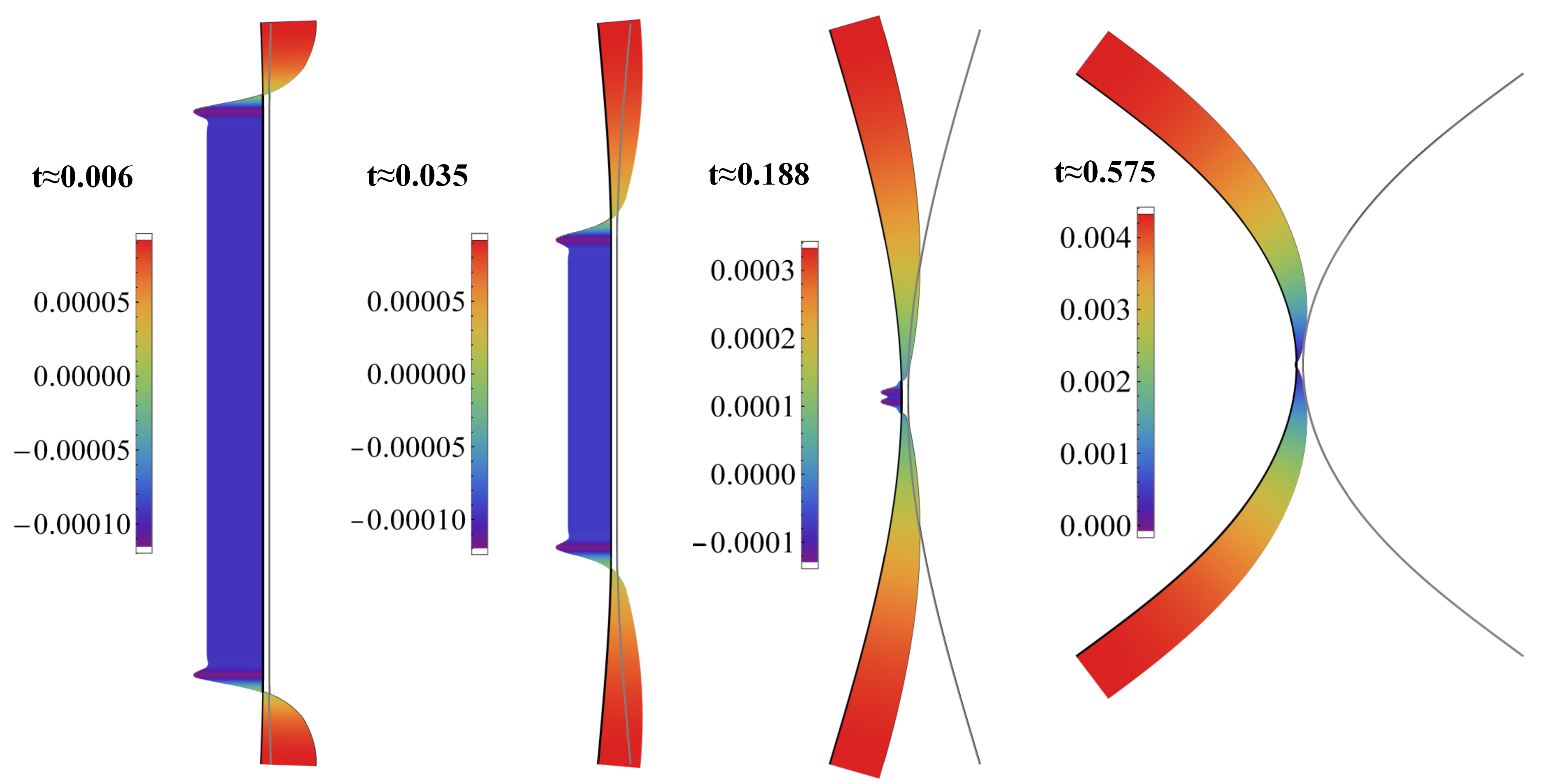}\centering
	\caption{Distributions of the stress resultant for four characteristic configurations. The animation is available in the online dataset, cf. \nameref{supsec}.}
	\label{fig:nforce}
\end{figure}

Finally, the distributions of the stress resultant and stress couple are investigated. Although these quantities are well-researched in the context of isogeometric BE theory \cite{2023borkoviće}, it is worth taking a closer look at their distributions for the case of LJ interaction forces. The distributions of the stress couples at four configurations are shown in Fig.~\ref{fig:bmoment}.
Evidently, during the peeling stage, the inner part of the fibers is straight and no moment occurs. As the peeling develops and transits to pull, the whole beam bends. It is quite interesting that, due to the strong adhesion, the maximum curvature does not occur in the middle of the beam until the final pull-off. Finally, distributions of stress resultant (i.e. axial force) at four instances are displayed in Fig.~\ref{fig:nforce}.
The stress resultant is compressive at the inner part of the beam due to the interaction forces, while it is tensile in the outer part, due to the reaction forces. It should be noted that the stress resultant is not zero in the middle of the beam at pull-off, but small compression exists. However, the stress resultant due to the interaction forces is approximately two orders of magnitude smaller than the one due to the reaction forces (note different scales in Fig.~\ref{fig:nforce}).

\subsection{Verification}
\label{verif}

\begin{figure}[!htb]
	\includegraphics[width=16cm]{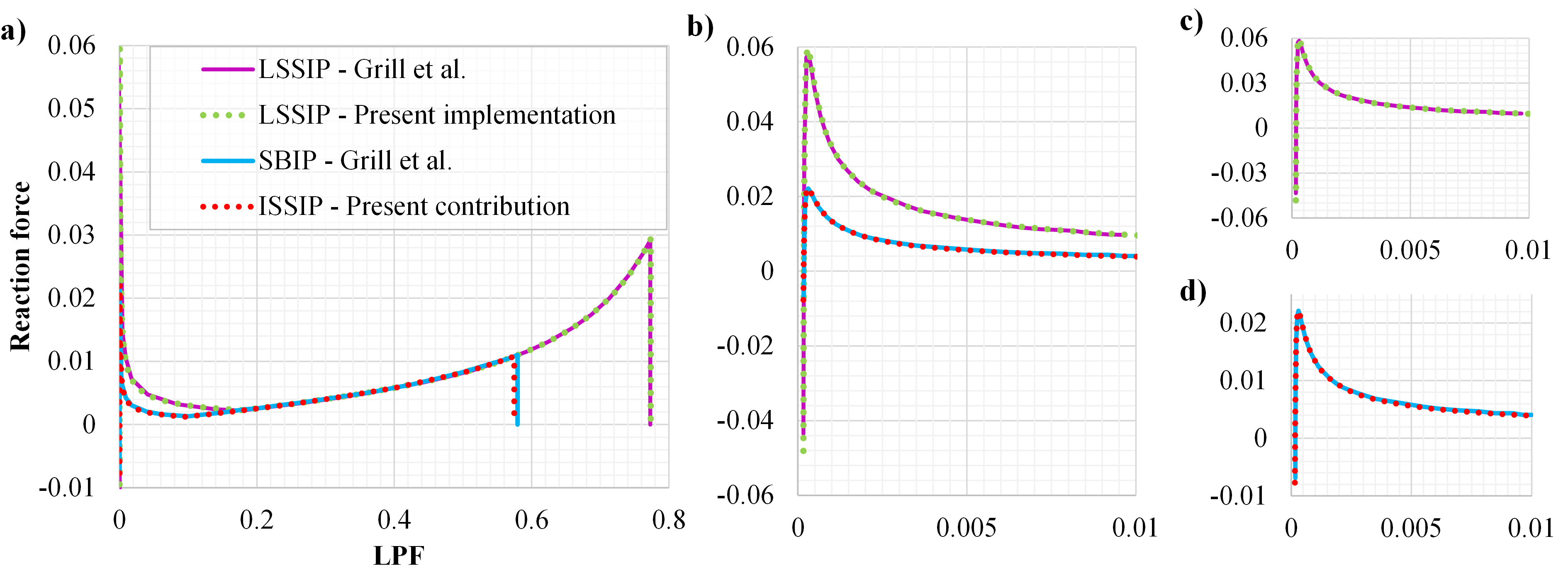}\centering
	\caption{Comparison of reaction forces vs. LPF for different formulations. a) Complete paths for all formulations. b) Initial part of the equilibrium path for all formulations. c) Initial part of the equilibrium path for the LSSIP implemented in \cite{2021grill} and in the presented research, see Appendix \ref{appendix:b}. d) Initial part of the equilibrium path for SBIP \cite{2022grill} and the present ISSIP.  }
	\label{fig:comp2}
\end{figure}

In this subsection, the novel ISSIP is verified by comparison with results reported in the literature. For this, the reaction forces obtained with existing LSSIP, SBIP, and novel ISSIP are compared in Fig.~\ref{fig:comp2}a.
There are three major observations here: First, the LSSIP results obtained in \cite{2021grill} and in the present research agree. A small difference exists at the initial configuration, which can be attributed to different spatial discretizations and integration techniques. Second, the results returned by the new ISSIP and the SBIP from \cite{2022grill,2023grill} are practically the same. Small differences at initial and final configurations are present, but it can be argued that they are negligible for the modeled problem. Third, the difference between the existing LSSIP on one side, and the SBIP and ISSIP on the other side in Fig.~\ref{fig:comp2} is evident. The initial values, maximum values, and pull-off points significantly differ as a consequence of the very coarse approximation used for LSSIP.

These results verify the accuracy of the novel ISSIP and confirm that it can be used reliably to model the peeling and pull-off of elastic fibers due to short-ranged interaction forces.

\begin{figure}[!htb]
	\includegraphics[width=16cm]{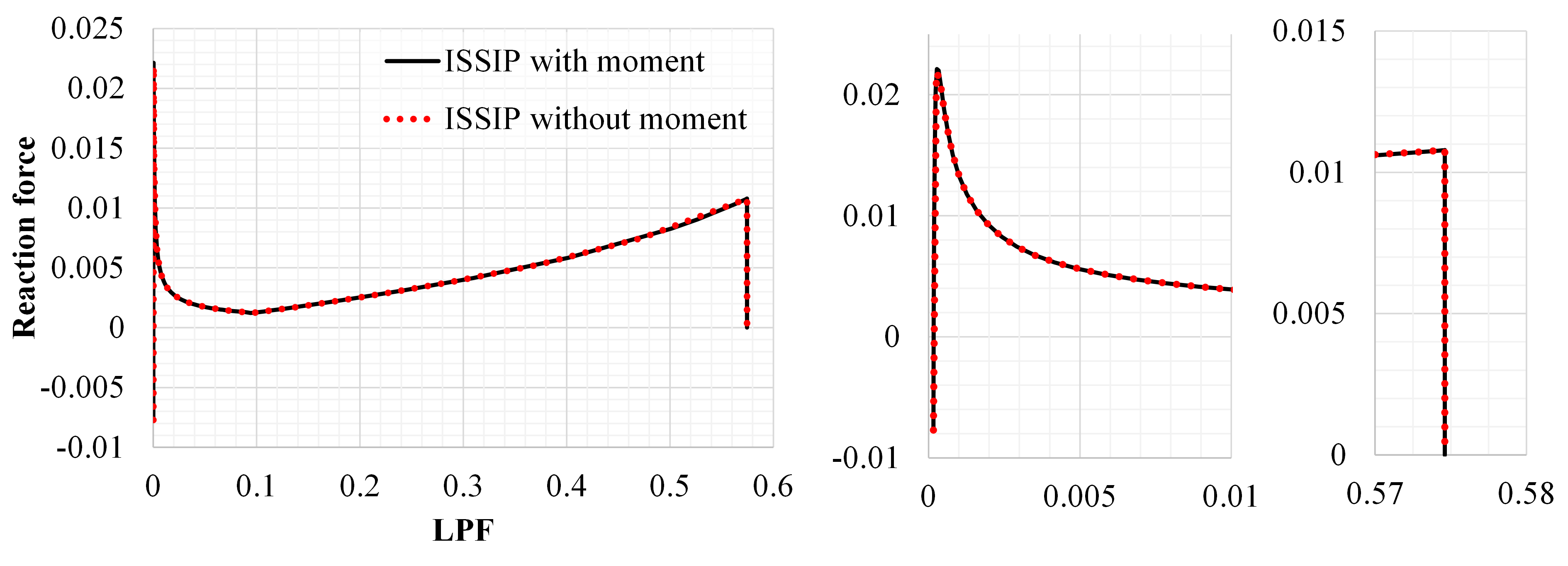}\centering
	\caption{Comparison of the reaction force obtained with two implementations of ISSIP: with and without the moment.}
	\label{fig:compmom}
\end{figure}

To close the subsection, let us compare the two variants of ISSIP, the one that implements interaction moments \eqqref{eq: full4xx} and the one without it \eqqref{eq: full3x}. The results in Fig.~\ref{fig:compmom} are practically the same and the use of the simplified formulation without the moments is justified.
 We can argue that the influence of the interaction moments is not significant since their effects mainly cancel each other. However, for a deeper insight, it would be valuable to investigate the effect of interaction moment more thoroughly.

\subsection{Cutoff distance}
\label{cutoff}

\begin{figure}[!htb]
	\includegraphics[width=\textwidth]{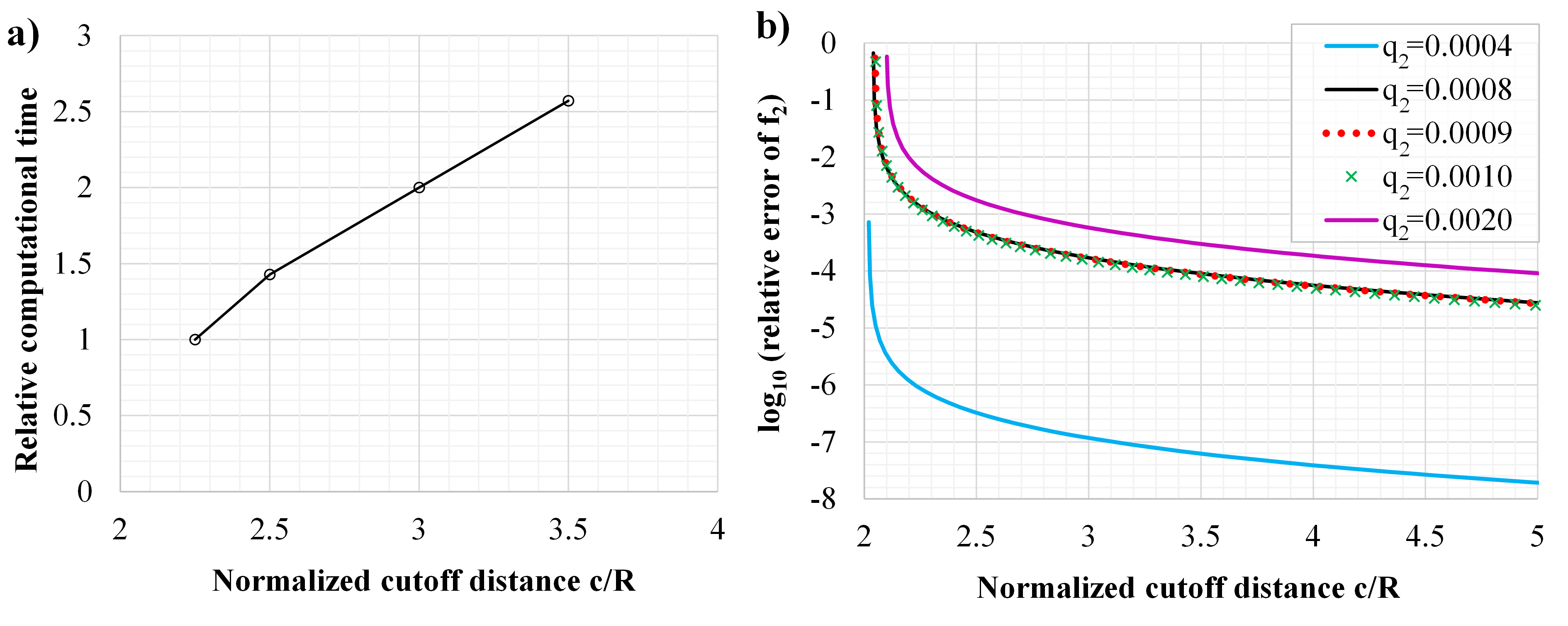}\centering
	\caption{a) Relative computational time vs.~cutoff distance. b) Relative error of the normal component of the interaction force between a cross-section and a straight beam vs. the normalized cutoff radius. Five characteristic values of gap $q_2$ are considered. }
	\label{fig:cutoff1}
\end{figure}

The influence of the cutoff distance on the simulation results is investigated in this subsection. The cutoff distance represents an adopted radius around one interacting particle, beyond which the interaction with other particles is neglected. Since the cutoff distance significantly affects both accuracy and computational time, it is necessary to find a balance between these two counteracting requirements. For example, the initial peeling phase is computationally most expensive due to a large number of interacting sections. For the adopted mesh, the relative computational times per iteration are calculated for different values of cutoff during the first few load steps. The results are displayed in Fig.~\ref{fig:cutoff1}a where a linear increase of the computational time w.r.t.~the cutoff distance is evident. 

\textbf{Remark} It is possible to speed up the computations by omitting the interaction in parts of the beams. For example, we could consider a simulation where $5-10 \%$ of the beams' ends are not interacting. This would allow us to avoid a sharp peak in the reaction force which is the most computationally expensive and still obtain correct values at pull-off.

Let us consider the integration of our developed ISSIP force law. Since the law explicitly depends on the offset $q_1$ and gap $q_2$, we can integrate section-section forces over one beam with arbitrary geometry, and find the force between a cross section and that beam. For simplicity, we will consider the normal component of the interaction force between a cross section and a straight beam. 

The physical and geometric properties are the same as in Fig.~\ref{fig:setup}. First, we integrate the interaction force from negative to positive infinity w.r.t.~$q_1$ analytically and mark this solution as exact. A similar integration is already done in \eqqref{eq: ip13} to find the interaction potential between a cross section and a beam. Next, we are varying the cutoff distance which gives us finite limits of integration w.r.t.~$q_1$. These limits are then utilized to calculate a set of approximately integrated interaction forces. The errors of such approximations for six values of the gap $q_2$ are shown in Fig.~\ref{fig:cutoff1}b. As expected, increasing the cutoff distance reduces the error, and we can make an informed decision on the adopted cutoff distance. It is also evident that the accuracy is much better for a very small gap (e.g., $q_2=0.0004$) due to the extremely short range of steric forces. For our present simulations of peeling and pull-off, gaps in the range of $q_2\in (0.0008,0.001)$ are most significant and the error is then practically invariant w.r.t.~the cutoff distance. Increasing the gap to $q_2=0.002$ leads to a significant error increase due to the small gradient of the interaction force for this separation.

In our simulations, the cutoff distance of $c=2.5 R=0.05$ is readily adopted. This value gives a relative error near $0.04 \%$ which represents a reasonable trade-off between accuracy and efficiency. Additionally, we have run similar simulation for the interaction between a single cross section and a circular beam, and the influence of the cutoff distance is quite similar. However, these results represent a coarse approximation and they are omitted in the paper for brevity.

\begin{figure}[!htb]
	\includegraphics[width=16cm]{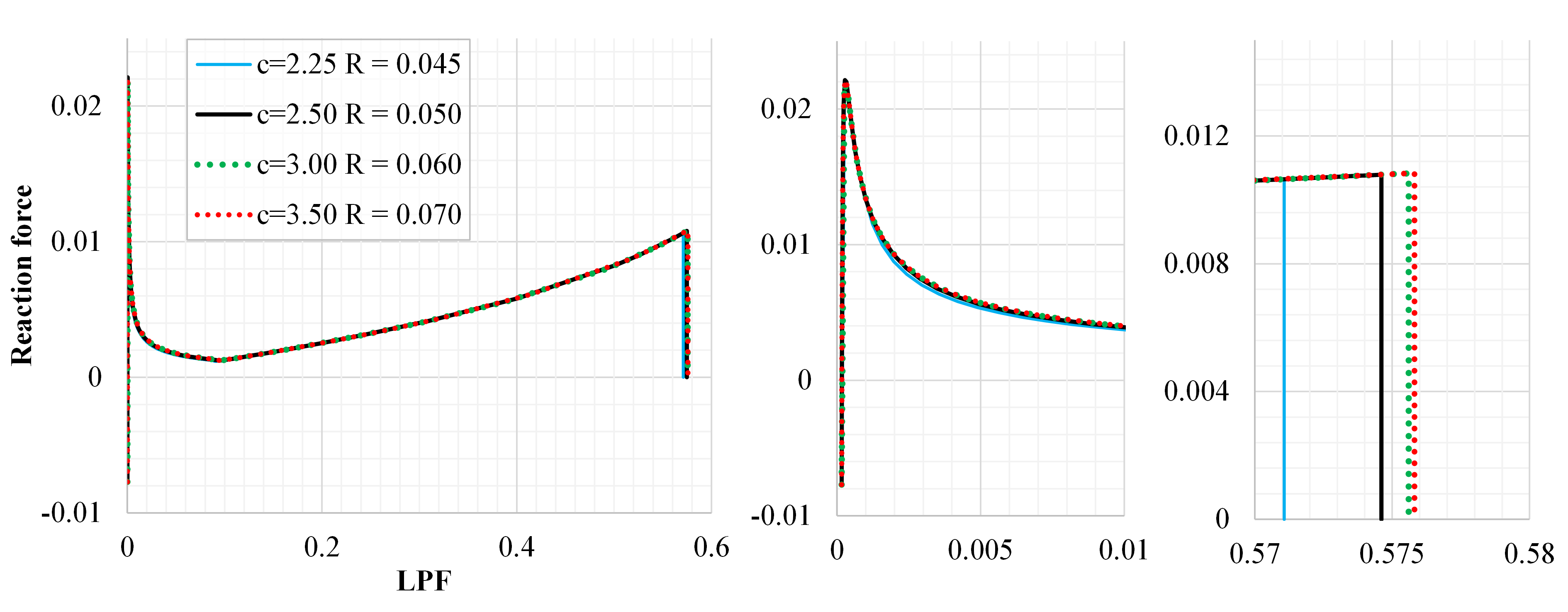}\centering
	\caption{Reaction force vs. LPF for four values of the cutoff distance.}
	\label{fig:cutoff2}
\end{figure}

Let us now turn to a macroscopic point of view, and check the influence of the cutoff distance on the reaction force. The results for four values of the cutoff distance are shown in Fig.~\ref{fig:cutoff2}. 
The largest difference occurs at the point of pull-off because the separations between beams are then at their maximum. The relative errors at pull-off for $c=0.045$, $c=0.05$ and $c=0.06$ w.r.t. $c=0.07$ are $0.82\%$, $0.21\%$, and $0.04\%$, respectively. Again, the adopted value of $c=0.05$ allows a good approximation of the considered problem.

\begin{figure}[!htb]
	\includegraphics[width=16cm]{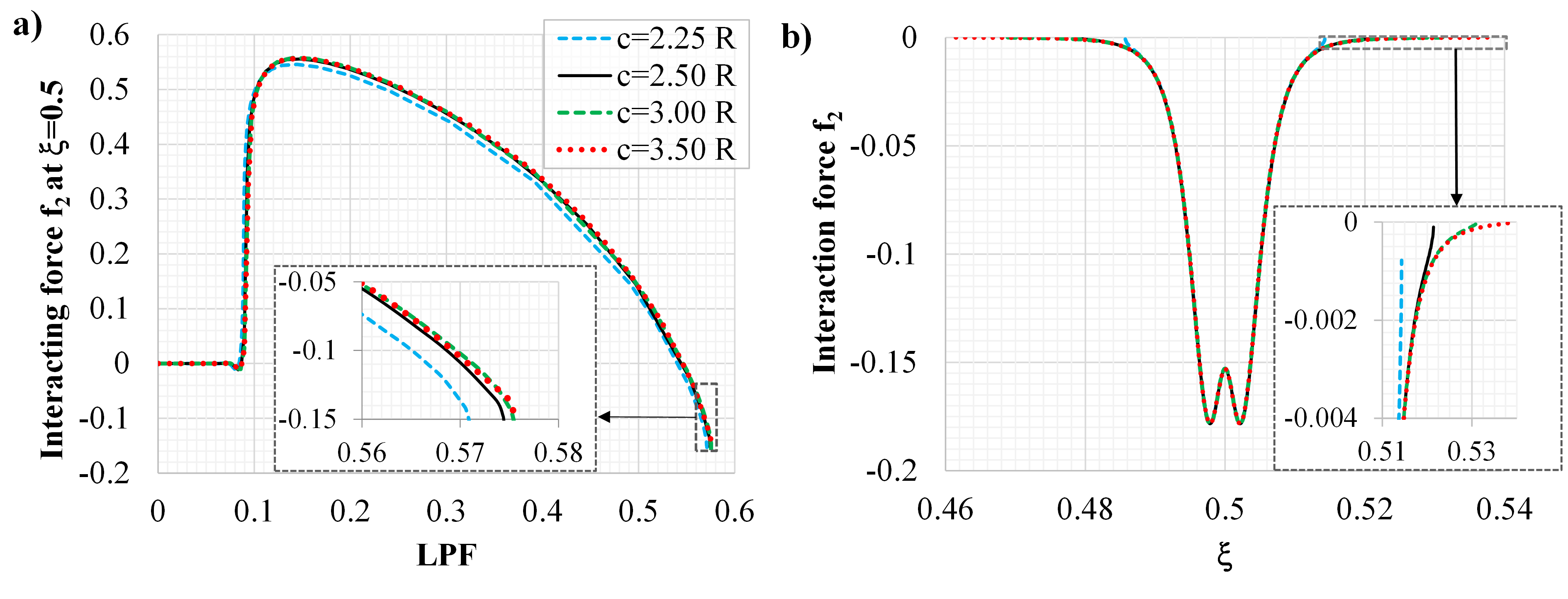}\centering
	\caption{Influence of different cutoff distances on the interaction force component $f_2$. a) $f_2$ in the middle of the beam vs. time. b) Distribution of $f_2$ at pull-off.}
	\label{fig:cutoff3}
\end{figure}

Finally, the normal component of the interaction force in the middle of the beam is considered. The same four cutoff values are considered and the results are shown in Fig.~\ref{fig:cutoff3}a.
A similar relation between the results for different cutoffs as for the reaction force is observed. An interaction force at a section of one beam is the result of the interaction between that section and the whole other beam. The distribution of the interaction force $f_2$ on a beam that results from the interaction with the middle of the other beam is shown in Fig.~\ref{fig:cutoff3}b. The source of error due to the cutoff distance becomes apparent by looking at the enlargement. Furthermore, the distribution of the normal force component at the instance of pull-off was already shown in Fig.~\ref{fig:forces}. The fact that the maximum value of the force $f_2$ does not occur in the middle of the beam is now seen more clearly.

To summarize, finding a balance between accuracy and efficiency due to the cutoff distance is crucial for the present simulation. The ISSIP allows us to estimate the cutoff distance in advance. The error increases with the separation between interacting beams, and the adopted value of $c=2.5 R$ represents a good balance between accuracy and efficiency.

\subsection{Convergence}
\label{convergence}

\begin{figure}[!htb]
	\includegraphics[width=0.5\textwidth]{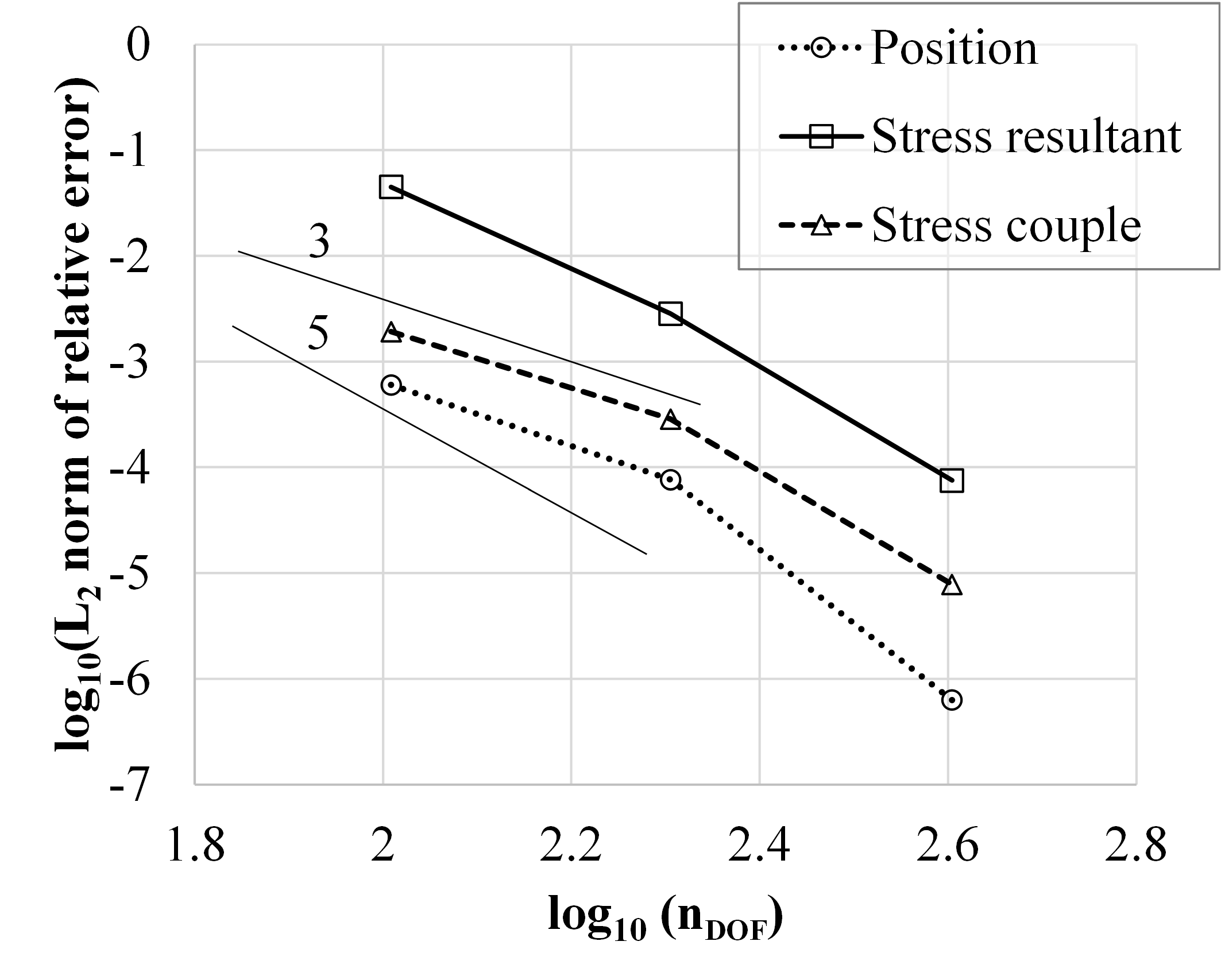}\centering
	\caption{Spatial convergence study w.r.t. number of DOFs.}
	\label{fig:conv}
\end{figure}

Here, we perform convergence studies of the proposed formulation. First, let us consider the influence that the spatial discretization has on the structural response. A reference solution is computed using $n_{\mathrm{DOF}}=1206$ and the total number of integration points is kept fixed $n_{\mathrm{GP}}=16000$. The convergences of the $L_2$ norm of the relative error for position, stress resultant, and stress couple at the instance of pull-off are shown in Fig.~\ref{fig:conv}.
Since we are using quartic B-splines with the strongly curved BE model, see \eqqref{eq:e11eq}, the theoretical asymptotic convergence rates are 5 for the position and 3 for the stress resultant and stress couple. The obtained rates are in the range of these predictions while the accuracy is highest for the position and lowest for the stress resultant. Note that the stress resultant is affected by both axial strain and bending curvature \cite{2023borkoviće}, and therefore very sensitive to mesh density. 

\begin{figure}[!htb]
	\includegraphics[width=\textwidth]{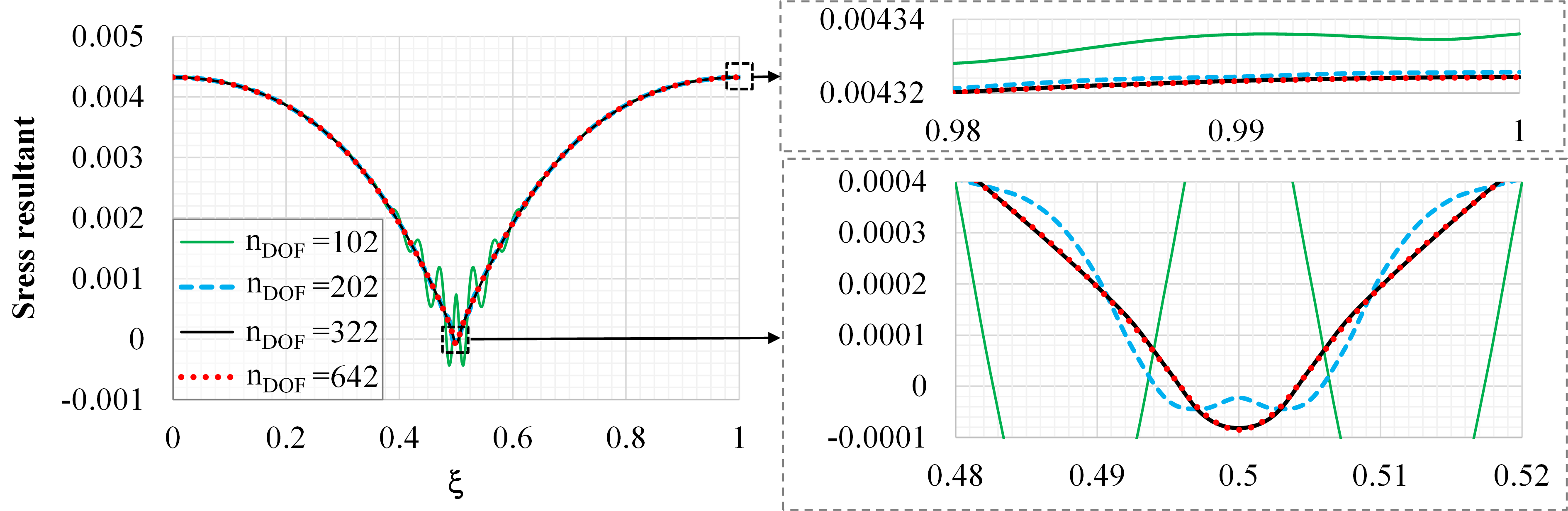}\centering
	\caption{Distribution of stress resultant for four spatial discretizations.}
	\label{fig:normalcon}
\end{figure}

Let us now consider the stress resultant at pull-off. Fig.~\ref{fig:normalcon} shows the distribution of the stress resultant for four levels of spatial discretization.
As seen, the mesh with $n_{\mathrm{DOF}}=102$ is too coarse and the results are unacceptable. Regarding the three denser meshes, they give quite similar results. However, by taking a closer look at the middle and at the end of the beam, it is evident that the mesh with $n_{\mathrm{DOF}}=202$ differs from the other two. This study confirms that the adopted quartic mesh with 161 control points and $C^3$-continuity (i.e., $n_{\mathrm{DOF}}=322$) returns reasonably accurate results.

\begin{figure}[!htb]
	\includegraphics[width=\textwidth]{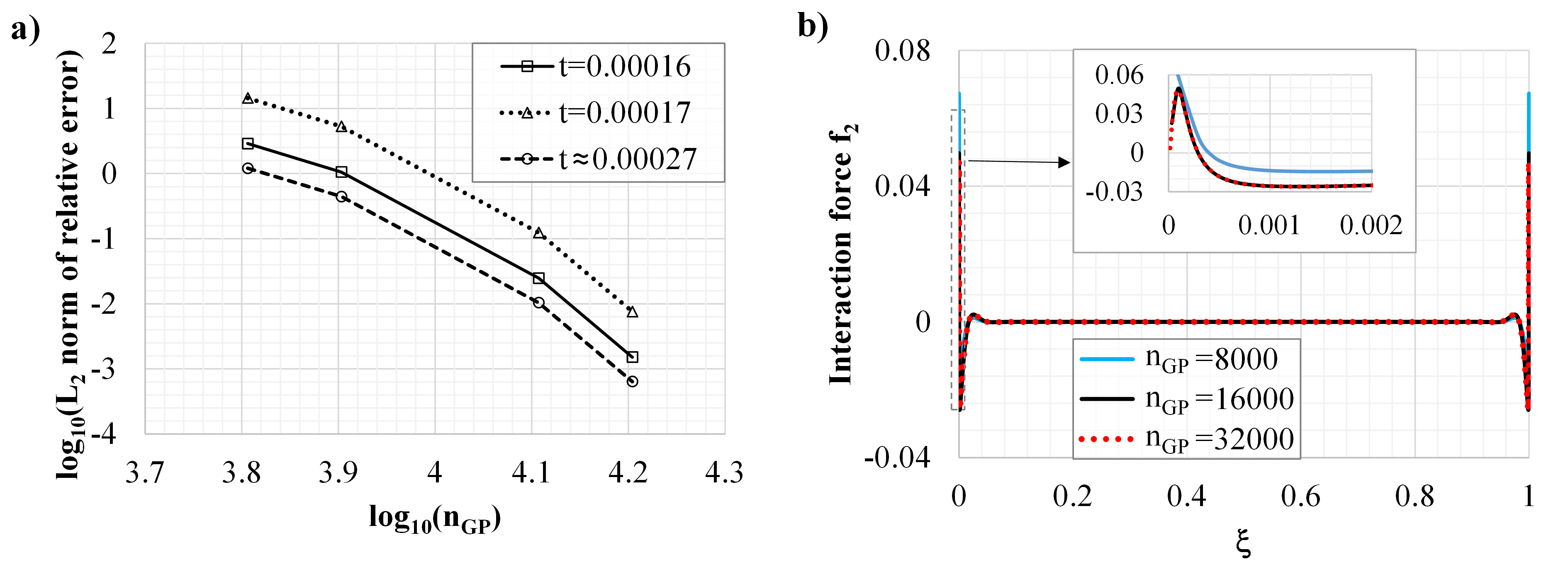}\centering
	\caption{a) Convergence of the interaction force component $f_2$ for three different load steps vs. the number of integration points. b) Distribution of the interaction force $f_2$ for three numbers of integration points at $t=0.00017$. }
	\label{fig:con1}
\end{figure}

Next, let us observe the convergence of the normal force component $f_2$ w.r.t.~the number of integration points. The reference solution is obtained with $n_{\mathrm{GP}}=32000$ and the results for the three initial load steps are shown in Fig.~\ref{fig:con1}a. 
The convergence rates are similar for all three load steps, but the accuracies differ. The relatively low accuracy of the first two increments is caused by the initial distance between supports that is below the equilibrium distance between beams. For these two load steps, the cross sections near the supports are in the repulsive regime of the LJ potential, which negatively impacts the accuracy due to the steep gradient there. Distributions of interaction force component $f_2$ for three values of $n_{\mathrm{GP}}$ are displayed in Fig.~\ref{fig:con1}b. The mentioned gradient is clearly depicted. Furthermore, it is seen that the adopted number of integration points $n_{\mathrm{GP}}=16000$ gives a very good solution.

\subsection{Integration of interaction forces}

\label{sectionIntegration}

The interaction force functions and their derivatives are difficult to integrate due to the oscillation of results w.r.t.~the number of integration points. After thorough numerical studies, we concluded that the mid-point rule is the most appropriate for the present analysis. In this subsection, we will elaborate on this issue.

\begin{figure}[!htb]
	\includegraphics[width=\textwidth]{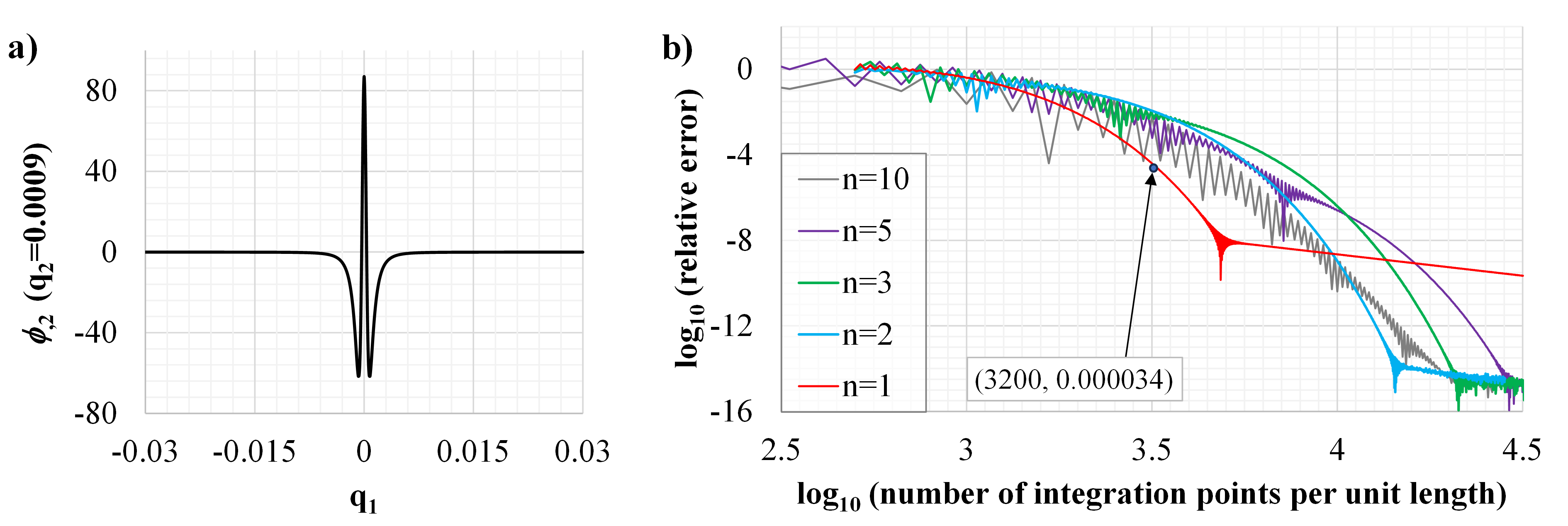}\centering
	\caption{a) Function $\phi_{,2}$ on the interval $q_1\in[-0.03,0.03]$ for fixed $q_2=0.0009$. b) Relative error of integration of function $\phi_{,2}$ on the interval [-0.03,0.03] for fixed $q_2=0.0009$ using five different integration orders $n$. The point (3200,0.000034) marks the adopted number of integration points and its error.}
	\label{fig:integration1}
\end{figure}

Let us consider an integration of the function $\phi_{,2}$ given with \eqqref{eq: ip15}. If we use the input data from Fig.~\ref{fig:setup} and fix the gap at $q_2=0.0009$, $\phi_{,2}$ becomes a univariate function of $q_1$ and its plot is given in Fig.~\ref{fig:integration1}a for the interval $q_1\in[-0.03,0.03]$. This interval approximates a realistic interval of integration for our studies with the adopted cutoff $c=2.5R=0.05$. 
This function is integrated using the Gauss integration rule with five different orders $n$ and the total number of integration points is varied. The relative error of such numerical integration w.r.t.~the exact analytical solution is plotted in Fig.~\ref{fig:integration1}b. Note that the oscillations of the graphs increase with the integration order. We attribute this behavior to the high values of the derivatives of the integrand which directly affect the error estimate. 

The present analysis suggests that the mid-point rule ($n=1$) is an optimal choice for the problem at hand. It is the most robust, gives small oscillations, and provides reasonable good accuracy for the ranges of integration points that are feasible for our calculations. Based on these considerations, we have chosen to model our beams with 3200 points per unit length, which returns an error of $0.0034\%$, see Fig.~\ref{fig:integration1}b. For the adopted finite element mesh, this corresponds to 100 integration points per element.

\begin{figure}[!htb]
	\includegraphics[width=\textwidth]{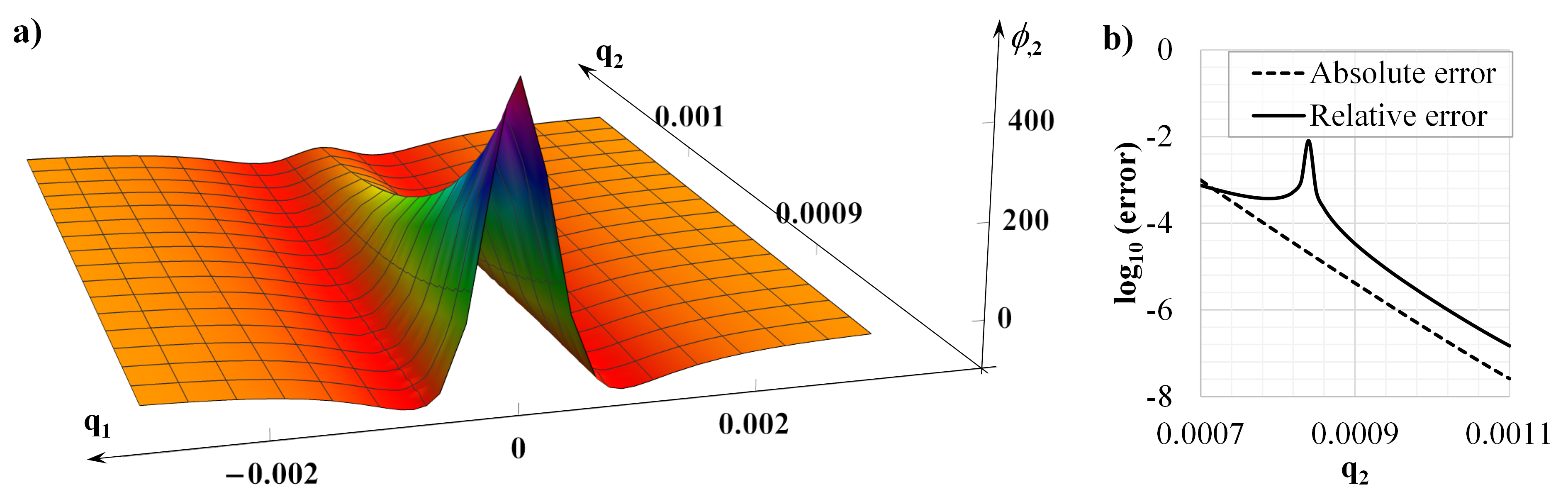}\centering
	\caption{a) 3D plot of function $\phi_{,2}$ over domain $q_1\in \left[-0.003,0.003\right]$ and $q_2\in  \left[0.0008,0.001\right]$. b) Error of integration of function $\phi_{,2}$ using the mid-point rule with 3200 integration points per unit length vs.~$q_2$.}
	\label{fig:integration2}
\end{figure}

It should be emphasized that this error of integration also depends on the value of the gap $q_2$. As the gap decreases, the derivatives of the observed function are increasing. This is shown in Fig.~\ref{fig:integration2}a where the function $\phi_{,2}(q_1,q_2)$ is plotted over a very narrow domain to clearly depict the described behavior.
The error of integration as a function of gap $q_2$ is shown in Fig.~\ref{fig:integration2}b. Both absolute and relative errors are plotted in order to clarify that the relative error measure increases near the equilibrium due to the division with near-zero values. The absolute error is consistent and it is evident that the error is reciprocal to the increase of gap $q_2$.

This analysis shows that we can reduce the number of integration points for large separations. On the other hand, the cutoff distance should be increased for large separations, see Fig.~\ref{fig:cutoff1}. This raises the question of how to gain an advantage in accuracy and efficiency for large separations. A promising approach is to implement an efficient adaptive integration method. We have conducted some preliminary work on this issue and it will be considered for future studies.

\section{Conclusions}

Short-range interactions due to intermolecular forces between in-plane fibers are studied here. A new law named \emph{improved section-section interaction potential} (ISSIP) is developed and verified. It is based on a coarse-graining procedure and existing section-section interaction potential law (LSSIP). The LSSIP is improved by refining the relative position of interacting sections w.r.t. local coordinate axes. Our analytical and numerical considerations show that the ISSIP provides improved accuracy in comparison with existing approaches.

Section-section interaction potential laws are important as a step towards the development of accurate and efficient computational models for the analysis of interactions between slender bodies at small scales. They stem from the basic ideas of beam theories -- assume rigid cross sections and preintegrate w.r.t.~the cross-sectional area.

Due to an explicit dependence on the distance components w.r.t.~local axes, the ISSIP allows an analytical estimation of the error introduced by the cutoff distance. Also, the nonlinear solver behaves well, and some difficulties described in the literature are avoided.

A section-beam interaction potential (SBIP) approach has been recently developed in \cite{2022grill,2023grill}. Due to the integration w.r.t.~only one beam, the method is more efficient than a section-section approach but has its own shortcomings. For example, the SBIP approach in \cite{2022grill} is always biased w.r.t.~the reference beam which can cause differences in the force response at the system level up to $1.5\%$.  Furthermore, the SBIP is based on the interaction between a section and a straight beam. Including the beam's curvature would be useful. To conclude, the investigation of both section-section and section-beam computational approaches is beneficial for the purpose of developing an optimal formulation. 

The ISSIP is tailored to small separations $q_2 \ll R$, where $q_2$ is the gap between cross sections and $R$ is the cross-sectional radius. On the other hand, the simple LSSIP law derived in \cite{2020grill} is suited for large separations $q_2 \gg R$. However, the transient range $q_2 \approx R$ has not been successfully tackled yet with a section-section approach and calls for further research. Future research should also consider beams with deformable cross sections, inertia, and topology changes.

\section*{Supplementary data}
\label{supsec}

The supplementary animations are curated in the online dataset with doi: \href{https://doi.org/10.3217/41tvr-6wr81}{10.3217/41tvr-6wr81}.

\section*{Acknowledgments}

	This research was funded by the Austrian Science Fund (FWF) P 36019-N. For the purpose of open access, the authors have applied a CC BY public copyright license to any Author Accepted Manuscript version arising from this submission.


\section*{Appendices}

\appendix

\section{Integration of a point-pair potential over two disks that belong to the same plane}
\label{appendix:a}

\setcounter{equation}{0}
\setcounter{figure}{0}
\renewcommand\theequation{A\arabic{equation}}
\renewcommand\thefigure{A\arabic{figure}} 

\begin{figure}[h]
	\includegraphics[width=11 cm]{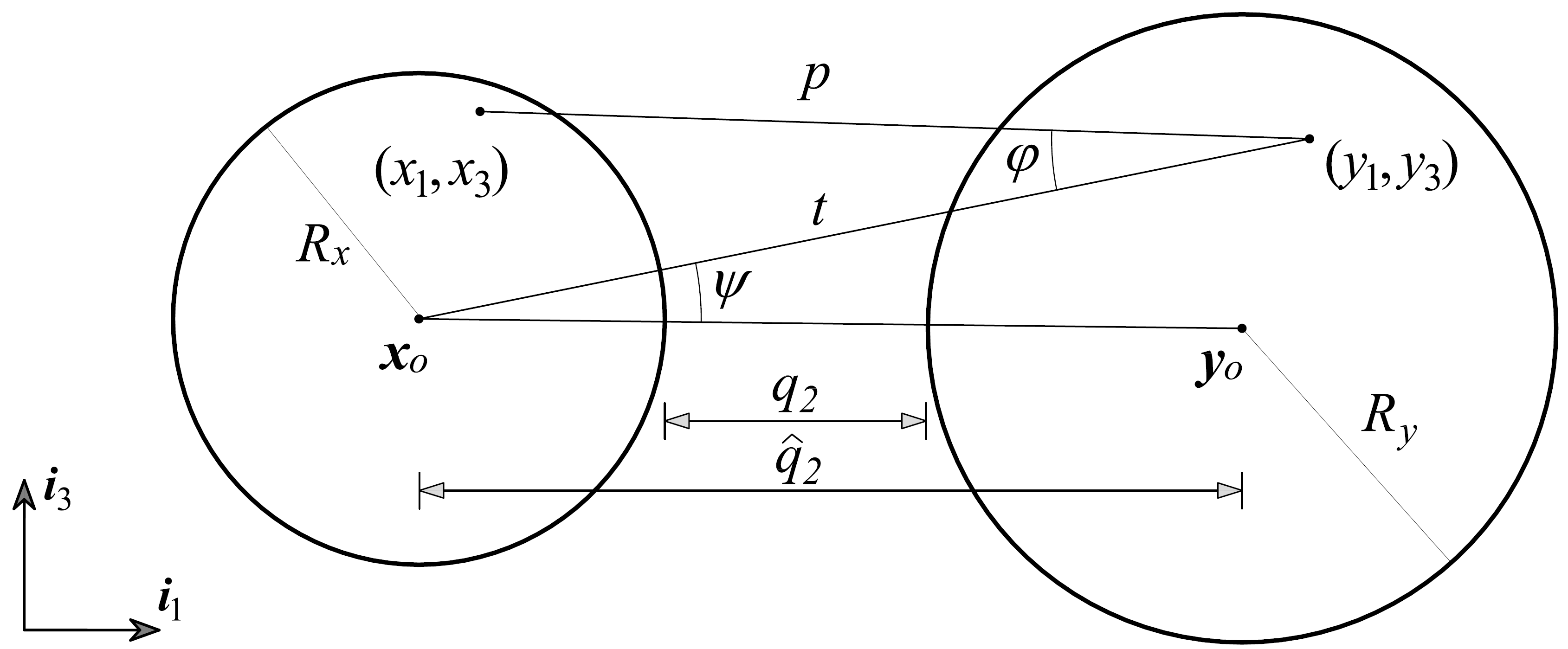}\centering
	\caption{Integration of a short-range interaction potential over two disks that lie in the same plane. Definition of the coordinate system introduced in \cite{1972langbein}.  }
	\label{fig:langcoor}
\end{figure}

In this Appendix, we will go through the integration procedure suggested in \cite{1972langbein} and take a closer look at each step. The aim is to integrate the point-pair interaction law $p^{-m}$ over two disks, $x_s$ and $y_s$, that lie in the same plane, cf.~Fig.~\ref{fig:langcoor}. 
The first step is to choose an appropriate coordinate system (CS). Starting from a Cartesian CS with unit vectors $\left(\ve{i}_1,\ve{i}_3\right)$, we can define the positions of two arbitrary interacting points with coordinates $\left(x_1,x_3\right)$ and $\left(y_1,y_3\right)$. Since the Cartesian CS does not allow an analytical integration, a new CS is utilized in \cite{1972langbein} from a geometrical viewpoint. Our aim is to explicitly define this new CS and to introduce an appropriate coordinate transformation.

The new CS consists of two polar CSs, the second one being relative to the first one. The first polar CS is defined with its origin at the center point of disk $x_s$, polar axis $t$, and polar angle $\psi$ that is measured w.r.t.~the line passing through the origins of the disks. This CS describes the position of the point $\left(y_1,y_3\right)$. The second, relative, polar CS is defined with its origin at $\left(y_1,y_3\right)$, polar axis $p$, and polar angle $\varphi$ that is measured w.r.t.~the axis $t$. This CS defines the relative position of $\left(x_1,x_3\right)$ w.r.t.~the point $\left(y_1,y_3\right)$.

By adopting the origin of the Cartesian CS at the center point of disk $x_s$, the coordinate transformation from the Cartesian $\left(x_1,x_3,y_1,y_3\right)$ to the new $\left(t,p,\psi,\varphi\right)$ CS is
\begin{equation}
	\label{eq: ip7711}
	\begin{aligned}
		y_1 \left(t,p,\psi,\varphi\right) &= t \cos \psi, \\
		y_3 \left(t,p,\psi,\varphi\right) &= t \sin \psi, \\
		x_1 \left(t,p,\psi,\varphi\right) &= p \cos \left(\psi-\varphi\right) + t \cos \psi, \\
		x_3 \left(t,p,\psi,\varphi\right) &= p \sin \left(\psi-\varphi\right) + t \sin \psi,
	\end{aligned}
\end{equation}
where the Jacobian of the transformation is $pt$. The main advantage of this approach is that the distance between point-pairs of interacting disks, $p$, is actually one coordinate of the new CS. The coordinate transformation \eqref{eq: ip7711} yields
\begin{equation}
	\label{eq: ip77carn}
	\begin{aligned}
		I_{\Phi_m} &= \int_{-R_x}^{Rx} \int_{-\sqrt{R_x^2-x_1^2}}^{\sqrt{R_x^2-x_1^2}} \int_{-R_y}^{Ry} \int_{-\sqrt{R_y^2-\left(\hat{q_2}-y_1\right)^2}}^{\sqrt{R_y^2-\left(\hat{q_2}-y_1\right)^2}} p^{-m}  \dd{y_3} \dd{y_1} \dd{x_3} \dd{x_1} \\
		&=   \int_{t_{min}}^{t_{max}} \int_{p_{min}\left(t\right)}^{p_{max}\left(t\right)} \int_{-\hat{\psi}\left(t\right)}^{\hat{\psi}\left(t\right)} \int_{-\hat{\varphi}\left(p,t\right)}^{\hat{\varphi}\left(p,t\right)} p^{-m} t\,p \dd{\varphi} \dd{\psi} \dd{p} \dd{t} ,
	\end{aligned}
\end{equation}
where the limits of integration w.r.t.~the new CS must be found. The limits of the polar angles are easily obtained from the cosine theorem, i.e.
\begin{equation}
	\label{eq: ip77car88a}
	\begin{aligned}
		\hat{\psi} &=\arccos \frac{t^2+\hat{q}_2^2-R_y^2}{2t\hat{q}_2} \quad \text{and} \quad \hat{\varphi}=\arccos \frac{t^2+p^2-R_x^2}{2tp}.
	\end{aligned}
\end{equation}
Since the coordinate $t$ is independent of $p$, its limits are fixed and follow from Fig.~\ref{fig:langcoor} as
\begin{equation}
	\label{eq: ip77car88ttnew1}
	\begin{aligned}
		q_2+R_x \le t \le q_2+R_x+2R_y.		
	\end{aligned}
\end{equation}
The limits of the relative coordinate $p$ are the functions of $t$
\begin{equation}
	\label{eq: ip77car88ttnew}
	\begin{aligned}
		t-R_x \le p \le t+R_x,	
	\end{aligned}
\end{equation}
which can be easily verified from geometrical considerations in Fig.~\ref{fig:langcoor}. Therefore, the final form of the integral to be solved is
\begin{equation}
	\label{eq: ip77car00}
	\begin{aligned}
		I_{\Phi_m} 
		&=   4 \int_{q_2+R_x}^{q_2+R_x+2R_y} \int_{t-R_x}^{t+R_x} 
		\arccos \frac{t^2+\hat{q}_2^2-R_y^2}{2t\hat{q}_2}
		\arccos \frac{t^2+p^2-R_x^2}{2tp}
		 p^{-m+1} t \dd{p} \dd{t}.
	\end{aligned}
\end{equation}

\begin{figure}[h]
	\includegraphics[width=11 cm]{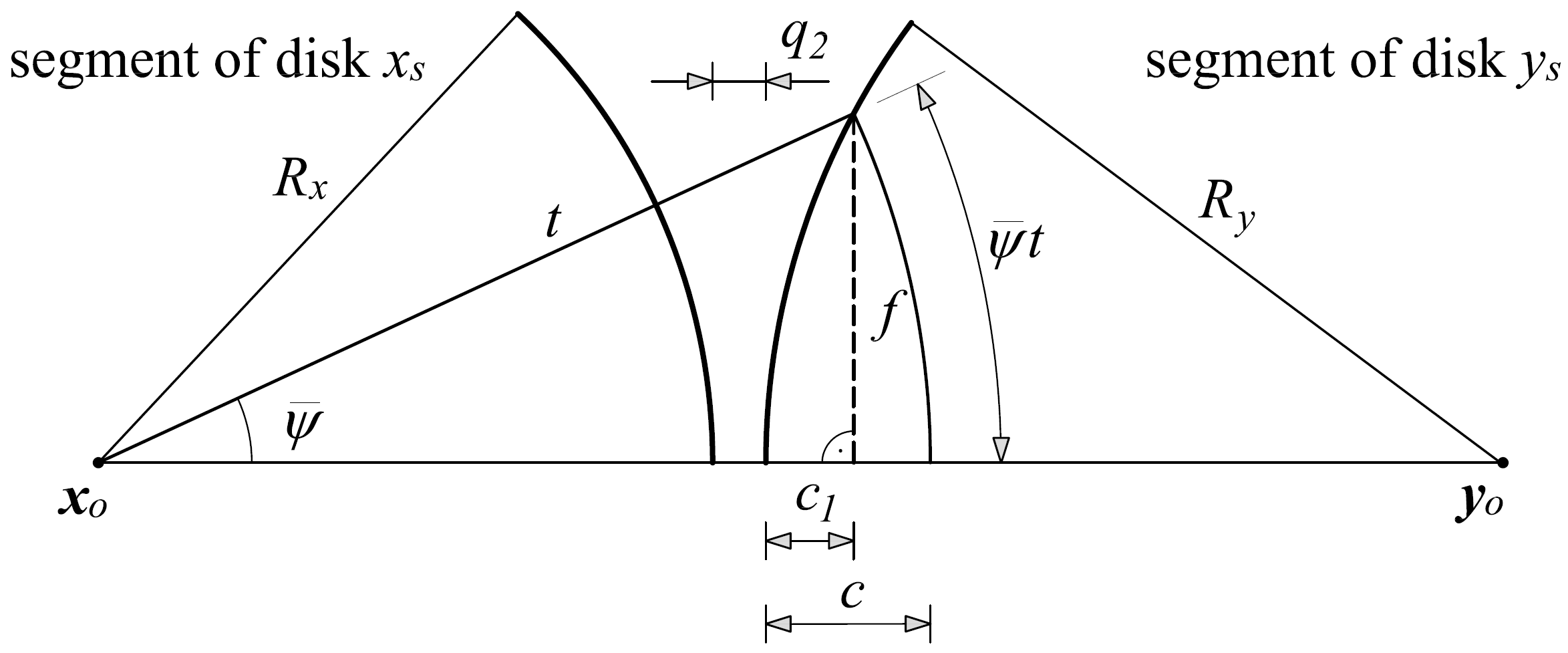}\centering
	\caption{Integration of a short-range interaction potential over two disks that lie in the same plane. For clarity, only segments of the disks are displayed. An arc length $\bar{\psi}t$ is approximated with length $f$. }
	\label{fig:psiapp}
\end{figure}

This form of integral is too complicated for an analytical integration and some approximations are required. Since we are dealing with short-ranged interactions at small separations, the major contribution to our potential comes from the closest point-pairs and quickly reduces with distance. Therefore, by considering only the closest interacting point-pairs, the first approximation is related to the limit angle $\hat{\psi}$. The cuts of interacting disks are shown in Fig.~\ref{fig:psiapp} and the idea is to approximate the arc length $\hat{\psi} t$ with the length $f$, which is a good approximation for the closest point-pairs.
 We can find $f$ as a function of $c$ from
\begin{equation}
	\label{eq: ip77caraaa}
	\begin{aligned}
		 f^2 = \left(R_x +q_2+c\right)^2 - \left(R_x+q_2+c_1\right)^2= R_y^2  - \left(R_y - c_1\right)^2 \implies c_1 = \frac{c^2 + 2c\left(R_x+q_2\right)}{2 \left(R_x+R_y+q_2\right)},
	\end{aligned}
\end{equation}
and then linearize $f^2$ w.r.t.~$c$, giving
\begin{equation}
	\label{eq: ip77caraaaff}
	\begin{aligned}
		f^2 = \frac{2 R_y c \left(R_x+q_2\right)}{R_x+R_y+q_2} +\mathcal{O}(c^2).
	\end{aligned}
\end{equation}
Finally, by letting $q_2 \rightarrow 0$, we obtain an approximation for $f$ and arc-length $t\bar{\psi}$,
\begin{equation}
	\label{eq: ip77caraaaffss}
	\begin{aligned}
		f^2 \approx  \frac{2 R_x R_y}{R_x + R_y} c \implies t\bar{\psi} \approx f \approx \sqrt{\frac{2 R_x R_y}{R_x + R_y} \left(t-R_x-q_2\right)}.
	\end{aligned}
\end{equation}

\begin{figure}[h]
	\includegraphics[width=11 cm]{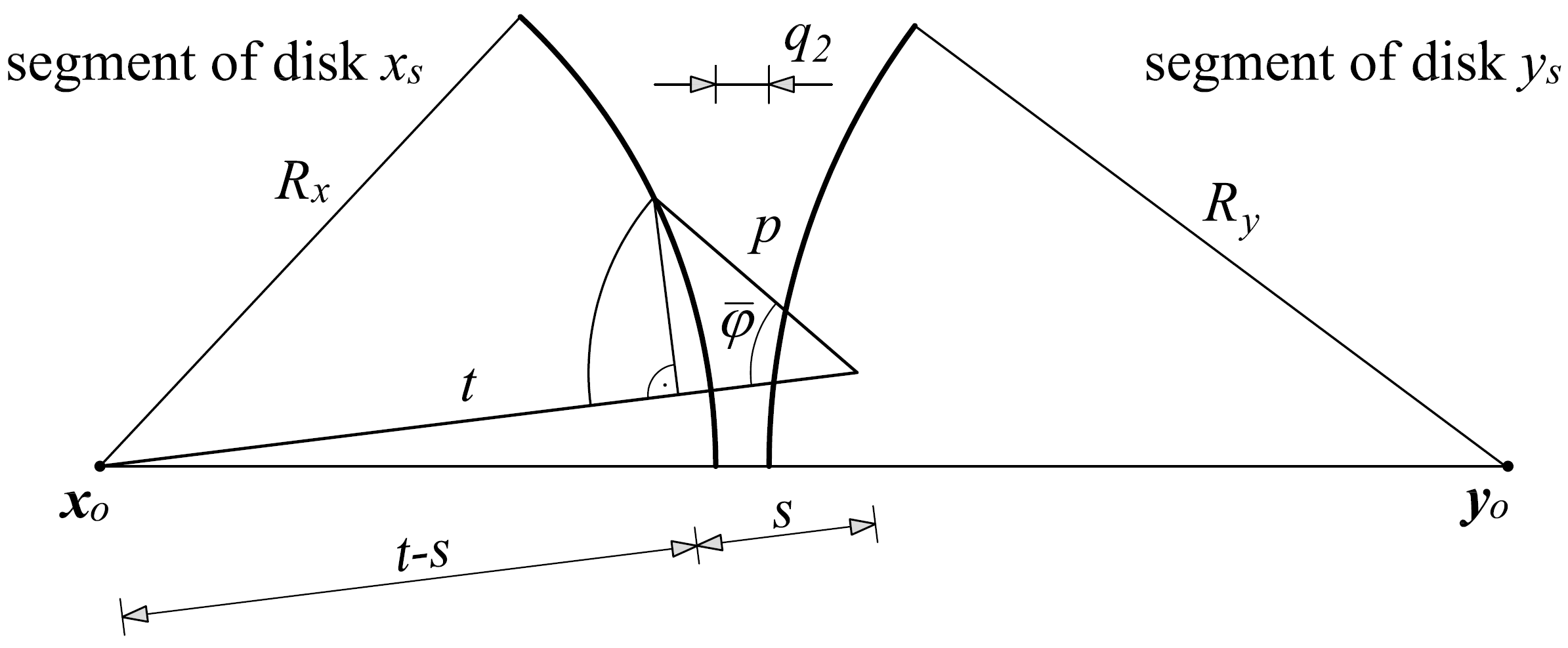}\centering
	\caption{Integration of a short-range interaction potential over two disks that lie in the same plane. For clarity, only segments of the disks are displayed. The length $s$ is approximated with $t-R_x$. }
	\label{fig:phiapp}
\end{figure}

Regarding the other limit angle $\bar{\varphi}$, its approximation is more straightforward. Again, the two cuts of interacting disks are shown in Fig.~\ref{fig:phiapp}.
For small values of $p$, we can approximate the cosine of the angle $\bar{\varphi}$ by assuming $s \approx t-R_x$, giving
\begin{equation}
	\label{eq: ip77car88ff}
	\begin{aligned}
	\cos \bar{\varphi} = \frac{t^2+p^2-R_x^2}{2tp} = \cos \frac{s}{p} \approx \cos \frac{t-R_x}{p}.
	\end{aligned}
\end{equation}
By inserting Eqs.~\eqref{eq: ip77caraaaffss} and \eqref{eq: ip77car88ff} into \eqqref{eq: ip77car00}, the integral is approximated as
\begin{equation}
	\label{eq: ip77xttapp}
	\begin{aligned}
		I_{\Phi_m} &=  4 \sqrt{\frac{2 R_x R_y}{R_x+R_y}}\int_{q_2+R_x}^{q_2+R_x+2R_y} \int_{t-Rx}^{t+R_x} p^{-m+1}   \arccos \frac{t-R_x}{p} \sqrt{t-R_x-q_2} \dd{p} \dd{t}.
	\end{aligned}
\end{equation}
The rest of the integration procedure is analogous to Subsection \ref{subexistingSSIP}. We introduce reduced variables $\hat{p}=p/q_2$ and $\hat{t} = \left(t-R_x\right)/q_2$, and assume $\left(t+R_x\right)/q_2=\infty$ and $\left(q_2+2R_y\right)/q_2=\infty$. The final form of the approximated integral is
\begin{equation}
	\label{eq: ip77xttapp99}
	\begin{aligned}
			I_{\Phi_m} &=  4 \sqrt{\frac{2 R_x R_y}{R_x+R_y}}  \int_{1}^{\infty} \int_{\hat{t}}^{\infty} \hat{p}^{-m+1} q_2^{-m+\frac{7}{2}} \arccos \frac{\hat{t}}{\hat{p}} \sqrt{\hat{t}  - 1} \dd{\hat{p}} \dd{\hat{t}},
	\end{aligned}
\end{equation}
and the closed-form result is
\begin{equation}
	\label{eq: ip13jkkap}
	\begin{aligned}
		I_{\Phi_{m}} =  2^{\frac{5}{2}-m} \pi^{\frac{3}{2}} \sqrt{\frac{R_x R_y}{R_x+R_y}} \frac{\Gamma \left(m-\frac{7}{2}\right)}{\Gamma \left(m/2\right)^2} \; \ii{q}{-m+\frac{7}{2}}{2}, \quad m > \frac{7}{2}.
	\end{aligned}
\end{equation}

The integration approach in \cite{1972langbein, 2020grill} is focused on a geometrical interpretation, rather than an explicit coordinate transformation. The adopted limits of integration are not defined consistently, and they return complex values for some ratios of $R_x$ and $R_y$. Nevertheless, by introducing the reduced variables $\hat{p}$ and $\hat{t}$, and by approximating the upper limits of integration with infinity, the final form of the integral \eqref{eq: ip77xttapp99} is the same.

\section{Gradient of the existing section-section interaction potential}

\label{appendix:b}

\setcounter{equation}{0}
\renewcommand\theequation{B\arabic{equation}}

Since the existing LSSIP \cite{2020grill} is used during the initial research, some basic details on this approach are given in this Appendix. The LSSIP is defined as
\begin{equation}
	\label{eq: ip021}
	\begin{aligned}
		\bar{\phi}_{m,\mathrm{ss}} &= \beta_{x} \, \beta_{y} \int_{A_x} \int_{A_y}  k_m \, r^{-m} \dd{A_y} \dd{A_x}
		= c_{m,\mathrm{ss}} q^{-m+\frac{7}{2}}_2,  \\
		c_{m,\mathrm{ss}} &= k_m \, \beta_{x} \, \beta_{y} \, 2^{\frac{5}{2}-m} \pi^{\frac{3}{2}} \sqrt{\frac{R_x R_y}{R_x+R_y}} \frac{\Gamma \left(m-\frac{7}{2}\right)}{\Gamma \left(m/2\right)^2}, \quad m>\frac{7}{2}, \\
		q_2&=\norm{\iv{x}{}{} - \iv{y}{}{}} -R_x - R_y.
	\end{aligned}
\end{equation}
The distance vector between axes, and its unit vector are
\begin{equation}
	\label{eq: ip031}	
	\ve{d} = \ve{x}-\ve{y}, \quad d=\norm{\ve{x} - \ve{y}}, \quad \hat{\ve{d}} = \frac{\ve{x} - \ve{y}}{\norm{\ve{x} - \ve{y}}}=\frac{\ve{d}}{d},
\end{equation}
and the gradients follow as
\begin{equation}
	\label{eqA: ip033}	
	\begin{aligned}
\nabla_x \hat{\ve{d}} &=\nabla_{\ve{x}} \hat{\ve{d}} = \frac{1}{d} \left(\ve{I} - \hat{\ve{d}} \otimes \hat{\ve{d}} \right), \quad 		\nabla_y \hat{\ve{d}} =  \nabla_{\ve{y}} \hat{\ve{d}} =-\frac{1}{d} \left(\ve{I} - \hat{\ve{d}} \otimes \hat{\ve{d}} \right), \\
\nabla_x \ve{d} &= \nabla_{\ve{x}} \ve{d}= \ve{I} , \quad \nabla_y \ve{d} = \nabla_{\ve{y}} \ve{d}= - \ve{I}. \\
	\end{aligned}
\end{equation}
The gradients of the gap w.r.t.~the positions of the interacting beams are
\begin{equation}
	\label{eq: ip033}	
	\begin{aligned}
		\nabla_{\ve{x}} q_2 &=\nabla_{\ve{x}} d= \nabla_{\ve{x}} \left(\iv{d}{}{} \cdot \iv{d}{}{}\right)^{1/2} = \frac{1}{2} \left(\iv{d}{}{} \cdot \iv{d}{}{}\right)^{-1/2} 2 \: \iv{d}{}{} = \hat{\ve{d}}, \\
		\nabla_{\ve{y}} q_2 &=\nabla_{\ve{y}} d= \nabla_{\ve{y}} \left(\iv{d}{}{} \cdot \iv{d}{}{}\right)^{1/2} = -\frac{1}{2} \left(\iv{d}{}{} \cdot \iv{d}{}{}\right)^{-1/2} 2 \: \iv{d}{}{} = -\hat{\ve{d}}. 
	\end{aligned}
\end{equation}
The section-section interaction forces are the gradients of the LSSIP. Since the LSSIP is expressed as a function of the gap $q_2$ only, the section-section interaction forces are
\begin{equation}
	\label{eq: ip04x}
	\begin{aligned}
		\iv{f}{}{x} &= \ve{f} =  -\nabla_{\ve{x}} \bar{\phi}_{m,\mathrm{ss}} =\frac{\partial \bar{\phi}_{m,\mathrm{ss}} \left(q\right)}{\partial q} \nabla_{\ve{x}} q_2= \left(m-\frac{7}{2}\right) c_{m,\mathrm{ss}}  q^{-m+\frac{5}{2}} \hat{\ve{d}}, \\ 
		\iv{f}{}{y} &= -\ve{f}=  -\nabla_{\ve{y}} \bar{\phi}_{m,\mathrm{ss}} =-\frac{\partial \bar{\phi}_{m,\mathrm{ss}} \left(q\right)}{\partial q} \nabla_{\ve{y}} q_2= -  \left(m-\frac{7}{2}\right) c_{m,\mathrm{ss}}  q^{-m+\frac{5}{2}} \hat{\ve{d}}. \\ 
	\end{aligned}
\end{equation}
Due to the symmetry of interacting sections w.r.t.~the distance vector $\ve{d}$, there is no resultant couple and the section-section forces act along the unit vector $\hat{\ve{d}}$.

\section{Detailed derivation of the straightforward approach}

\label{appendix:c}

\setcounter{equation}{0}
\renewcommand\theequation{C\arabic{equation}}

By adopting the LCS of beam $x$ as the reference LCS, $\left(\ve{t}_{\mathrm{ref}},\ve{n}_{\mathrm{ref}}\right)=\left(\ve{t}_x,\ve{n}_x\right)$, the gap and the offset are
\begin{equation}
	\label{eqB: g1}
	\begin{aligned}
		q_1 &= \ve{d}\cdot \ve{t}_x,   \\
		q_2 &= \abs{\ve{d}\cdot \ve{n}_x} - R_1 - R_2 , \quad \hat{q_2} = \abs{\ve{d}\cdot \ve{n}_{\ve{x}}} = q_2 + R_1 + R_2.
	\end{aligned}
\end{equation}
In order to find their gradients, the auxiliary relations
\begin{equation}
	\label{eqB: g2help}
	\begin{aligned}
		\nabla_x \ve{t}_x &= \nabla_{\iv{x}{}{,1}} \ve{t}_x =  \frac{1}{\sqrt{g_x}} \left(\ve{I} - \ve{t}_x \otimes \ve{t}_x \right) = \frac{1}{\sqrt{g_x}}  \ve{n}_x \otimes \ve{n}_x, \\
		\nabla_x \ve{n}_x &= \nabla_{\iv{x}{}{,1}}  \ve{n}_x = \nabla_{\iv{x}{}{,1}}   \left(\ve{\Lambda} \ve{t}_x \right) = \ve{\Lambda} \nabla_{\iv{x}{}{,1}}  \ve{t}_x = \frac{1}{\sqrt{g_x}} \ve{\Lambda} \left(\ve{n}_x \otimes \ve{n}_x\right), \\
		\nabla_{\iv{x}{}{,1}} \sqrt{g_x} &= \ve{t}_x, \quad
		\nabla_{\iv{x}{}{,1}} \frac{1}{\sqrt{g_x}} =-\frac{ 1}{2 g_x^{3/2}} 2 \iv{g}{}{1x} = - \frac{ 1}{g_x} \ve{t}_x,
	\end{aligned}
\end{equation}
are required. With these expressions at hand, we can find the gradients of the gap and the offset
\begin{equation}
	\label{eqB: g2}
	\begin{aligned}
		\nabla_{\iv{x}{}{}} q_1 &= -\nabla_{\iv{y}{}{}} q_1 =  \ve{t}_x = \nabla_{\iv{d}{}{}} q_1,\\
		\nabla_{\iv{x}{}{,1}} q_1 &= \frac{s_\alpha}{\sqrt{g_x}} \hat{q}_2  \ve{n}_x,\\
		\nabla_{\iv{y}{}{,1}} q_1 &=\ve{0}, \\
		\nabla_{\iv{x}{}{}} q_2 &=  s_\alpha \ve{n}_x = - \nabla_{\iv{y}{}{}} q_2 = 		\nabla_{\iv{d}{}{}} q_2,   \\
		\nabla_{\iv{x}{}{,1}} q_2 &=  -	\frac{s_\alpha }{\sqrt{g_x}} q_1 \ve{n}_x,\\
		\nabla_{\iv{y}{}{,1}} q_2 &= 0.
	\end{aligned}
\end{equation}
Now, the variation of the ISSIP w.r.t.~beam $x$ can be written as
\begin{equation}
	\label{eqB: full1}
	\begin{aligned}
		\delta_{x} \phi &= \phi_{,1} \left( \nabla_{\iv{x}{}{}} q_1 \cdot \delta \iv{u}{}{} + \nabla_{\iv{x}{}{,1}} q_1 \cdot \delta \iv{u}{}{,1} \right)  + \phi_{,2} \left( \nabla_{\iv{x}{}{}} q_2 \cdot \delta \iv{u}{}{}  +  \nabla_{\iv{x}{}{,1}} q_2 \cdot \delta \iv{u}{}{,1} \right) \\
		&=  \phi_{,1} \left( \ve{t}_x \cdot \delta \iv{u}{}{}  +   \frac{s_\alpha}{\sqrt{g_x}} \hat{q_2}  \ve{n}_x \cdot \delta \iv{u}{}{,1} \right) + \phi_{,2} s_\alpha  \left( \ve{n}_x \cdot \delta \iv{u}{}{} -	\frac{1 }{\sqrt{g_x}} q_1 \ve{n}_x \cdot \delta \iv{u}{}{,1} \right) \\
		&= \left(\phi_{,1}  \ve{t}_x + \phi_{,2} s_\alpha \ve{n}_x \right) \cdot \delta \iv{u}{}{} +	\frac{s_\alpha}{\sqrt{g_x}} \left(   \phi_{,1} \hat{q_2} - \phi_{,2} q_1 \right)  \ve{n}_x \cdot \delta \iv{u}{}{,1} = \ve{f} \cdot \delta \iv{u}{}{} +\hat{\ve{f}} \cdot \delta \iv{u}{}{,1}.
	\end{aligned}
\end{equation}

\section{Linearization of the variation of the interaction potential for the averaged LCS approach}

\label{appendix:d}
\setcounter{equation}{0}
\renewcommand\theequation{D\arabic{equation}}

The equilibrium equation \eqref{eq: ip14xtt} is highly nonlinear both due to the large deformations and the interaction potential that is configuration-dependent. Since there is no coupling between the interacting beams in the strain energy and the external potential, they can be simply derived as in reference \cite{2023borkoviće}. Let us focus on the interaction term $\delta \Phi_{\mathrm{IP}}$. The calculation of this term requires the integration and the spatial discretization of $\delta \phi$. This is the standard procedure of the finite element method and we will skip these steps for brevity. Finally, the solution of the obtained nonlinear equation requires linearization. Therefore, the main steps in the linearization process of $\delta \phi$ in the continuum setting are presented in this Appendix.

For the averaged LCS approach we adopt $\left(\ve{t}_{\mathrm{ref}},\ve{n}_{\mathrm{ref}}\right)=\left(\hat{\ve{t}}_{xy},\hat{\ve{n}}_{xy}\right)$, where the basis vectors are defined in \eqqref{eq: g1x2}. The gradients of the gap and the offset w.r.t.~the positions are given in \eqqref{eq: full3x66}. For the linearization, we need to also find the gradients of the gap and the offset w.r.t.~the tangent vectors. Let us first define the gradients of magnitude $t_{xy}=\hat{t}_{xy} \cdot \hat{t}_{xy}$ and its square root by
\begin{equation}
	\label{eqC: ip17x}
	\begin{aligned}
		\nabla_x t_{xy} &= 2 \nabla_x \ve{t}_x \left(\ve{t}_x+\ve{t}_y\right) = 2 \nabla_x \ve{t}_x \cdot \ve{t}_y = 2 \frac{1}{\sqrt{g_x}} \left(\ve{t}_y - \left(\ve{t}_x \cdot \ve{t}_y\right) \ve{t}_x \right),\\
		\nabla_x \sqrt{t_{xy}} &= \frac{1}{2 \sqrt{t_{xy}}} 2  \nabla_x \ve{t}_x \cdot \ve{t}_y = \frac{1}{ \sqrt{t_{xy}}}  \nabla_x \ve{t}_x \cdot \ve{t}_y  =\frac{1}{ \sqrt{t_{xy}}} \frac{1}{\sqrt{g_x}}  \left(\ve{t}_y - \left(\ve{t}_x \cdot \ve{t}_y\right) \ve{t}_x \right),\\
		\nabla_y \sqrt{t_{xy}} &=  \frac{1}{2 \sqrt{t_{xy}}} 2  \nabla_y \ve{t}_y \cdot \ve{t}_x = \frac{1}{ \sqrt{t_{xy}}}  \nabla_y \ve{t}_y \cdot \ve{t}_x =\frac{1}{ \sqrt{t_{xy}}} \frac{1}{\sqrt{g_y}}  \left(\ve{t}_x - \left(\ve{t}_x \cdot \ve{t}_y\right) \ve{t}_y \right).
	\end{aligned}
\end{equation}
Now, the gradients of the basis vectors are
\begin{equation}
	\label{eqC: ip17x88}
	\begin{aligned}
		\nabla_x \hat{\ve{t}}_{xy} &=\frac{\sqrt{t_{xy}} \nabla_x \ve{t}_x - \ve{t}_{xy} \otimes \nabla_x \sqrt{t_{xy}}}{t_{xy}} = \frac{\sqrt{t_{xy}} \nabla_x \ve{t}_x - \ve{t}_{xy} \otimes \frac{1}{ \sqrt{t_{xy}}}  \nabla_x \ve{t}_x \cdot \ve{t}_y}{t_{xy}}\\
		&= \frac{\sqrt{t_{xy}} \nabla_x \ve{t}_x - \hat{\ve{t}}_{xy} \otimes   \ve{t}_y \cdot  \nabla_x \ve{t}_x }{t_{xy}} = \frac{  1 }{t_{xy}} \left( \sqrt{t_{xy}} \ve{I} - \hat{\ve{t}}_{xy} \otimes   \ve{t}_y \right)  \nabla_x \ve{t}_x,\\
		\nabla_y \hat{\ve{t}}_{xy} &=\frac{  1 }{t_{xy}} \left( \sqrt{t_{xy}} \ve{I} - \hat{\ve{t}}_{xy} \otimes   \ve{t}_x \right)  \nabla_y \ve{t}_y,\\
		\nabla_x \hat{\ve{n}}_{xy} &= \nabla_x \left( \pmb{\Lambda} \hat{\ve{t}}_{xy} \right) = \pmb{\Lambda}\nabla_x \hat{\ve{t}}_{xy}, \\
		\nabla_y \hat{\ve{n}}_{xy} &= \nabla_y \left( \pmb{\Lambda} \hat{\ve{t}}_{xy} \right) = \pmb{\Lambda}\nabla_y \hat{\ve{t}}_{xy}. \\
	\end{aligned}
\end{equation}
The gap and the offset are defined in \eqqref{eq: g1x}
and their gradients w.r.t.~the tangent vectors follow as
\begin{equation}
	\label{eqC: ipxxf}
	\begin{aligned}
		\nabla_{\iv{x}{}{,1}} q_1 &=  \nabla_{\iv{x}{}{,1}} \left( \ve d \cdot \hat{\ve{t}}_{xy}\right) =   \left(\nabla_{\iv{x}{}{,1}}  \hat{\ve{t}}_{xy}\right)^T \ve{d} = \nabla_x \ve{t}_x \frac{1}{t_{xy}} \left(\sqrt{t_{xy}} \ve{d} - q_1  \ve{t}_y\right),\\
		\nabla_{\iv{y}{}{,1}} q_1 &=  \nabla_{\iv{y}{}{,1}}  \left(\ve d \cdot \hat{\ve{t}}_{xy}\right) =   \left(\nabla_{\iv{y}{}{,1}}  \hat{\ve{t}}_{xy}\right)^T \ve{d} =  \nabla_y \ve{t}_y \frac{1}{t_{xy}} \left(\sqrt{t_{xy}} \ve{d} - q_1  \ve{t}_x\right),\\
		\nabla_{\iv{x}{}{,1}} q_2 &=\nabla_{\iv{x}{}{,1}}   \abs{\ve d \cdot \hat{\ve{n}}_{xy}} = s_{\alpha}  \left(\nabla_{\iv{x}{}{,1}} \hat{\ve{n}}_{xy}\right)^T \ve{d}  =\nabla_x \ve{t}_x \frac{1}{t_{xy}}\left( s_{\alpha} \sqrt{t_{xy}} \pmb{\Lambda}^T \ve{d} -  \hat{q}_2 \ve{t}_y\right),\\ 
		\nabla_{\iv{y}{}{,1}} q_2 &=\nabla_{\iv{y}{}{,1}}   \abs{\ve d \cdot \hat{\ve{n}}_{xy}} = s_{\alpha}  \left(\nabla_{\iv{y}{}{,1}} \hat{\ve{n}}_{xy}\right)^T \ve{d}    = \nabla_y \ve{t}_y \frac{1}{t_{xy}}\left( s_{\alpha} \sqrt{t_{xy}} \pmb{\Lambda}^T \ve{d} -  \hat{q}_2 \ve{t}_x\right).\\
	\end{aligned}
\end{equation}
Now, we have all the relations needed for the linearization. Since the ISSIP is a function of positions of both beams, we need to linearize both $\delta \phi_x$ and $\delta \phi_y$. The linearized increment of the section-section interaction force \eqref{eq: full3xff} w.r.t.~beam $x$ is
\begin{equation}
	\label{eq: full3vfxuu}
	\begin{aligned}
		\Delta_x  \ve{f} &= \hat{\ve{t}}_{xy} \otimes \left(\phi_{,11} \Delta_x q_1 +\phi_{,12} \Delta_x q_2  \right) + s_\alpha \hat{\ve{n}}_{xy} \otimes \left(\phi_{,21} \Delta_x q_1   + \phi_{,22}  \Delta_x q_2 \right) + \left(\phi_{,1} \ve{I}   + \phi_{,2} s_\alpha \pmb{\Lambda} \right) \Delta_x \hat{\ve{t}}_{xy} \\
		&= \left(\phi_{,11} \hat{\ve{t}}_{xy} + s_\alpha \phi_{,21} \hat{\ve{n}}_{xy} \right) \otimes \left( \Delta_x q_1  \right) + \left(\phi_{,12} \hat{\ve{t}}_{xy} + s_\alpha \phi_{,22} \hat{\ve{n}}_{xy} \right) \otimes \left( \Delta_x q_2  \right)  + \left(\phi_{,1} \ve{I}   + \phi_{,2} s_\alpha \pmb{\Lambda} \right) \Delta_x \hat{\ve{t}}_{xy}\\
		&= \left(\phi_{,11} \hat{\ve{t}}_{xy} + s_\alpha \phi_{,21} \hat{\ve{n}}_{xy} \right) \otimes \left( \nabla_{\ve{x}} q_1 \Delta \ve{u} + \nabla_{\iv{x}{}{,1}} q_1 \Delta \ve{u}_{,1} \right) \\
		&+ \left(\phi_{,12} \hat{\ve{t}}_{xy} + s_\alpha \phi_{,22} \hat{\ve{n}}_{xy} \right) \otimes \left( \nabla_{\ve{x}} q_2 \Delta \ve{u} + \nabla_{\iv{x}{}{,1}} q_2 \Delta \ve{u}_{,1} \right)  \\
		&+ \left(\phi_{,1} \ve{I}   + \phi_{,2} s_\alpha \pmb{\Lambda} \right)  \nabla_{\iv{x}{}{,1}} \hat{\ve{t}}_{xy} \Delta \iv{u}{}{,1} \\
		&= \left[ \left(\phi_{,11} \hat{\ve{t}}_{xy} + s_\alpha \phi_{,21} \hat{\ve{n}}_{xy} \right) \otimes  \hat{\ve{t}}_{xy} + \left(\phi_{,12} \hat{\ve{t}}_{xy} + s_\alpha \phi_{,22} \hat{\ve{n}}_{xy} \right) \otimes s_{\alpha} \hat{\ve{n}}_{xy}\right] \Delta \ve{u} \\
		&+  \left[ \left(\phi_{,11} \hat{\ve{t}}_{xy} + s_\alpha \phi_{,21} \hat{\ve{n}}_{xy} \right) \otimes   \nabla_{\iv{x}{}{,1}} q_1 + \left(\phi_{,12} \hat{\ve{t}}_{xy} + s_\alpha \phi_{,22} \hat{\ve{n}}_{xy} \right) \otimes  \nabla_{\iv{x}{}{,1}} q_2\right] \Delta \iv{u}{}{,1} \\  
		&+ \left(\phi_{,1} \ve{I}   + \phi_{,2} s_\alpha \pmb{\Lambda} \right)  \nabla_{\iv{x}{}{,1}} \hat{\ve{t}}_{xy} \Delta \iv{u}{}{,1} \\
		& = D_{\ve{x}}\ve{f}  \Delta \ve{u} + D_{\iv{x}{}{,1}}\ve{f} \Delta \iv{u}{}{,1},
	\end{aligned}
\end{equation}
where
\begin{equation}
	\label{eq: full3vfxuu11}
	\begin{aligned}
		D_{\ve{x}}\ve{f} & =  \ve{a}_1 \otimes  \hat{\ve{t}}_{xy} + \ve{a}_2 \otimes s_{\alpha} \hat{\ve{n}}_{xy}, \\
		D_{\iv{x}{}{,1}}\ve{f} &=  \ve{a}_1 \otimes   \nabla_{\iv{x}{}{,1}} q_1 + \ve{a}_2 \otimes  \nabla_{\iv{x}{}{,1}} q_2 +  \ve{A} \nabla_{\iv{x}{}{,1}} \hat{\ve{t}}_{xy}, \\
		\ve{a}_1 &= \left(\phi_{,11} \hat{\ve{t}}_{xy} + s_\alpha \phi_{,21} \hat{\ve{n}}_{xy} \right), \\
		\ve{a}_2 &= \left(\phi_{,12} \hat{\ve{t}}_{xy} + s_\alpha \phi_{,22} \hat{\ve{n}}_{xy} \right),\\
		\ve{A} &= \left(\phi_{,1} \ve{I}   + \phi_{,2} s_\alpha \pmb{\Lambda} \right).
	\end{aligned}
\end{equation}
Linearized increment of the interaction force w.r.t.~beam $y$ is then
\begin{equation}
	\label{eq: full3vfxvv}
	\begin{aligned}
		\Delta_y  \ve{f} & = - D_{\ve{x}}\ve{f}  \Delta \ve{v} + D_{\iv{y}{}{,1}}\ve{f} \Delta \iv{v}{}{,1}, \\
		D_{\iv{y}{}{,1}}\ve{f} &=  \ve{a}_1 \otimes   \nabla_{\iv{y}{}{,1}} q_1 + \ve{a}_2 \otimes  \nabla_{\iv{y}{}{,1}} q_2 +  \ve{A} \nabla_{\iv{y}{}{,1}} \hat{\ve{t}}_{xy}.
	\end{aligned}
\end{equation}
With these expressions, the linearized increments of $\delta \phi_x$ and $\delta \phi_y$ can be written as
\begin{equation}
	\label{eq: full3vfx}
	\begin{aligned}
		\Delta \delta_{x} \phi &= \Delta \ve{f} \cdot \delta \iv{u}{}{} = \left(\Delta_x \ve{f} + \Delta_y \ve{f}\right) \cdot \delta \iv{u}{}{} \\
		&= \delta \ve{u} \cdot \left(D_{\ve{x}} \ve{f} \cdot \Delta \iv{u}{}{} + D_{\iv{x}{}{,1}} \ve{f} \cdot \Delta \iv{u}{}{,1} + D_{\ve{y}} \ve{f} \cdot \Delta \iv{v}{}{}+ D_{\iv{y}{}{,1}} \ve{f} \cdot \Delta \iv{v}{}{,1}\right),\\
		\Delta \delta_{y} \phi &= - \Delta \ve{f} \cdot \delta \iv{v}{}{}\\
		&= -\delta \ve{v} \cdot \left(D_{\ve{x}} \ve{f} \cdot \Delta \iv{u}{}{} + D_{\iv{x}{}{,1}} \ve{f} \cdot \Delta \iv{u}{}{,1} + D_{\ve{y}} \ve{f} \cdot \Delta \iv{v}{}{}+ D_{\iv{y}{}{,1}} \ve{f} \cdot \Delta \iv{v}{}{,1} \right),
	\end{aligned}
\end{equation}
or in the matrix form,
\begin{equation}
	\begin{aligned}
		\label{eq:matlin1xy}
		\Delta \delta \phi &= \Delta \left( \begin{bmatrix}
			\delta	 \iv{u}{}{} & \delta \iv{v}{}{} 
		\end{bmatrix} 
		\begin{bmatrix}
			\ve{f}_x 	\\
			\ve{f}_y 
		\end{bmatrix}
		\right) = 
		\begin{bmatrix}
			\delta	 \iv{u}{}{} & \delta \iv{v}{}{} 
		\end{bmatrix}  \Delta \begin{bmatrix}
			\ve{f}	\\
			-\ve{f} 
		\end{bmatrix} \\
		&= \begin{bmatrix}
			\delta	 \iv{u}{}{} & \delta \iv{v}{}{} 
		\end{bmatrix}  \left( \begin{bmatrix}
			D_{\ve{x}} \ve{f} & D_{\ve{y}} \ve{f}	\\
			-D_{\ve{x}} \ve{f} & -D_{\ve{y}} \ve{f}
		\end{bmatrix}
		\begin{bmatrix}
			\Delta \iv{u}{}{} 	\\
			\Delta \iv{v}{}{}  
		\end{bmatrix}+\begin{bmatrix}
			D_{\iv{x}{}{,1}} \ve{f} & D_{\iv{y}{}{,1}}  \ve{f}	\\
			-D_{\iv{x}{}{,1}}  \ve{f} & -D_{\iv{y}{}{,1}}  \ve{f}
		\end{bmatrix}
		\begin{bmatrix}
			\Delta \iv{u}{}{,1} 	\\
			\Delta \iv{v}{}{,1}  
		\end{bmatrix} \right).
	\end{aligned}
\end{equation}
Evidently, the resulting stiffness matrix is not symmetric. This is a consequence of neglecting the interaction moments.
The linearization of the interaction moments for the averaged LCS approach is quite complicated and we will not pursue it here. Nevertheless, our numerical simulations show that the derived tangent operator yields good convergence rates, even for the formulation that includes the interaction moments. This finding actually suggests that the influence of the interaction moment is not significant in the overall response, as already noted in Subsection \ref{verif}.

\section{Tangent stiffness accuracy test}

\label{appendix:e}
\setcounter{equation}{0}
\renewcommand\theequation{E\arabic{equation}}

An accurate tangent stiffness is crucial for the efficient calculation of nonlinear problems. Therefore, it is highly desirable to have a reliable testing procedure. Here, we have tested the accuracy of the tangent stiffness by using the complex-step method \cite{2023gangl}. It is briefly explained here for the consistency of presentation and as a reference.

The residual $\pmb{\Psi}$ is a function of the vector $\ven{q}{}{}$, which is stacked with all the degrees of freedom. A second-order approximation of the residual in some unknown configuration is
\begin{equation}
	\label{eq: compltay}
	\begin{aligned}
		\pmb{\Psi} \left(\ivpre{q}{}{} + \Delta \ven{q}{}{}\right) \approx \pmb{\Psi} \left(\ivpre{q}{}{} \right) + \left(\nabla_{\ven{q}{}{}} \pmb{\Psi}\right)^\sharp \Delta \ven{q} + \frac{1}{2} \left(\nabla_{\ven{q}{}{}}^2 \pmb{\Psi}\right)^\sharp \left(\Delta \ven{q} \right)^2,
	\end{aligned}
\end{equation}
where $\sharp$ designates the previously calculated configuration. Let us introduce an incremental vector of unknowns, where all but the $k^{th}$ element are zero,
\begin{equation}
	\label{eq: compltay1}
	\Delta \iveq{q}{}{k} = 
	\begin{bmatrix}
		0 & 0 & ... & 1 & ...& 0
	\end{bmatrix}.
\end{equation}
Furthermore, let us define an infinitesimal positive quantity $\epsilon: 0 < \epsilon \ll 1$. Now, we can find the second-order approximation of the residual in a new configuration, incremented by $\Delta \iveq{q}{}{k}\epsilon i$, where $i$ is the imaginary unit. i.e.
\begin{equation}
	\label{eq: compltay2}
	\begin{aligned}
		\pmb{\Psi} \left(\ivpre{q}{}{} + \Delta \iveq{q}{}{k}\epsilon i\right) \approx \pmb{\Psi} \left(\ivpre{q}{}{} \right) + \left(\nabla_{\ven{q}{}{}} \pmb{\Psi}\right)^\sharp \Delta \iveq{q}{}{k} \epsilon i + \frac{1}{2} \left(\nabla_{\ven{q}{}{}}^2 \pmb{\Psi}\right)^\sharp \left(\Delta \iveq{q}{}{k} \epsilon i \right)^2.
	\end{aligned}
\end{equation}
If we take the imaginary part of this expression,
\begin{equation}
	\label{eq: compltay3}
	\begin{aligned}
		\Im \left[\pmb{\Psi} \left(\ivpre{q}{}{} + \Delta \iveq{q}{}{k}\epsilon i\right) \right] \approx  \left(\nabla_{\ven{q}{}{}} \pmb{\Psi}\right)^\sharp \Delta \iveq{q}{}{k} \epsilon, 
	\end{aligned}
\end{equation}
and divide the equation with $\epsilon$,
\begin{equation}
	\label{eq: compltay4}
	\begin{aligned}
		\frac{1}{\epsilon} \Im \left[\pmb{\Psi} \left(\ivpre{q}{}{} + \Delta \iveq{q}{}{k}\epsilon i\right) \right] \approx  \ven{K}_{T} \Delta \iveq{q}{}{k} , \quad \textrm{where} \quad \ven{K}_{T} = \left(\nabla_{\ven{q}{}{}} \pmb{\Psi}\right)^\sharp,
	\end{aligned}
\end{equation}
we obtain a second-order accurate approximation of one column of stiffness matrix $\ven{K}_{T}$. Namely, the left-hand side of this expression is equal to the $k^{th}$ column of the tangent stiffness matrix.

This is a straightforward (but expensive) way to test the accuracy of the stiffness matrix. Numerical experiments show that our linearization is accurate.

\textbf{Remark} If the residual vector includes operations such as \emph{Norm}, \emph{Sign}, and \emph{Abs}, special care is required due to the complex arguments in \eqqref{eq: compltay2}.


\bibliography{papersIP} 
\bibliographystyle{ieeetr}


\end{document}